\documentclass[10pt,aps,prd,twocolumn,superscriptaddress,preprintnumbers,floatfix,notitlepage]{revtex4-1}

\usepackage[justification=raggedright,singlelinecheck=false]{caption}

\usepackage{hyperref}
\hypersetup{
    colorlinks=true,
    linkcolor=blue,
    filecolor=magenta,      
    urlcolor=blue,
}
\usepackage{amsmath,amsfonts,amssymb}
\usepackage{fancyhdr}
\usepackage{graphicx}
\usepackage{xspace}
\usepackage{rotating}
\usepackage[normalem]{ulem}
\usepackage{braket}
\usepackage{verbatim}
\usepackage{xcolor}
\usepackage{hyperref}
\usepackage[utf8]{inputenc}
\usepackage{slashed}
\usepackage{subcaption}
\usepackage{hyperref}

\usepackage{tikz}
\usetikzlibrary{arrows.meta,calc,decorations.pathmorphing}

\def\lsim{\mathrel{\raise.3ex\hbox{$<$\kern-.75em\lower1ex\hbox{$\sim$}}}}
\def\gsim{\mathrel{\raise.3ex\hbox{$>$\kern-.75em\lower1ex\hbox{$\sim$}}}}

\newcommand{\be}{\begin{equation}}
\newcommand{\ee}{\end{equation}}
\newcommand{\bea}{\begin{equation}\begin{aligned}}
\newcommand{\eea}{\end{aligned}\end{equation}}

\usepackage{tikz}
\usepackage{tikz-feynman}
\tikzfeynmanset{compat=1.1.0}

\begin{document}


\title{Axion-Induced Casimir Interaction Between Graphene Plates}

\author{Ahmad Alachkar}
\affiliation{Institut de Physique Théorique, Université Paris-Saclay,CEA, CNRS, F-91191 Gif-sur-Yvette Cedex, France}


\author{Philippe Brax}
\affiliation{Institut de Physique Théorique, Université Paris-Saclay,CEA, CNRS, F-91191 Gif-sur-Yvette Cedex, France}

\author{Pierre Brun}

\affiliation{Irfu/Département de Physique des Particules, Université Paris-Saclay, \\ CEA, F-91191 Gif-sur-Yvette Cedex, France}

\begin{abstract}

Axion dark matter may induce observable electromagnetic effects in resonant cavity systems, and potentially lead to modifications of the Casimir interaction.
In this context, graphene represents a particularly attractive platform owing to its distinctive and tunable electromagnetic properties, and the fact that its electromagnetic response can be modelled microscopically from first principles within quantum field theory.
The electromagnetic response induced by axion dark matter is investigated in a planar cavity geometry consisting of parallel graphene interfaces in the presence of a homogeneous external magnetic field, incorporating finite temperature, chemical potential and dissipation effects through the graphene conductivity. Closed analytical expressions are obtained for the induced electric field and the resulting pressure exerted on the interfaces. The pressure exhibits resonant enhancement, at a series of plate separations satisfying $
d_n=\frac{2\pi n-\phi(r)}{m_a}$, where $m_a$ is the axion mass and the phase $\phi(r)$ is determined by the reflection coefficient $r$, which depends on the graphene conductivity evaluated at the axion frequency $\omega=m_a$. The resonant structure is strongly influenced by the graphene chemical potential and damping parameter. In particular, increased doping, for example via a gate voltage, sharpens the resonances and amplifies the axion-induced signal.
By comparing the resonantly enhanced signal with the conventional Casimir background, the parametric regimes in which the effect could become experimentally relevant are identified, with the strongest sensitivity obtained for highly doped low-dissipation graphene configurations operated near resonance.
These results demonstrate that graphene-based Casimir-type configurations may provide a sensitive framework for probing axion-induced electromagnetic phenomena and highlight the interplay between axion electrodynamics, cavity resonances, and material properties in low-dimensional systems.
\end{abstract}

\maketitle

\section{INTRODUCTION}

The existence of dark matter \cite{ParticleDataGroup:2022pth} is supported by a wide range of observations, from galaxy rotation curves and gravitational lensing to the dynamics of galaxy clusters and measurements of the cosmic microwave background. Despite this compelling evidence, the nature of dark matter remains unknown, with viable candidates spanning many orders of magnitude in mass. Among the proposed candidates, axions \cite{preskill1983cosmology,abbott1983cosmological,dine1983not,weinberg1978new,wilczek1978problem} are particularly well motivated. Originally introduced to solve the strong CP problem \cite{peccei2008strong,peccei1977constraints}, the QCD axion provides a predictive framework in which the particle mass and coupling are related. More generally, string theory and other ultraviolet completions generically predict a plethora of light pseudoscalar fields \cite{svrcek2006axions}, giving rise to the so-called axiverse \cite{arvanitaki2010string}. Like the QCD axion, these axion-like particles (ALPs) couple weakly to electromagnetism, providing a basis for experimental searches, but need not obey the mass-coupling relation of the QCD axion. Unless otherwise specified, we use the term \emph{axion} throughout this work as a generic shorthand encompassing both the QCD axion and ALPs.

Beyond traditional experimental searches, one may ask whether precision measurements of electromagnetic phenomena may offer sensitivity to such couplings. In particular, the Casimir effect—arising from quantum vacuum fluctuations of the electromagnetic field—may provide a sensitive probe of modifications to the electromagnetic sector. Since the Casimir force depends on the electromagnetic response of the system and the boundary conditions, any new physics that modifies the electromagnetic sector may, in principle, affect it. In this sense, it may provide sensitivity to particles that couple to photons.

In this work, we investigate the possibility of detecting axion dark matter through its coupling to electromagnetism in a magnetised Casimir setup. In the presence of an external magnetic field, the axion field induces electromagnetic currents, leading to a modification of the pressure between two closely spaced surfaces. We focus on graphene, whose tunable electronic properties and well-understood electromagnetic response make it an attractive platform for such studies.
Its peculiar electronic structure leads to unusual optical and electrodynamic properties. 
In particular, graphene is characterised by a universal conductivity of order $e^2/\hbar$, and its interaction with electromagnetic fields can be captured by the polarisation tensor of the associated Dirac quasiparticles. 
The response depends on several physical parameters, including the temperature $T$, the chemical potential $\mu$ (which can be tuned experimentally through doping or gating), and a possible mass gap $\Delta$ arising from symmetry breaking or substrate effects.
These properties make graphene an especially attractive system for studying Casimir and related fluctuation-induced phenomena, such as Casimir–Polder and van der Waals interactions \cite{bordag2015advances,klimchitskaya2014casimir,bordag2009electromagnetic}. 
Unlike ordinary metallic or dielectric materials, whose response is typically described by macroscopic dielectric functions, the electromagnetic response of graphene is intrinsically two-dimensional and can be treated directly within quantum field theory. 
This allows the Casimir interaction involving graphene sheets to be expressed in terms of the polarisation tensor of the Dirac quasiparticles, providing a direct link between quantum electrodynamics in $(2+1)$ dimensions and measurable dispersion forces.
Furthermore, the tunability of graphene’s electronic properties through doping, temperature, and external fields offers a unique opportunity to explore how quantum vacuum forces can be modified by controllable material parameters. 
For these reasons, graphene-based systems have attracted considerable interest in both theoretical and experimental studies of Casimir physics.

By exploiting resonant enhancement in a cavity, we derive the resulting pressure between graphene plates induced by the axion field and assess the sensitivity of Casimir experiments to axion–photon couplings.
Physically, the pressure in the cavity may be understood as arising from Lorentz forces between currents induced on the interfaces. Electromagnetic fluctuations, sourced by the axion background, generate surface currents $\mathbf{J}_s=\sigma \mathbf{E}$ on the conducting sheets, which interact with the magnetic field inside the cavity through the Lorentz force density $\mathbf{f} = \mathbf{J}_s \times \mathbf{B}$.
Within this framework, we find that the axion-induced pressure exhibits resonant enhancement when the cavity separation satisfies a resonance condition set primarily by the axion Compton wavelength, with a small shift determined by the phase of the graphene reflection coefficient, leading to sharp increases in the electromagnetic response. The magnitude and width of these resonances are strongly controlled by the graphene conductivity. In particular, increasing the chemical potential enhances the resonant response, while larger damping broadens the resonances and reduces the peak pressure. Comparing the resonantly enhanced signal with the conventional Casimir background, we identify the parametric regimes in which the effect could become experimentally relevant. The signal is maximised for highly doped, low-dissipation graphene configurations operated near resonance, whereas away from resonance the induced pressure becomes strongly suppressed.

The paper is organised as follows. In Sec.~\ref{sec:axion_graphene}, we introduce the axion--electrodynamics framework and discuss the electromagnetic response of graphene, including its conductivity within the Kubo formalism. In Sec.~\ref{sec:casimir_effect}, we review the Lifshitz formalism for the Casimir effect and discuss the corresponding reflection coefficients and polarisation-tensor description for graphene systems. In Sec.~\ref{sec:classical_casimir_pressure}, we derive the electromagnetic boundary conditions and construct the Green’s function describing the axion-induced cavity response. The resulting resonant electric field and pressure are then obtained analytically. In Sec.~\ref{sec:resonances_and_experimental_forecast}, the resonance structure is analysed in terms of the graphene conductivity and cavity reflection properties. We then present projected sensitivity estimates and and discuss their  implications for axion-induced modifications of Casimir-type systems.
Finally, Sec.~\ref{sec:conclusions} summarises our main results. Unless explicitly stated otherwise, we work in natural units where $c$ and  $\hbar$ are taken to be unity.

\section{Axion Electrodynamics and Graphene Conductivity} \label{sec:axion_graphene}

\subsection{The Model}

The interaction between the axion-like field $\phi$ and the electromagnetic field is described by the axion–photon coupling term added to the standard Lagrangian of electrodynamics in vacuum. The effective Lagrangian density takes the form
\begin{equation}
\mathcal{L} =
-\frac{1}{4}F_{\mu\nu}F^{\mu\nu}
-\frac{\phi}{4M}F_{\mu\nu}\tilde{F}^{\mu\nu},
\end{equation}
where $F_{\mu\nu}$ is the electromagnetic field tensor and $\tilde{F}_{\mu \nu}=\frac{1}{2} \epsilon_{\mu \nu \rho \sigma} F^{\rho \sigma}$ is its dual, with the convention $\epsilon_{0123}=+1$ for the antisymmetric Levi–Civita tensor. 
The parameter $M$ denotes the characteristic high-energy scale associated with the axion--photon interaction and is related to the axion--photon coupling by $g_{a\gamma\gamma}\equiv 1/M$. In the case of the QCD axion, this scale is related to the Peccei--Quinn (PQ) symmetry-breaking scale.
Here, the axion kinetic, mass, and higher-order self-interaction terms are omitted, as we focus on the conventional electromagnetic field modes, and these terms do not contribute to the electromagnetic equations obtained by varying the action with respect to the electromagnetic field.

The interaction term can be written in terms of the electric and magnetic fields as
\begin{equation}
\mathcal{L}_{a\gamma} = g_{a\gamma\gamma}\phi\, \mathbf{E}\cdot\mathbf{B}.
\end{equation}

\noindent Varying the action with respect to the electromagnetic four-potential leads to modified Maxwell's equations. In particular, the Ampère-Maxwell law acquires an additional source term,
\begin{equation}
\nabla\times\mathbf{B}-\frac{\partial \mathbf{E}}{\partial t}
= \mathbf{J}
+ g_{a\gamma\gamma}\left(
\mathbf{B}\,\dot{\phi}
+\nabla\phi\times\mathbf{E}
\right)
\end{equation}
\noindent 
where $\mathbf{J}$ is an external current. Following Ref.~\cite{brax2024classical}, the gauge field is decomposed into a background field and a perturbation $A_\nu=\bar{A}_\nu+a_\nu$
and the Lorenz gauge $\partial_\mu A^{\mu}=0$ is imposed. In terms of the electromagnetic fields this corresponds to
\begin{equation}
\mathbf{B} = \mathbf{B}_0 + \mathbf{b}, \qquad
\mathbf{E} = \mathbf{e},
\end{equation}
where $\mathbf{e}$ and $\mathbf{b}$ denote the electric and magnetic field perturbations, respectively, and $\mathbf{B}_0$ is a homogeneous external magnetic field.

In the presence of $\mathbf{B}_0$,
the oscillating axion dark matter field acts as an effective source for the electric field. 
Owing to its large occupation number as an ultralight boson, the axion dark matter can be treated as a classical coherent field. Neglecting spatial gradients, we assume
\begin{equation}
\partial_\mu \phi = \dot{\phi}\,\delta_\mu^{0},
\end{equation}
corresponding to the spatially homogeneous axion field associated with the dark matter background in the local environment. Linearising in the small perturbations then leads to 
\begin{equation} \label{eq:eq_of_b}
\nabla \times \mathbf{b}-\frac{\partial \mathbf{e}}{\partial t}=\mathbf{J}+g_{a \gamma \gamma} \dot{\phi} \mathbf{B}_0,
\end{equation}
so that the axion field acts as an effective current source. The axion–photon interaction therefore provides an effective current source that drives the electromagnetic modes in the cavity formed by the parallel plates. Combining Eq.~\ref{eq:eq_of_b} with Faraday’s law
\begin{equation}
\nabla \times \mathbf{e}+\frac{\partial \mathbf{b}}{\partial t}=0,
\end{equation}
one readily obtains the corresponding wave equations for the electromagnetic perturbations:
\begin{align}
\Box\mathbf{b} &= -\frac{\dot{\phi}}{M}\,\nabla \times (\mathbf{B}_0 + \mathbf{b}), \\
\Box \mathbf{e} &= \frac{1}{M}\!\left[\ddot{\phi}\,(\mathbf{B}_0 + \mathbf{b})
+ \dot{\phi}\,\frac{\partial}{\partial t}(\mathbf{B}_0 + \mathbf{b})\right],
\end{align}
where $\Box$ is the d’Alembertian operator and $\mathbf{J}=0$ from now on. At leading order in $1/M$ (i.e leading order in the axion-photon coupling) they reduce to:
\begin{align} \label{eq:e_pert}
\Box\mathbf{b} &= -\frac{\dot{\phi}}{M}\,\nabla \times \mathbf{B}_0 , \\
\Box \mathbf{e} &= \frac{1}{M}\!\left[\ddot{\phi}\, \mathbf{B}_0 
+ \dot{\phi}\,\frac{\partial}{\partial t}\mathbf{B}_0 \right].
\end{align}
Denoting the axion-induced source term driving the electric-field perturbation by
\begin{equation}
\mathbf{J}_{\rm eff}(x)=\frac{1}{M}\left[\ddot{\phi}(x)\,\mathbf{B}_0(x)
+\dot{\phi}(x)\,\frac{\partial \mathbf{B}_0(x)}{\partial t}\right],
\end{equation}
the electric-field perturbation can be written formally as
\begin{equation}
\mathbf{e}(x)=\int d^{4}u \, G(x,u)\,\mathbf{J}_{\rm eff}(u),
\end{equation}
in terms of the retarded Green's function $G(x,u)$ associated with the relevant wave operator and boundary conditions, which encodes the geometry of the experimental configuration. Here, \(x^\mu=(t,\mathbf{x})\) and \(u^\mu=(t',\mathbf{u})\) denote four-vectors. For the static external magnetic field considered in this work, the second term in \(\mathbf J_{\rm eff}\) vanishes. Note that the effective current is parallel to the external magnetic field. This implies that the electric field in Eq.~\ref{eq:e_pert} generated by this current is also aligned with the external magnetic field.

\subsection{Graphene}
It has long been recognised that carbon-based nanostructures exhibit remarkable mechanical, electrical, and optical properties \cite{dresselhaus2011past}. Among these materials, graphene \cite{geim2007rise} has attracted particular attention both experimentally and theoretically, becoming a central topic of research in condensed matter physics and related fields. It is a two-dimensional sheet consisting of a single layer of carbon atoms arranged in a hexagonal lattice and was the first material of this kind to be experimentally isolated \cite{peres2010colloquium,castro2009electronic}. Its principal feature is that at low energies (or frequencies) below a few eV \cite{zhu2021dynamical}, its electronic excitations 
satisfy a linear dispersion relation with respect to momentum characterised by the Fermi velocity $v_F \approx c/300$, and are described by relativistic quantum electrodynamics (QED) \cite{katsnelson2007graphene}. In particular, graphene can be described within the framework of the Dirac model as a set of massless (or very light) electronic quasiparticles whose dynamics and interactions with an electromagnetic field are governed by the Dirac equation in $2+1$ dimensions \cite{das2011electronic}, with the speed of light in vacuum $c$ replaced by $v_{\mathrm{F}}$. This microscopic description successfully captures a variety of phenomena, including the optical properties of graphene \cite{nair2008fine} and the observation of giant Faraday rotation \cite{crassee2011giant}. However, it contrasts with phenomenological approaches, such as the hydrodynamic model \cite{barton2004casimir,bordag2006casimir,bordag2006lifshitz}, which treat the charge carriers as a two-dimensional electron fluid, as well as other descriptions based on the Kubo formalism \cite{falkovsky2007space,falkovsky2007optical} or random-phase approximation density correlations \cite{gomez2009thermal,sernelius2011casimir}.

Owing to the relative simplicity of graphene as a physical system (see \cite{castro2009electronic} for a review), its electromagnetic response can be derived from first principles within finite-temperature quantum electrodynamics, as described by the polarisation tensor of the Dirac quasiparticles. This generally depends on physical parameters such as the temperature $T$, the chemical potential $\mu$, and a possible mass gap $\Delta$ induced by symmetry breaking or substrate effects. As a result, the dielectric response relevant for Casimir interactions can be determined microscopically, making graphene a particularly attractive system for the study of the Casimir force. Unlike ordinary metallic or dielectric materials, whose electromagnetic response is usually modelled through phenomenological dielectric functions, the response of graphene can be incorporated directly into the Lifshitz theory of the Casimir effect through the polarisation tensor. Moreover, the possibility of tuning the electronic properties of graphene, for example through doping or gating \cite{gusynin2007ac,falkovsky2007space}, temperature, or external fields offers a promising route to exploring how quantum vacuum forces may be modified by controllable material parameters.

\subsection{Electric Conductivity}
Graphene has a simple band structure characterised by Dirac cones, in which the conduction and valence bands touch at a single point (the Dirac point), around which the dispersion relation is linear. This structure can be derived either from symmetry considerations or within the (nearest neighbour) tight-binding approximation \cite{stauber2008optical,falkovsky2008optical}. The corresponding electromagnetic response of graphene can be described within the Kubo formalism \cite{kubo1957statistical,kubo1957statistical2,falkovsky2007space}, in which the conductivity tensor $\sigma_{\mu \nu}$, relating the applied electric field $E_\nu$ to the induced
electric current $J_\mu$, according to Ohm’s law, is obtained from linear response theory. More precisely, the Kubo formalism relates the conductivity tensor to the retarded current--current correlation function of the graphene Dirac quasiparticles. Explicitly,
\begin{equation}
\left\langle J_\mu\right\rangle=\sigma_{\mu \nu} E^\nu,
\end{equation}
where $\sigma_{\mu\nu}$ generally depends on the (real or imaginary)
frequency $\omega$, the in-plane momentum $k_\perp$ (accounting for spatial dispersion), the chemical potential $\mu$, the temperature $T$, the dissipation
rate $\Gamma = \tau^{-1}$ of the electronic quasiparticles, as well as the mass gap $\Delta$ corresponding to an energy gap between the valence and conduction bands of the graphene Dirac quasiparticles. The finite relaxation time $\tau$ arises from the loss of coherence of charge carriers mainly due to scattering off charged impurities \cite{ando2006screening,nomura2007quantum}, either in the substrate or within the graphene sheet itself. In the Born approximation, the corresponding scattering rate can be estimated as
$\tau^{-1} \approx 2 \pi^2 e^4 n_{\mathrm{imp}} / \left( \epsilon_g^2 \varepsilon \right)$ \cite{falkovsky2008optical},
where $n_{\text {imp }}$ is the charged impurity density, $\epsilon_g$ is the effective dielectric constant, and $\varepsilon$ is the characteristic carrier energy, typically of order the Fermi energy or temperature. In this work, we neglect any energy dependence of the dissipation rate $\Gamma(\omega)$, restrict to the gapless case $\Delta=0$, and do not consider finite mass-gap effects. 
Although a nonzero gap modifies the low-frequency electromagnetic response and leads to quantitative corrections to the graphene conductivity, these effects are expected to remain relatively small for realistic values of the gap parameter \cite{klimchitskaya2013van} and are not expected to qualitatively alter the main conclusions of this work.



The conductivity tensor can be conveniently decomposed into longitudinal $\sigma_L$, transverse $\sigma_T$, and Hall $\sigma_H$ components \cite{wooten2013optical}, 
\begin{equation}
\begin{aligned}
\sigma_{ij}(\omega,\boldsymbol{k}_\perp)
&=
\frac{k_{\perp i}k_{\perp j}}{k_\perp^2}
\,\sigma_{\mathrm{L}}(\omega,k_\perp)
\\
&\quad
+
\left(
\delta_{ij}
-
\frac{k_{\perp i}k_{\perp j}}{k_\perp^2}
\right)
\sigma_{\mathrm{T}}(\omega,k_\perp)
\\
&\quad
+
\epsilon_{ij}\,\sigma_{\mathrm{H}}(\omega,k_\perp), 
\end{aligned}
\end{equation}
where \(k_{\perp i}\) are the components of the in-plane wave vector
\(\boldsymbol{k}_\perp\),  \(k_\perp  \equiv |\boldsymbol{k}_\perp|\), $\delta_{i j}$ is the Kronecker delta function and $\epsilon_{i j}$ is the 2D Levi-Civita symbol. For  brevity, the dependence on the other parameters is left implicit here.

In many applications, spatial dispersion is neglected and the local limit $k_\perp \rightarrow 0$ is adopted. In the present setup, the axion field is spatially homogeneous, such that only the $k_\perp = 0$ mode is excited, and the local limit therefore applies exactly (see Ref. \cite{rodriguez2018nonlocal,rodriguez2025electric,falkovsky2007space,drosdoff2012effects} for the effects of spatial dispersion on the graphene conductivity and Casimir pressure between graphene plates).
In this case, or equivalently in the limit of high frequencies $\omega \gg k_\perp v_F$, the conductivity tensor becomes diagonal and proportional to the identity matrix $\sigma(\omega) \delta_{i j}$,  with equal longitudinal and transverse components, and within the Kubo formalism \cite{kubo1957statistical,kubo1957statistical2} can be written as
\begin{equation}\label{eq:interband_intraband}
\sigma=  \sigma_{\text {intra }}+\sigma_{\text {inter }},
\end{equation}
where the first term corresponds to intraband electron-photon scattering processes (electron transitions within one band) \cite{falkovsky2008optical,rodriguez2025electric}, and is given by
\begin{equation}
\begin{aligned}
\sigma_{\text{intra}}(\omega,T,\mu,\Gamma)
&= \frac{e^2 \omega}{i \pi  }
\int_{-\infty}^{+\infty} d\varepsilon\,
\frac{|\varepsilon|}{\omega^2}
\frac{d f_0(\varepsilon)}{d \varepsilon},
\\
&= \frac{2 e^2 T}{\pi  }
\frac{i}{\omega + i \Gamma}
\ln \left[ 2 \cosh \left( \frac{\mu}{2 T} \right) \right].
\end{aligned}
\end{equation}
Here, $f_0(\varepsilon) \equiv\left(e^{(\varepsilon-\mu) / T}+1\right)^{-1}
$ denotes the Fermi-Dirac distribution. This contribution is analogous to the Drude–Boltzmann conductivity, which in the limit $\mu \gg T$, takes the form $\sigma_{\text{intra}}(\omega)= i e^2|\mu|/(\pi \left(\omega+i \Gamma\right))$. Note that the chemical potential $\mu$, measured from the Dirac point, is positive for electron (n-type) doping and negative for hole (p-type) doping.
At $\mu=0$, the valence band is fully occupied while the conduction band is empty. Through the application of a gate voltage (or via chemical doping) \cite{novoselov2004electric,jabbarzadeh2019comparative}, the chemical potential can be shifted, allowing electrons to populate the conduction band or, for opposite bias, holes to be introduced in the valence band.
Owing to particle-hole symmetry, many quantities depend only on $|\mu|$. Hereafter, we restrict to $\mu>0$ for simplicity.

The chemical potential is in general fixed by the carrier density in the system, which in turn can be written as
\begin{equation}
n_0=\frac{2}{\pi v_F^2} \int_0^{\infty} \varepsilon\left[f_0(\varepsilon)-f_0(-\varepsilon)\right] d \varepsilon,
\end{equation}
where the density is defined relative to charge neutrality at the Dirac point. It follows that, in
the presence of carriers ($n_0 \neq 0$), the chemical potential
at zero temperature is determined by the carrier density according to $\mu = v_F \sqrt{\pi n_0}$ (while at high temperatures, it tends to zero inversely proportionally to temperature).


The second term in Eq. \ref{eq:interband_intraband} owes its origin to the direct interband electron transitions and is given by \cite{falkovsky2008optical,rodriguez2025electric}

\begin{equation}
\sigma_{\text{inter}}(\omega)=\frac{e^2 \omega}{i \pi } \left[-\int_0^{\infty}  \frac{f_0(-\varepsilon)-f_0(\varepsilon)}{(\omega+i \delta)^2-4 \varepsilon^2}d \varepsilon \right],
\end{equation}
where the infinitesimal parameter $\delta \rightarrow0 $ determines how the integration contour bypasses the pole of the integrand, and interband dissipation is neglected. This can be rewritten into a form more convenient for numerical calculations:
\begin{equation}
\begin{aligned}
\sigma_{\text{inter}}(\omega,T,\mu,0)
&= \frac{e^2 }{4 } \, \Bigg[
\Theta\!\left( \frac{ \omega}{2}\right) 
\\
&\quad
-  \frac{4\omega }{ i \pi}\,
\int_0^{\infty} d \varepsilon \,
\frac{\Theta(\varepsilon)-\Theta\!\left(\frac{ \omega}{2} \right)}
{ \omega^2 - 4 \varepsilon ^2} \Bigg] \, ,
\end{aligned}
\end{equation}
where 
\begin{equation}
\begin{aligned}
\Theta(\epsilon)
&\equiv f_0(-\epsilon) - f_0(\epsilon), \\
&= \frac{\sinh(\epsilon/T)}{\cosh(\mu/T) + \cosh(\epsilon/T)}. 
\end{aligned}
\end{equation}
This follows from the zero-temperature, gapless result in the nondissipative limit for the interband term, generalised to finite temperature via the Maldague formula \cite{maldague1978many} which incorporates the thermal smearing of the Fermi surface by relating the finite-temperature response function to a thermal convolution of the corresponding zero-temperature response. At $T=0$, this evaluates to 
\begin{equation} \label{eq: inter T=0}
\sigma_{\text {inter}}(\omega,T,\mu,0)=\frac{e^2}{4 }\left[\theta(\omega-2 \mu)-\frac{i}{2 \pi} \ln \left(\frac{(\omega+2 \mu)^2}{(\omega-2 \mu)^2}\right)\right].
\end{equation}
At finite but low temperature, the result can be obtained from Eq. \ref{eq: inter T=0} by applying the substitutions \cite{falkovsky2008opticaloriginal}:
\begin{equation}    
\begin{aligned}
\theta(\omega-2 \mu) & \rightarrow \frac{1}{2}+\frac{1}{\pi} \arctan [(\omega-2 \mu) / 2 T], \\
(\omega-2 \mu)^2 & \rightarrow(\omega-2 \mu)^2+(2 T)^2.
\end{aligned}
\end{equation}
This replacement implements thermal broadening of the interband absorption threshold at $\omega=2 \mu$, smoothing the zero-temperature singular behaviour over an energy scale set by the temperature.

\begin{figure}[h]
    \centering
    \includegraphics[width=1\linewidth]{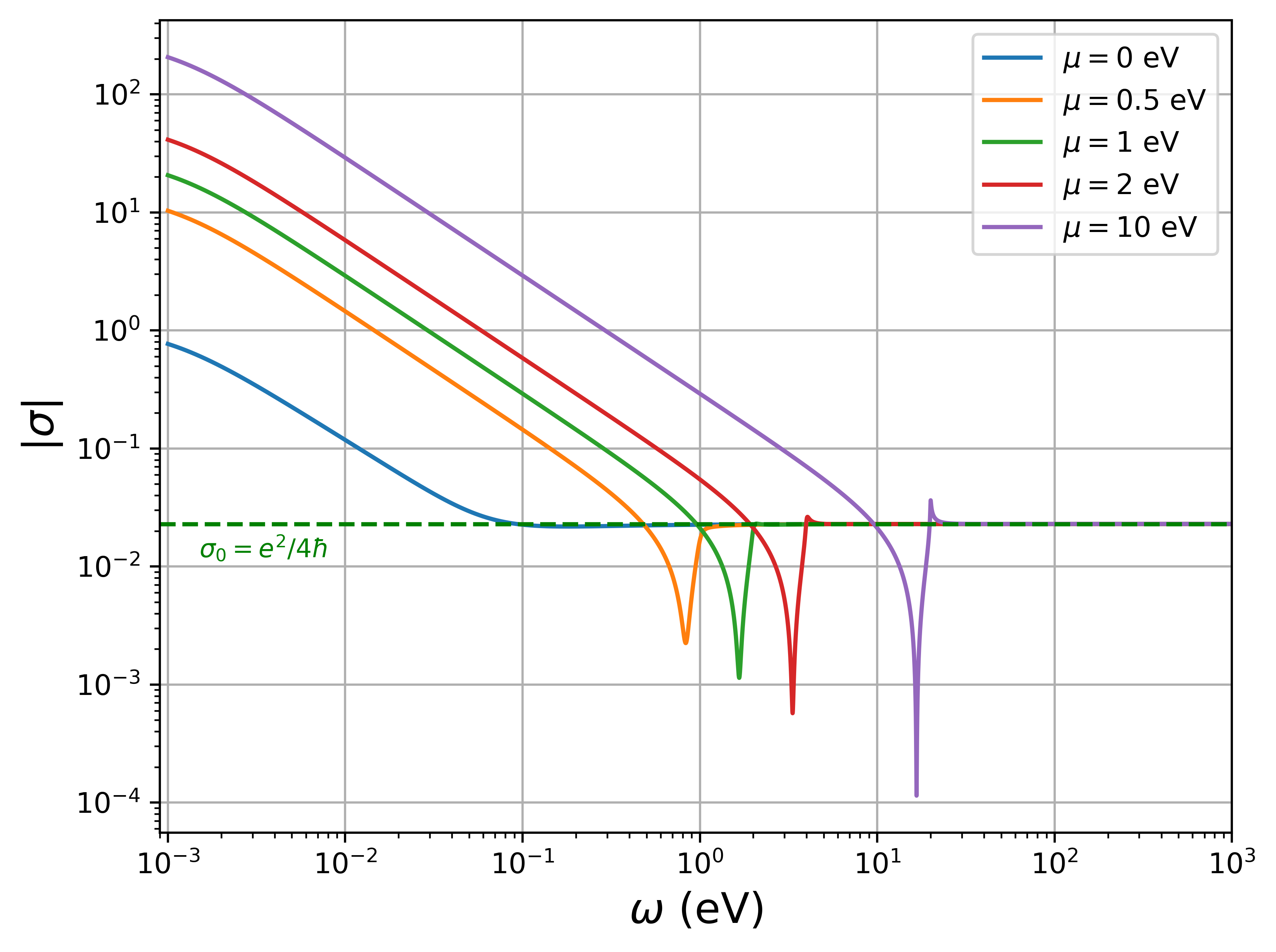}
    \caption{Magnitude of the conductivity in the Kubo formalism as a function of frequency, at $T=300$ K and $\Gamma=10^{-3}$ eV.}
   \label{fig:modulus_sigma}
\end{figure}

Figure~\ref{fig:modulus_sigma} shows the magnitude of the graphene conductivity $|\sigma(\omega)|$ as a function of frequency for several values of the chemical potential. In all cases, we take a phenomenological quasiparticle scattering rate of $\Gamma=10^{-3}~ \mathrm{eV}$, representative of a moderate level of dissipation in the graphene response. The behaviour can be understood in terms of the interplay between the intraband (Drude-like) and interband contributions introduced in Eq.~\ref{eq:interband_intraband}. At low frequencies, $\omega \ll \mu$, the response is dominated by the intraband term, which scales as $|\sigma_{\mathrm{intra}}| \propto \mu/(\omega^2+\Gamma^2)^{1/2}$. This explains the large conductivity at small $\omega$, with a magnitude that increases with chemical potential. As the frequency increases, the intraband contribution decreases, while interband transitions remain suppressed due to Pauli blocking for $\omega < 2\mu$. In this intermediate regime, the two contributions compete: the intraband term is already diminishing, while the interband term has not yet become active. This results in a pronounced minimum in the conductivity around $\omega \sim 2\mu$, whose position shifts to higher frequencies with increasing chemical potential, as observed in the figure. For $\mu=0$, the conductivity does not exhibit a corresponding minimum. In this case, the Fermi level, which separates occupied and unoccupied electronic states, lies at the Dirac point, and interband transitions are not Pauli blocked, so they contribute at arbitrarily small frequencies. As a result, there is no intermediate frequency range in which both the intraband and interband contributions are simultaneously suppressed.

For $\omega \gtrsim 2\mu$, interband transitions across the Dirac point become kinematically allowed and begin to dominate the response. In the high-frequency regime, $\omega \gg \mu,T,\Gamma$, the conductivity approaches the universal value $\sigma_0 = e^2/(4\hbar)$, indicated by the dashed line, independently of the chemical potential. This limiting behaviour reflects the scale-invariant Dirac spectrum of graphene and has been extensively studied both theoretically and experimentally \cite{PhysRevLett.94.176803}. Overall, the frequency dependence of the graphene conductivity is governed by the competition between intraband and interband processes. At low frequencies, the response is dominated by Drude-like intraband transport, while at high frequencies it approaches the universal interband conductivity of graphene. Between these two regimes, Pauli blocking suppresses interband transitions for $\omega \lesssim 2\mu$, leading to the characteristic minimum in the conductivity near $\omega \sim 2\mu$.


\section{Casimir Effect}
\label{sec:casimir_effect}

\subsection{Lifshitz Formalism}
The Casimir interaction between material surfaces can be described within the Lifshitz formalism \cite{lifshitz2013statistical}, which generalises the original Casimir result \cite{casimir1948attraction} for ideal perfectly conducting plates to arbitrary dispersive and dissipative media at finite temperature. In this framework, the force arises from electromagnetic fluctuations in the presence of material boundaries and is determined entirely by the reflection properties of the interfaces. At finite temperature, the Casimir pressure is expressed as a sum over Matsubara frequencies, with the material dependence entering through the transverse electric (TE) and transverse magnetic (TM) reflection coefficients. For graphene systems, these reflection coefficients are determined by the electromagnetic response of the Dirac quasiparticles and may be expressed either in terms of the QFT polarisation tensor formulation, discussed below, or through the conductivity obtained within the Kubo formalism, Eq.~\ref{eq:interband_intraband}. In the local regime considered here, the nonlocal Kubo and QFT descriptions consistently reduce to the local conductivity model derived in Ref. \cite{falkovsky2007space,falkovsky2007optical} (see Refs.~\cite{klimchitskaya2025temperature,rodriguez2025electric,bordag2025comment,rodriguez2026reply} for discussion of the nonlocal case).

For a two-dimensional conducting sheet at real frequency $\omega$, the TE and TM reflection coefficients can be written generally in terms of the longitudinal and transverse conductivities as \cite{koppens2011graphene,klimchitskaya2023casimir,klimchitskaya2025temperature}
\begin{equation}
r_{\rm TE}(\omega,k_\perp)
=
-\frac{ 2\pi \omega \sigma_{\mathrm{T}}(\omega,k_\perp)}
{2 \pi \omega\sigma_{\mathrm{T}}(\omega,k_\perp) +iq},
\end{equation}
and
\begin{equation}
r_{\rm TM}(\omega,k_\perp)
=
\frac{2\pi iq \sigma_{\mathrm{L}}(\omega,k_\perp)}
{2\pi iq \sigma_{\mathrm{L}}(\omega,k_\perp)+\omega},
\end{equation}
where
\begin{equation}
q=\sqrt{k_\perp^2 -\omega^2}.
\end{equation}
The reflection coefficients on the real frequency axis are the physically relevant quantities for describing real propagation effects, such as the response of the cavity to the axion field.

On the other hand, within the Lifshitz formalism for the Casimir effect, these reflection coefficients are evaluated on the imaginary frequency axis through the analytic continuation
$\omega\rightarrow i\xi_l$, where $\xi_l=2\pi l  T$ are the Matsubara frequencies, with $l=0,1,2, \dots$. This continuation follows from the analytic structure of the response functions, which is a consequence of causality and is encoded in the Kramers--Kronig relations \cite{landau1984electrodynamics}. In this sense, the real- and imaginary-frequency descriptions correspond to the same electromagnetic response of the system. Thus, the same microscopic response governing the real-frequency axion-induced fields also determines the fluctuation-induced Casimir interaction.
The axion-induced signal and the Casimir background are computed using the same underlying material response, evaluated in different frequency domains appropriate to each physical effect. 

Before deriving the axion-induced pressure, we first compute the conventional Casimir pressure within the Lifshitz formalism, which provides the dominant background contribution in the cavity system considered here. We consider the pressure $P(d,T)$ between two parallel planar structures separated by a distance $d$ and in thermal equilibrium at temperature $T$, given by the Lifshitz formula in Eq.~\ref{eq: Lifshitz_formula}. The formalism can be written in a general form in terms of the electromagnetic reflection coefficients of the two surfaces, allowing for different physical configurations. In particular, we specialise to (i) two thick dielectric plates (semispaces) and (ii) two freestanding graphene sheets.

Introducing dimensionless variables for computational convenience, and retaining the fundamental constants explicitly for dimensional transparency, the Lifshitz formula can be expressed in terms of appropriately defined reflection coefficients $r_{\mathrm{TM}}^{(n)}$ and $r_{\mathrm{TE}}^{(n)}$ on the two boundary planes $n=1,2$, for the independent TM and TE polarisations of
the electromagnetic field \cite{bordag2009advances,klimchitskaya2014observability}:
\begin{equation}\label{eq: Lifshitz_formula}
\begin{aligned}
P(d, T)= & -\frac{k_B T}{8 \pi d^3} \sum_{l=0}^{\infty}{'} \int_{\zeta_l}^{\infty} y^2 d y \\
& \times\left[\frac{r_{\mathrm{TM}}^{(1)}\left(i \zeta_l, y\right) r_{\mathrm{TM}}^{(2)}\left(i \zeta_l, y\right)}{e^y-r_{\mathrm{TM}}^{(1)}\left(i \zeta_l, y\right) r_{\mathrm{TM}}^{(2)}\left(i \zeta_l, y\right)}\right. \\
& \left.+\frac{r_{\mathrm{TE}}^{(1)}\left(i \zeta_l, y\right) r_{\mathrm{TE}}^{(2)}\left(i \zeta_l, y\right)}{e^y-r_{\mathrm{TE}}^{(1)}\left(i \zeta_l, y\right) r_{\mathrm{TE}}^{(2)}\left(i \zeta_l, y\right)}\right].
\end{aligned}
\end{equation}
This follows directly from differentiating the Helmholtz free energy of interaction per unit area with respect to the separation between the two plates. The free energy is given by 
\begin{equation}
\begin{aligned}
\mathcal{F}(d, T)= & \frac{k_B T}{8 \pi d^2} \sum_{l=0}^{\infty}{'} \int_{\zeta_l}^{\infty} y d y \\
& \times\left\{\ln \left[1-r_{\mathrm{TM}}^{(1)}\left(i \zeta_l, y\right) r_{\mathrm{TM}}^{(2)}\left(i \zeta_l, y\right) e^{-y}\right]\right. \\
& \left.+\ln \left[1-r_{\mathrm{TE}}^{(1)}\left(i \zeta_l, y\right) r_{\mathrm{TE}}^{(2)}\left(i \zeta_l, y\right) e^{-y}\right]\right\},
\end{aligned}
\end{equation}
 from which the Casimir pressure is obtained as 
\begin{equation}
P(d, T)=-\frac{\partial \mathcal{F}(d, T)}{\partial d}.
\end{equation}
Here $k_B$ is the Boltzmann constant, $\zeta_l$ are the dimensionless Matsubara frequencies defined in terms of the dimensional ones as $\zeta_l = \xi_l/\omega_c$, where $\omega_c = c/(2d)$. The magnitude of the projection of the wave vector on the plane of a plate, $k_\perp$, is used to define the dimensionless wave vector $y = 2dq_l \equiv 2d\left(k_\perp^{2} + \xi_l^{2} /c^2 \right)^{1/2}$. The prime on the summation sign indicates that the term with $l=0$ is halved.
Note the distinction between the reflection amplitude, $r$, and the reflectance (power reflection), sometimes denoted as $R \equiv |r|^2$. Depending on the physical configuration under consideration, these quantities correspond to the electromagnetic reflection coefficients of metallic or dielectric plates, freestanding graphene sheets, or graphene-coated substrates. 

The reflection coefficients of electromagnetic oscillations on (the boundary between vacuum and) a thick metallic plate have the standard form \cite{bordag2009advances,bordag2012thermal}
\begin{equation}\label{eq:reflection_coeff_seimspace}
\begin{aligned}
r_{\mathrm{TM}}^{(p)}\left(i \zeta_l, y\right) & =\frac{\varepsilon_l y-\left[y^2+\zeta_l^2\left(\varepsilon_l \mu_l-1\right)\right]^{1 / 2}}{\varepsilon_l y+\left[y^2+\zeta_l^2\left(\varepsilon_l \mu_l-1\right)\right]^{1 / 2}}, \\
r_{\mathrm{TE}}^{(p)}\left(i \zeta_l, y\right) & =\frac{\mu_l y-\left[y^2+\zeta_l^2\left(\varepsilon_l \mu_l-1\right)\right]^{1 / 2}}{\mu_l y+\left[y^2+\zeta_l^2\left(\varepsilon_l \mu_l-1\right)\right]^{1 / 2}},
\end{aligned}
\end{equation}
where the dielectric permittivity $\varepsilon_l \equiv \varepsilon\left(i \omega_c \zeta_l\right)$ and the magnetic permeability $\mu_l \equiv \mu\left(i \omega_c \zeta_l\right)$ (not to be confused with the chemical potential) are computed along the imaginary axis. This is what is required by the Lifshitz formula, i.e. an integration along the imaginary axis in the frequency complex plane. 

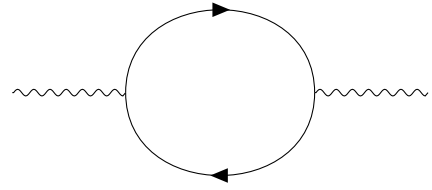
\begin{figure}[h]
    \centering
    \begin{tikzpicture}
    \begin{feynman}

    \vertex (a);
    \vertex [right=2.5cm of a] (b);

    \vertex [left=1.5cm of a] (i1);
    \vertex [right=1.5cm of b] (i2);

    \diagram*{
        (i1) -- [photon] (a),
        (b) -- [photon] (i2),

        (a) -- [half left, fermion, looseness=1.5] (b),
        (b) -- [half left, fermion, looseness=1.5] (a),
    };


    \end{feynman}
    \end{tikzpicture}
    \caption{
    One-loop Feynman diagram contributing to the quadratic effective action for the external electromagnetic field, corresponding to the polarisation tensor of graphene Dirac quasiparticles.
    }
    \label{fig:polarization_tensor}
\end{figure}

We now consider the case in which both boundary planes consist of freestanding graphene sheets, such that 
$r^{(1)}_{\mathrm{TM,TE}} = r^{(2)}_{\mathrm{TM,TE}} \equiv r^{(g)}_{\mathrm{TM,TE}}$. 
Within the framework of the Dirac model at finite temperature, the corresponding reflection coefficients, first computed in Ref. \cite{fialkovsky2011finite}, can be expressed in terms of the components of the polarisation tensor in $(2+1)$-dimensional spacetime as \cite{klimchitskaya2013van,bordag2012thermal,chaichian2012thermal}
\begin{align}
r_{\mathrm{TM}}\left(i \zeta_l, y\right)
&=\frac{y \tilde{\Pi}_{00}}{y \tilde{\Pi}_{00}+2\left(y^2-\zeta_l^2\right)}, \\
r_{\mathrm{TE}}\left(i \zeta_l, y\right)
&=-\frac{\left(y^2-\zeta_l^2\right) \tilde{\Pi}_{\mathrm{tr}}-y^2 \tilde{\Pi}_{00}}
{\left(y^2-\zeta_l^2\right)\left(\tilde{\Pi}_{\mathrm{tr}}+2 y\right)-y^2 \tilde{\Pi}_{00}},
\end{align}
where the dimensionless components (denoted by a tilde) $\tilde{\Pi}_{00,\mathrm{tr}}$ are defined as $\tilde{\Pi}_{00, \mathrm{tr}}=2 a \Pi_{00, \mathrm{tr}}$, with $\Pi_{\mathrm{tr}} \equiv \Pi^{\mu}_{\ \mu}$ denoting the trace. The polarisation tensor 
$\Pi_{\beta \gamma, l} \equiv \Pi_{\beta \gamma}\left(i \xi_l, k_{\perp}, T, \Delta, \mu\right)$  arises from the effective action for the external electromagnetic field generated by the quantised graphene Dirac quasiparticles \cite{fialkovsky2012quantum,valenzuela2015graphene}. To quadratic order in the electromagnetic field, it corresponds diagrammatically to the one-loop Feynman diagram shown in Fig.~\ref{fig:polarization_tensor}, consisting of an electronic quasiparticle loop with two external photon legs, and encodes the linear electromagnetic response of the graphene quasiparticles. 


\noindent It generally depends on the energy gap $\Delta = 2mv^{2}_{\mathrm{F}}$,  where $m$ here is the effective Dirac mass of the quasiparticles and $v_F \approx c/300$ is the Fermi velocity,
as well as on the chemical potential $\mu$. Due to gauge invariance $k_\mu \Pi^{\mu \nu}\left(k_0, \boldsymbol{k}_{\perp}\right)=0$ (transversality condition) and rotational symmetry in the graphene plane, it is characterised by two independent scalars, which may be chosen as $\Pi_{00}$ and $\Pi_{\mathrm{tr}}$.
The explicit expressions for arbitrary $\Delta$ and $\mu$ can be found in Ref.~\cite{fialkovsky2011finite,klimchitskaya2020casimir}. We verified that these general expressions consistently reduce to the known undoped graphene limit $\mu=0$, yielding the simpler analytic forms \cite{chaichian2012thermal,klimchitskaya2013van}: 

\begin{widetext}
\begin{multline}
\tilde{\Pi}_{00}(i\zeta_l,y)
= 8 e^{2} (y^2-\zeta_l^2)\int_0^1 dx \,\frac{x(1-x)}{D_l(x)}
+ \frac{8 e^{2}  }{\tilde v_F^2}\int_0^1 dx \Bigg\{
\frac{\tau}{2\pi}
\ln\!\Big[1 + 2\cos(2\pi l x) \exp(-g_l) + \exp(-2g_l)\Big]
\\
-\frac{\zeta_l}{2}(1-2x)\,
\frac{\sin(2\pi l x)}{\cosh(g_l) + \cos(2\pi l x)}
+\frac{\tilde{\Delta}^2+\zeta_l^2 x(1-x)}{D_l(x)}\,
\frac{\cos(2\pi l x)+\exp(-g_l)}{\cosh (g_l) + \cos(2\pi l x)}
\Bigg\},
\label{eq:Pi00}
\end{multline}

\begin{multline}
\tilde{\Pi}_{\mathrm{tr}}(i\zeta_l,y)
= 8 e^{2}  \big[y^2 + f_l(y)\big]\int_0^1 dx \,\frac{x(1-x)}{D_l(x)}
+ \frac{8 e^{2} }{\tilde v_F^2}\int_0^1 dx \Bigg\{
\frac{\tau}{2\pi}
\ln\!\Big[1 + 2\cos(2\pi l x) \exp(-g_l) + \exp(-2g_l) \Big]
\\
-\frac{\zeta_l(1-2\tilde v_F^2)}{2}(1-2x)
\frac{\sin(2\pi l x)}{\cosh g_l + \cos(2\pi l x)}
+\frac{
\tilde{\Delta}^2
+ x(1-x)\!\left[(1-\tilde v_F^2)^2\zeta_l^2 - \tilde v_F^4 y^2\right]
}{D_l(x)}
\frac{\cos(2\pi l x)+\exp(-g_l)}{\cosh g_l + \cos(2\pi l x)}
\Bigg\}.
\label{eq:Pitr}
\end{multline}
\end{widetext}

\noindent Here, $e$ denotes the elementary electric charge,
$\tilde{\Delta} \equiv \Delta/ \omega_c$ is the dimensionless mass-gap parameter, 
$\tilde{v}_F \equiv v_F/c$ represents the dimensionless Fermi velocity, 
and $\tau \equiv 4\pi aT $. We further introduce the dimensionless functions 
$f_l(y) \equiv f(\zeta_l,y) = \tilde{v}_F^2 y^2 + (1-\tilde{v}_F^2)\zeta_l^2$, 
$D_l(x) \equiv \sqrt{\tilde{\Delta}^2 + x(1-x) f_l(y)}$, 
and 
$g_l \equiv g(\tau,\zeta_l,y) = \frac{2\pi}{\tau} D_l(x)$. It is worth noting that $\tilde{\Pi}_{00}$, and consequently the reflection coefficients, depend on $T$ both implicitly through the Matsubara frequencies and explicitly as a parameter. For the case of graphene-coated dielectric substrates, the corresponding reflection coefficients are given in Ref.~\cite{klimchitskaya2014observability} and were originally derived in Refs.~\cite{PhysRevB.85.195427,PhysRevB.78.085432,PhysRevB.76.153410}.
The expressions for a freestanding graphene sheet follow from these results by setting $\varepsilon_l^{(n)} = 1$ and $k_l^{(n)} = q_l$. Conversely, the standard Fresnel reflection coefficients for plates made of ordinary materials are recovered by setting $\Pi_{00,l} = \Pi_l = 0$.

\begin{figure}[h]
    \centering
    \includegraphics[width=1\linewidth]{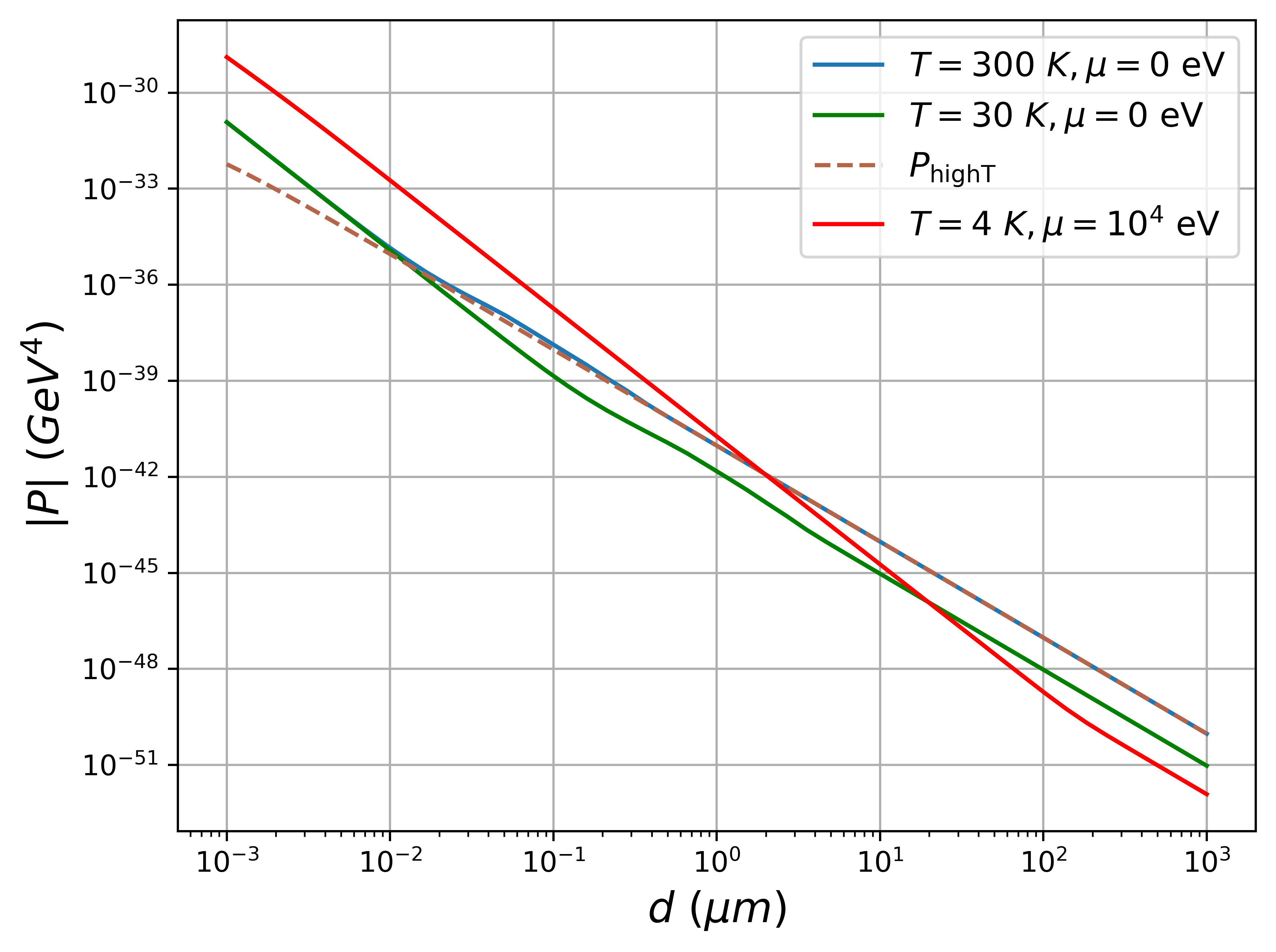}
\caption{
Magnitude of the Casimir pressure between two freestanding graphene sheets as a function of separation for several temperatures and chemical potentials.}
    \label{fig:Casimirbkg_highT}
\end{figure}

Using the reflection coefficients above, the Casimir pressure can be evaluated numerically from the Lifshitz formula in Eq.~\ref{eq: Lifshitz_formula}. Figure~\ref{fig:Casimirbkg_highT} shows the magnitude of the resulting Casimir pressure as a function of separation, for several temperatures and chemical potentials. Also shown is the corresponding high-temperature asymptotic expression for $T=300$ K, which accurately describes the large-separation behaviour of the numerical results. In the high-temperature (equivalently, large-separation) limit, the Casimir pressure between two graphene sheets admits an analytic asymptotic representation \cite{klimchitskaya2025temperature}
\begin{equation}\label{eq:highTCasimir}
P_{\text{highT}}(d,T)=-\frac{ T \zeta(3)}{8 \pi d^3}\left(1-\frac{3 v_F^2 }{8  e^2 d  T \ln 2}\right),
\end{equation}
where $\zeta(z)$ is the Riemann zeta function. For a non-zero gap, the factor $\ln 2$ is replaced by $\ln\!\left[2\cosh\!\left(\Delta/2k_B T\right)\right]$ \cite{klimchitskaya2013van}. The asymptotic behaviour of the Casimir free
energy and pressure is determined by the zero-frequency
contribution to the Lifshitz formula, and will also prove useful later when estimating the Casimir background entering our sensitivity forecasts. The $T=4 \mathrm{~K}$, $\mu=10^4 \,\mathrm{eV}$ curve is included for reference, as it serves as a benchmark parameter choice for the subsequent analysis. Chemical potentials well above $\mathcal{O}(10) ~ \mathrm{eV}$ lie outside the regime relevant for realistic graphene systems and are included solely to illustrate how the Casimir response would evolve if the graphene conductivity continued to follow the low-energy model as the chemical potential increased. In the sensitivity forecasts of Fig. \ref{fig:sensitivity_gamma_comparison}, we emphasise that realistic situations involve chemical potentials $\mu \lesssim\ 10 ~ \mathrm{eV}$.

The pressure interpolates between the zero-temperature regime, characterised by the scaling $|P|\propto d^{-4}$ (red dashed line), and the high-temperature regime described by the asymptotic expression $P_{\mathrm{highT}}$ (purple dashed line). The temperature controls the location of the crossover between these regimes: lowering the temperature shifts the onset of the thermal $d^{-3}$ behaviour to larger separations and suppresses the large-separation thermal contribution, while leaving the short-distance behaviour nearly unchanged. Increasing the chemical potential enhances the pressure over the entire separation range shown. 

\begin{figure}[h]
    \centering
    \includegraphics[width=1\linewidth]{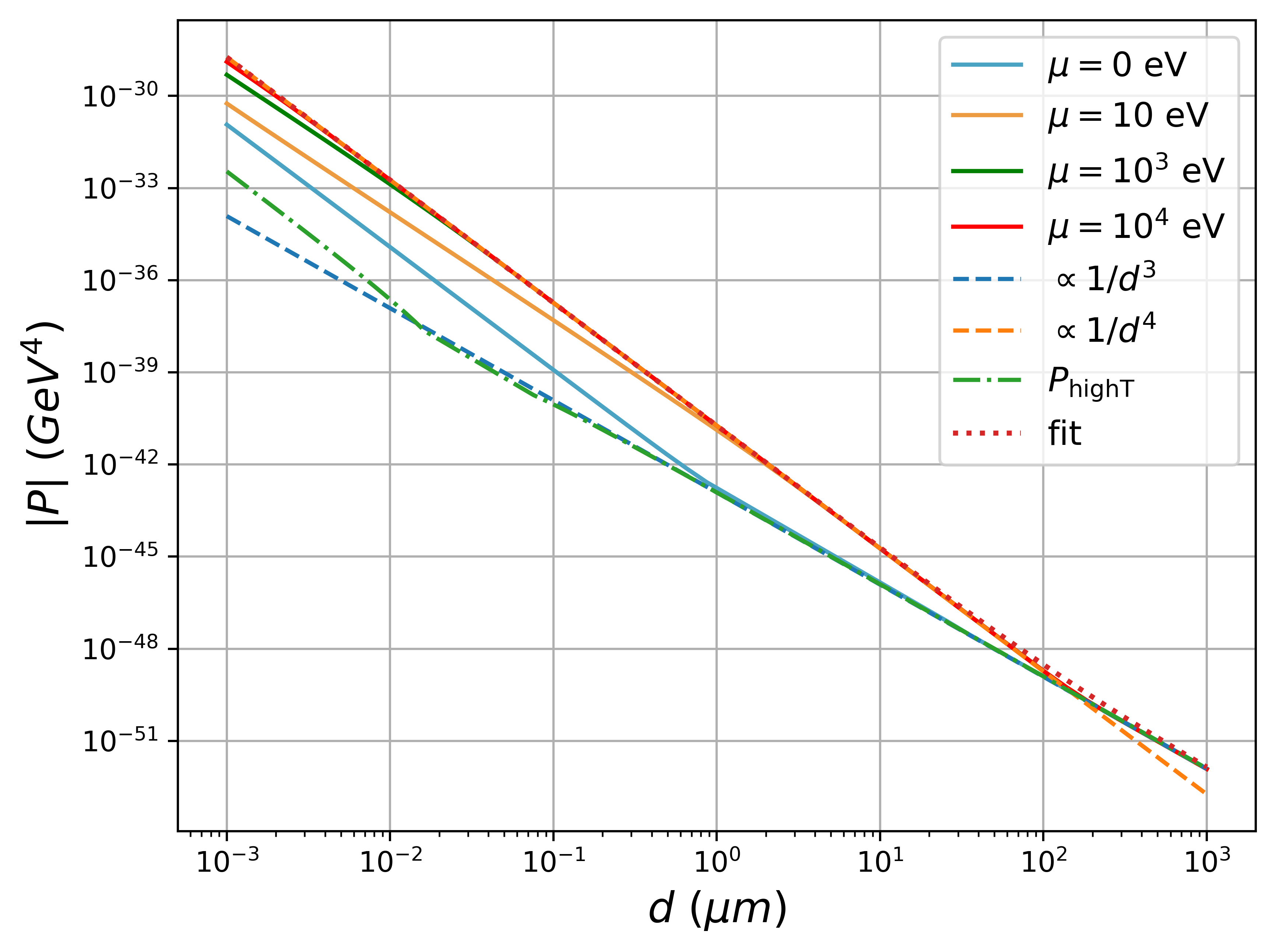}
    \caption{Magnitude of the Casimir pressure between two freestanding graphene sheets, at $T=4$ K, for a range of chemical potentials. Dashed lines show the asymptotic short- and large-distance behaviours.}
    \label{fig:Casimirbkg_T4K}
\end{figure}

The temperature dependence of the Casimir pressure 
arises from two distinct sources: the Matsubara 
summation and the intrinsic temperature dependence 
of the response functions. At small separations, $d 
\ll 1/T$, many Matsubara terms 
contribute and the sum can be approximated by an 
integral over frequencies. In this regime, the 
dominant frequencies are large compared to the 
thermal scale associated with graphene, and the 
response functions become effectively temperature-independent, leading to the zero-temperature scaling 
$P \sim d^{-4}$ (quantum regime). 
In contrast, at large separations, $d \gg  1/T$, the contribution from higher Matsubara 
modes is exponentially suppressed, and the pressure 
is dominated by the zero-frequency term, giving rise to the scaling $P \sim T/d^3$. In this regime, 
the explicit temperature dependence of the graphene 
response functions becomes important. Due to the 
small Fermi velocity in graphene, the associated 
thermal scale is significantly reduced compared to 
conventional 3D materials, as shown in 
\cite{gomez2009thermal}. 
In other words, thermal effects in the graphene 
response become important already at relatively 
short separations, because $T_{\mathrm{eff}}^{g} = 
 v_F/(2 d)$ is much smaller than the 
electromagnetic scale $T_{\mathrm{eff}} = 1/\left(2 
d \right)$.

Fig.~\ref{fig:Casimirbkg_T4K} shows the magnitude of the Casimir pressure as a function of separation at $T=4\,\mathrm{K}$ (corresponding to liquid-helium cryogenic conditions), for a range of chemical potentials. The grey dashed curve labelled "fit" shows a simple phenomenological fit to the numerical Casimir pressure. This interpolation is introduced later in Eq.~\ref{eq:fit} and is subsequently used to model the Casimir background in the projected sensitivity analysis. At small separations, the pressure exhibits a pronounced dependence on $\mu$, with larger chemical potentials leading to a stronger interaction. As the separation increases, the curves gradually converge, indicating that the pressure becomes progressively less sensitive to the chemical potential.

At large separations, the Casimir pressure is governed by the zero-frequency contribution to the Lifshitz formula, where the reflection coefficients depend on the static polarisation tensor, $\Pi(0, T, \mu) \sim T \ln \left[2 \cosh \left(\frac{\mu}{2 T}\right)\right]$. For $l=0$, the TM reflection coefficient reduces to
\begin{equation}
r_{\rm TM}(0,y)=
\frac{\tilde{\Pi}_{00}(0,y)}
{\tilde{\Pi}_{00}(0,y)+2y},
\end{equation}
such that the response saturates when $\tilde{\Pi}_{00}(0,y)\gg 2y$. The Lifshitz integral is dominated by values of $y \sim 1$, as the exponential factor $e^{-y}$ suppresses contributions from large $y$, while the phase-space factor in the integration measure suppresses small $y$, corresponding to modes with characteristic wave vector of order $1/d$. As the chemical potential increases, the static polarisation tensor grows, thereby enhancing the zero-frequency response and driving the reflection coefficients towards unity. As a result, the pressure approaches the universal thermal scaling $P \sim T/d^3$ and becomes effectively independent of $\mu$.

The convergence of the pressure curves at large separations occurs at different values of $d$, as it requires not only the suppression of higher Matsubara modes but also the saturation of the graphene response. For larger $\mu$, the polarisation tensor is enhanced, and larger separations are required for the reflection coefficients to reach their asymptotic values, thereby shifting the onset of the universal thermal regime.

\section{Classical Casimir Pressure on Graphene}
\label{sec:classical_casimir_pressure}

\subsection{Boundary Conditions}
\begin{figure}[h]
\centering 
\resizebox{\columnwidth}{!}{%
\begin{tikzpicture}[>=Latex,line cap=round,line join=round]

\def\d{5.5}      
\def\W{0.9}      
\def\S{0.55}     

\tikzset{
  axlbl/.style={font=\large},
  Blbl/.style={font=\large,red}
}

\coordinate (O) at (0,0);
\draw[->,thick] (O) -- (0,3.6) node[above,axlbl] {$x$};
\draw[->,thick] (O) -- (7.2,0) node[right,axlbl] {$z$};
\draw[->,thick] (O) -- (-2.2,-1.5) node[below left,axlbl] {$y$};
\draw[thick] (-0.2,0) -- (7.0,0);

\coordinate (L1) at (0,-1.2);
\coordinate (L2) at (0, 1.4);
\coordinate (L3) at (\W,1.4+\S);
\coordinate (L4) at (\W,-1.2+\S);

\fill[gray!20] (L1)--(L2)--(L3)--(L4)--cycle;
\draw[gray!60,thick] (L1)--(L2)--(L3)--(L4)--cycle;

\foreach \yy in {-1.1,-0.85,...,1.25}{
  \draw[gray!60] (0,\yy) -- ++(-0.25,0.12);
}
\node[font=\large] at (0,-1.65) {$z=0$};

\coordinate (R1) at (\d,-1.2);
\coordinate (R2) at (\d, 1.4);
\coordinate (R3) at (\d+\W,1.4+\S);
\coordinate (R4) at (\d+\W,-1.2+\S);

\fill[gray!20] (R1)--(R2)--(R3)--(R4)--cycle;
\draw[gray!60,thick] (R1)--(R2)--(R3)--(R4)--cycle;

\foreach \yy in {-1.1,-0.85,...,1.25}{
  \draw[gray!60] (\d,\yy) -- ++(+0.25,0.12);
}
\node[font=\large] at (\d,-1.65) {$z=d$};

\node[font=\large,fill=white,inner sep=1pt] at ({\d/2},-1.3) {Vacuum};

\draw[->,very thick,red] (-0.6,-1.0) -- (-0.6,2.2); 
\foreach \xx in {1.5,3.0,4.5}{ \draw[->,very thick,red] (\xx,-1.0) -- (\xx,2.2); } 
\draw[->,very thick,red] (6.6,-1.0) -- (6.6,2.2); 
\node[Blbl] at (3.05,2.6) {$\vec{B}_0$};

\def\Ain{0.33}
\def\Nin{5}
\def\marg{1.0}
\pgfmathsetmacro{\kin}{2*pi*\Nin/(\d-\marg)}
\draw[blue,thick,domain=0.5:\d-0.5,samples=320]
  plot (\x, {\Ain*sin(\kin*(\x-0.5) r)});

\def\AoutL{0.30}
\def\koutL{8.0}
\def\kappaL{1.25}
\pgfmathsetmacro{\phioutL}{1.2}
\draw[blue,thick,domain=-1.6:0,samples=280]
  plot (\x, {\AoutL*exp(\kappaL*\x)*sin(\koutL*\x r + \phioutL r)});

\def\AoutR{0.30}
\def\koutR{8.0}
\def\kappaR{1.25}
\pgfmathsetmacro{\phioutR}{2.2}
\draw[blue,thick,domain=\d:\d+1.6,samples=280]
  plot (\x, {\AoutR*exp(-\kappaR*(\x-\d))*sin(\koutR*(\x-\d) r + \phioutR r)});

\begin{scope}[shift={(5.15,3.63)}, scale=0.58]

\fill[white,rounded corners=2pt] (-2.1,-1.6) rectangle (2.1,1.15);
\draw[black!50,rounded corners=2pt] (-2.1,-1.6) rectangle (2.1,1.15);

\draw[thick] (0,-0.4) -- (0,0.5);

\draw[dashed,thick,->] (-1.4,0.05) -- (-0.1,0.05);
\node[left] at (-1.45,0.05) {\scriptsize $a$};

\draw[thick,->] (-0.35,0.05) -- (0.15,0.65);
\draw[thick,->] (-0.35,0.05) -- (0.15,-0.60);

\draw[blue,thick,decorate,decoration={snake,amplitude=1.5pt,segment length=6pt}]
(0,0.5) -- (1.4,0.5);
\node[right] at (1.45,0.55) {\scriptsize $\gamma$};

\draw[red,thick,decorate,decoration={snake,amplitude=1.5pt,segment length=6pt}]
(0,-0.4) -- (0,-1.2);

\draw (-0.08,-1.35) -- (0.08,-1.2);
\draw (-0.08,-1.2) -- (0.08,-1.35);

\node[right] at (0.15,-1.25) {\scriptsize $B_0$};

\node at (0.62,0.13) {\scriptsize $g_{a\gamma\gamma}$};

\end{scope}

\end{tikzpicture}%
} 
\caption{Coordinate system and cavity geometry considered in this work, consisting of two parallel plates separated along the $z$ direction in the presence of a homogeneous static external magnetic field $\mathbf{B}_0\parallel\hat{x}$. The plates may represent graphene sheets or dielectric/metallic half-spaces. The inset illustrates the Primakoff-type axion--photon conversion. } \label{fig:cavity}
\end{figure}
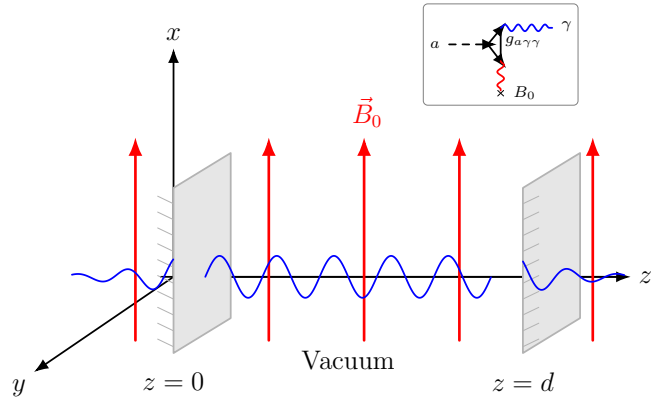

In the experimental configurations of interest, we assume a background magnetic field that permeates all space. The boundary conditions appropriate for the classical Casimir setup correspond to a cavity bounded by two parallel plates, with the external magnetic field aligned parallel to the plates along the $x$-direction, and the normal to the planes chosen to define the 
$z$-axis, as shown in Fig.~\ref{fig:cavity}.

\noindent The electrodynamic boundary conditions at a two-dimensional (graphene) sheet follow from Maxwell’s equations in the presence of a surface current. 
The tangential component of the electric field is continuous across the sheet,
\begin{equation}
(\mathbf{E}_2 - \mathbf{E}_1)\times \mathbf{n} = 0,
\end{equation}
where $\mathbf{n}$ is the unit normal to the surface. In contrast, the tangential component of the magnetic field exhibits a discontinuity determined by the induced surface current density $\mathbf{K}$, 
\begin{equation}
(\mathbf{H}_2 - \mathbf{H}_1)\times \mathbf{n} 
= -\mathbf{K}
= -\sigma\, \mathbf{n}\times(\mathbf{E}\times \mathbf{n}),
\end{equation}
where $\mathbf{H}_1$ and $\mathbf{H}_2$ are the magnetic fields evaluated immediately on either side of the sheet, and 
 $\sigma$ is the surface conductivity of graphene.

In general, graphene is characterised by both transverse and longitudinal conductivities, which give rise to two independent boundary conditions of the second type. However, since the axion-induced field considered here is polarised along the external magnetic field, $e\equiv E_x$, and propagation occurs only along the $z$ direction, the problem reduces effectively to a single electromagnetic mode involving a single conductivity component. At normal incidence, the distinction between TE and TM polarisations becomes degenerate.
The relevant boundary conditions therefore become
\begin{equation}\label{eq: BC1} 
\left.  e\right|_{z=s^{+}}= \left.   e\right|_{z=s^{-}},
\end{equation}
and 
\begin{equation}\label{eq: BC2}
\left. \frac{d e}{d z}\right|_{z=s^+}- \left. \frac{d e}{d z}\right|_{z=s^-}=\left.\frac{ \sigma(\omega) k^{2}_{z} }{i \omega} e\right|_{z=s},
\end{equation}
where $s=\{0,d\}$ and $k_z=\sqrt{ \omega^2-(k_x^2 + k_y^2)}$. 
The axion-induced pressure derived below is therefore governed by the graphene conductivity evaluated at real frequencies through its appearance in the electromagnetic boundary conditions. In contrast, the conventional Casimir pressure is computed within the Lifshitz formalism using the material response evaluated on the imaginary frequency axis. These two descriptions are nevertheless consistent, since both the conductivity and the polarisation tensor encode the same underlying electromagnetic response of graphene and are related by analytic continuation.

Using Green's function techniques, we derive the expression for the electric field by solving the following propagation equation: 
\begin{equation}
\left[
\partial_z^2+k_z^2
\right]
\tilde G(\mathbf k_\parallel;z,z_0,\omega)
=
\delta(z-z_0).
\end{equation}
Here, time-translation invariance has been used implicitly to fix the source time at $t_0=0$, while working in frequency space. We have also Fourier transformed with respect to the in-plane coordinates $(x,y)$.


\subsection{The Green's Function}
We solve for the induced electric-field perturbation $e$ using the Green’s-function method applied to the wave equation derived above. After performing a Fourier transform with respect to the in-plane coordinates $\mathbf{r}_\parallel=(x,y)$, the Green’s function may be written as
\begin{equation}
G(\mathbf r,\mathbf r';\omega)
=
\int \frac{d^2k_\parallel}{(2\pi)^2}
\,e^{i\mathbf k_\parallel\cdot(\mathbf r_\parallel-\mathbf r_\parallel')}
\,\tilde G(\mathbf k_\parallel;z,z_0,\omega),
\end{equation}
where $z_0$ denotes the source coordinate.
The electric-field perturbation in mixed Fourier space is then given by 
\begin{equation}
\tilde e(\mathbf k_\parallel,z;\omega)
=
\int_{-\infty}^{\infty} dz_0\,
\tilde G(\mathbf k_\parallel;z,z_0,\omega)\,
\tilde j(\mathbf k_\parallel,z_0;\omega),
\end{equation}
where $\tilde G(\mathbf k_\parallel;z,z_0,\omega)$ is the Green’s function associated with the wave equation in the presence of the graphene sheets. For the axion-induced source considered here, the effective current is spatially homogeneous and monochromatic,
\begin{equation}
j(\mathbf{r}_\parallel,z,t)=j_0 e^{-i\omega t},
\end{equation}
where the oscillation frequency is set by the axion mass $m_a$. Its mixed Fourier transform is therefore
\begin{equation}
\tilde \jmath(\mathbf{k}_\parallel,z;\omega)
=
(2\pi)^3 j_0\,
\delta(\omega-m_a)\,
\delta^{(2)}(\mathbf{k}_\parallel).
\end{equation}
For a constant external magnetic field $B_0$, the amplitude is given by
\begin{equation}    
j_0=\frac{m_a^2\phi_0 B_0}{M},
\end{equation}
where $\phi_0$ denotes the amplitude of the oscillating axion field, related to the local dark matter density through $\rho_0= \frac{1}{2}m_a^{2} \phi_{0}^2$, assuming that the axion constitutes all the local dark matter.
The delta function $\delta^{(2)}(\mathbf{k}_\parallel)$ implies that only the $k_\parallel=0$ mode contributes to the induced field. Substituting the source into the Green’s-function representation gives
\begin{equation}
\begin{aligned}
\tilde e(\mathbf k_\parallel,z;\omega)
&=
(2\pi)^3 j_0\,
\delta(\omega-m_a)\,
\delta^{(2)}(\mathbf k_\parallel)
\\
&\quad \times
\int_{-\infty}^{\infty} dz_0\,
\tilde G(\mathbf k_\parallel;z,z_0,\omega).
\end{aligned}
\end{equation}
Performing the inverse Fourier transform then yields
\begin{equation}
e(\mathbf{r}_\parallel,z,t)
=
\Re\!\left[e(z)e^{-im_at}\right],
\end{equation}
where the spatial profile is
\begin{equation}
\label{eq:e_spatial}
e(z)
=
j_0
\int_{-\infty}^{\infty} dz_0\,
\tilde G(0;z,z_0,m_a).
\end{equation}

\noindent Thus, the induced electric field is entirely determined by the $k_\parallel=0$ component of the Green’s function. In each spatial region, the Green’s function is expressed as a superposition of plane waves propagating in the $\pm z$ directions. 

\subsection{Region I: $z_0<0$}
For a source located to the left of the cavity, the Green’s function takes the form
\begin{equation}
\tilde{g} \left(z, z_0\right)=\left\{\begin{array}{ll}
A_0 e^{i k_z z} + B_0 e^{-i k_z z}, & z<z_0, \\
C_0 e^{i k_z z}+D_0 e^{-i k_z z}, & z_0<z<0, \\
E_0 e^{i k_z z}+F_0 e^{-i k_z z}, & 0<z<d, \\
G_0 e^{i k_z z} + H_0 e^{-i k_z z}, & z>d.
\end{array} \right. 
\end{equation}

\subsection{Region II: $0<z_0<d$}
For a source located inside the cavity:

\begin{equation}
\tilde{g}\left(z, z_0\right)= \begin{cases} I_0~ e^{i k_z z}  + J_0 e^{-i k_z z}, & z<0, \\ K_0 e^{i k_z z}+L_0 e^{-i k_z z}, & 0<z<z_0, \\ M_0 e^{i k_z z}+N_0 e^{-i k_z z}, & z_0<z<d,  \\ O_0 e^{i k_z z} + P_0 e^{-i k_z z} , & z>d.\end{cases} 
\end{equation}

\subsection{Region III: $z_0>d$}
For a source located to the right of the cavity:

\begin{equation}
\tilde{g} \left(z, z_0\right)=\left\{\begin{array}{ll}
Q_0 e^{i k_z z} + R_0 e^{-i k_z z}, & z>z_0, \\
S_0 e^{i k_z z}+T_0 e^{-i k_z z}, & d<z<z_0, \\
U_0 e^{i k_z z}+V_0 e^{-i k_z z}, & 0<z<d, \\
W_0 e^{i k_z z} + X_0 e^{-i k_z z}, &  z<0.
\end{array} \right. 
\end{equation}

The corresponding coefficients, which are given explicitly in Appendix \ref{appendix:A}, are fixed by imposing continuity, jump, and radiation conditions. At the source position $z=z_0$, the Green’s function is continuous,
\begin{equation}
\tilde G(z_0^+)=\tilde G(z_0^-),
\end{equation}
while its derivative satisfies the discontinuity condition obtained by integrating the differential equation across the delta-function source,
\begin{equation}
\left.\frac{\partial \tilde G}{\partial z}\right|_{z=z_0^+}
-
\left.\frac{\partial \tilde G}{\partial z}\right|_{z=z_0^-}
=1.
\end{equation}

\noindent At the graphene sheets located at $z=s$, with $s\in\{0,d\}$, the Green’s function satisfies the boundary conditions implied by the electrodynamic matching conditions,
\begin{equation}
\tilde G(s^+)=\tilde G(s^-),
\end{equation}
and
\begin{equation}
\left.\frac{\partial \tilde G}{\partial z}\right|_{s^+}
-
\left.\frac{\partial \tilde G}{\partial z}\right|_{s^-}
=
\frac{\sigma(\omega)k_z^2}{i\omega}\,\tilde G(s).
\end{equation}
Finally, the Green’s function satisfies the Sommerfeld radiation condition,
\begin{equation}
\lim_{z\to\pm\infty} \tilde G(z,z_0)=0,
\end{equation}
which ensures that no radiation is incoming from spatial infinity, such that only outgoing waves are present far from the source. As a consistency check, we note that the cavity is symmetric under exchange of the two interfaces at $z=0$ and $z=d$. Because the exterior media on both sides are identical, the same surface conductivity $\sigma$ and radiation conditions (and hence same boundary conditions) apply at both boundaries. The Green's function is therefore invariant under the reflection $
(z,z_0) \rightarrow (d-z,\, d-z_0)$, 
and we have explicitly verified that the three piecewise solutions corresponding to the different source regions satisfy this symmetry.

\subsection{Resonance Condition}
In the following, we define the denominator factor 
\begin{equation} \label{eq:D+D-D}
\mathcal{D} \equiv -\mathcal{D}_{+} \mathcal{D}_{-},
\end{equation}
with
\begin{equation}
\mathcal{D}_{ \pm} \equiv k_z \sigma\left(1 \pm e^{i k_z d}\right)+2 \omega.
\end{equation}
This factor appears repeatedly in the explicit expressions derived in Appendix~\ref{appendix:A}, where the factorised structure of the Green's-function denominators becomes manifest. The resonant structure of the cavity follows from the poles of the Green's function, which occur when
\begin{equation}
\mathcal{D}(\omega,k_z)=0.
\end{equation}
Using the factorised form of the denominator, this condition can be written equivalently as 
\begin{equation}
\mathcal{D}_{+}(\omega,k_z)=0
\qquad \text{or} \qquad
\mathcal{D}_{-}(\omega,k_z)=0,
\end{equation}
corresponding to the two cavity eigenmode branches. Introducing the reflection coefficient,
\begin{equation}\label{eq:r_graphene}
r(\omega,k_z)\equiv
-\frac{k_z\sigma}{2\omega+k_z\sigma},
\end{equation}
the denominator may be written in terms of the two branches
\begin{equation}
\frac{\mathcal{D}_{\pm}}{2\omega+k_z\sigma}
=
1\mp r e^{ik_z d}.
\end{equation}
The resonance condition is therefore
\begin{equation}
r e^{ik_z d}=\pm 1.
\label{eq:branch_resonance_condition}
\end{equation}
Writing
\begin{equation}
r=|r|e^{i\phi(r)},
\qquad
k_z=k_z^{(R)}+ik_z^{(I)},
\end{equation}
where $\phi(r)\equiv \arg(r)$ and $k_z^{(R)}$ and $k_z^{(I)}$ are the real and imaginary parts of $k_z$, respectively,
Eq.~\ref{eq:branch_resonance_condition} becomes
\begin{equation}
|r|e^{-k_z^{(I)}d}
e^{i(k_z^{(R)}d+\phi(r))}
=\pm 1.
\end{equation}
The magnitude condition gives
\begin{equation}
|r|e^{-k_z^{(I)}d}=1,
\end{equation}
and hence
\begin{equation}
k_z^{(I)}=\frac{1}{d}\ln |r| .
\end{equation}
For finite conductivity, $0<|r|<1$, so that $k_z^{(I)}<0$. The resonances lie below the real axis in the $k_z$ plane for both positive and negative real parts. This is the structure one expects for retarded Green's functions, as closing the contour of the Fourier integrals in the upper half plane vanishes due to the absence of poles. This is at the origin of the retarded nature of the Green's function. This describes damping due to leakage through the conducting sheets.

For the positive branch, the phase condition is
\begin{equation}
\label{eq:positive_branch_resonance}
k_z^{(R)}d=2\pi n -\phi(r),
\qquad n\in\mathbb{Z}.
\end{equation}
For the negative branch, one instead obtains
\begin{equation}
k_z^{(R)}d+\phi(r)=(2n+1)\pi.
\end{equation}
This form is equivalent to the positive-branch condition up to the convention used for defining the resonance branches and the reflection phase. In the present setup, the axion-induced field excites only the normal mode with vanishing transverse momentum, $k_\parallel=0$, so that $ k_z=\omega=m_a$. 
Thus, for the positive branch, the resonance condition becomes
\begin{equation}
m_a d=2\pi n-\phi(r).
\label{eq:mass_resonance_condition}
\end{equation}
In the perfect conductor limit, $r\to -1$ and hence $\phi(r)\to \pi$, recovering the resonance condition obtained
in Ref. \cite{brax2024classical} for ideal metallic plates.

\subsection{Pressure}
For a surface whose unit normal is $\hat z$, corresponding to plates located at
$z=0$ and $z=d$, the mechanical pressure on the sheet is determined by the
discontinuity of the normal-normal component of the electromagnetic stress-energy tensor,
\begin{equation}
P(z_s)=\left.\langle T_{zz}\rangle\right|_{z_s^{+}}
      -\left.\langle T_{zz}\rangle\right|_{z_s^{-}},
\end{equation}
where $z_s\in\{0,d\}$ and $\langle\cdots\rangle$ denotes a time average. The
time-averaged $zz$-component in vacuum is given by
\begin{equation}
\begin{aligned}
\langle T_{zz}\rangle
&=\frac12 \mathrm{Re}\!\left(
E_zE_z^*
-\frac12 \delta_{zz}\,\mathbf E\!\cdot\!\mathbf E^*
\right) \\
\quad
&+\frac12 \mathrm{Re}\!\left(
B_zB_z^*
-\frac12 \delta_{zz}\,\mathbf B\!\cdot\!\mathbf B^*
\right).
\end{aligned}
\end{equation}

\noindent In the present configuration the only nonvanishing field components are
$E_x$ and $B_y$, so that
\begin{equation}
\langle T_{zz}\rangle
=-\frac14|E_x|^2-\frac14|B_y|^2 .
\end{equation}
Using Maxwell's equation in vacuum,
$\nabla\times\mathbf E=-\partial_t\mathbf B$,
together with the harmonic time dependence $e^{i\omega t}$,
the magnetic field can be written as
\begin{equation}
B_y(z)=\frac{1}{i\omega}\frac{dE_x}{dz}.
\end{equation}
Consequently,
\begin{equation}
\langle T_{zz}\rangle
=-\frac14|E_x|^2
-\frac{1}{4\omega^2}\left|\frac{dE_x}{dz}\right|^2 .
\end{equation}
In the following, we identify $E_x(z)$ with the axion-induced electric-field perturbation and denote it by $e(z)\equiv E_x(z)$. Introducing the notation
\begin{equation}
\begin{aligned}
e'(z) &= \frac{de}{dz}, &
\Delta e' &= e'^+ - e'^-, \\
\bar e' &= \frac{e'^+ + e'^-}{2}, &
e_s &= e(z_s),
\end{aligned}
\end{equation}
the graphene boundary condition gives
\begin{equation}
\Delta e'=\frac{\sigma(\omega)k_z^2}{i\omega}e_s .
\end{equation}
Since $e$ is continuous across the sheet, the $|E_x|^2$ term cancels in the
stress-tensor discontinuity and the pressure reduces to
\begin{equation}
P(z_s)
=-\frac{1}{4\omega^2}
\left(|e'^{+}|^2-|e'^{-}|^2\right).
\end{equation}
Using
\begin{equation}
|e'^{+}|^2-|e'^{-}|^2
=\mathrm{Re}\!\left[\Delta e'\,(2\bar e')^*\right] ,
\end{equation}
one obtains
\begin{equation}
P(z_s)
=-\frac{1}{2\omega^2}
\mathrm{Re}\!\left[\Delta e'\,\bar e'^*\right].
\end{equation}
Substituting the jump condition yields
\begin{equation}
P(z_s)
=-\frac{1}{2\omega^2}
\mathrm{Re}\!\left[
\frac{\sigma(\omega)k_z^2}{i\omega}
e_s\,\bar e'^*
\right].
\end{equation}
For $k_\parallel=0$ one has $k_z=\omega$, and therefore
\begin{equation}
P(z_s)
=-\frac12
\mathrm{Re}\!\left[
\frac{\sigma(\omega)}{i\omega}
e_s\,\bar e'^*
\right],
\qquad \omega=m_a,
\label{eq:pressure}
\end{equation}
where in the case of axion-induced driving considered here, the frequency is fixed by the axion mass $m_a$. Physically, the pressure arises from the Lorentz force exerted by the electromagnetic field on the induced surface current in the graphene sheet, $
\mathbf K=\sigma(\omega)\mathbf E_\parallel$. Note that in natural units the pressure has mass dimension four, corresponding to an energy density. Since the electric field
has mass dimension two, whereas the graphene conductivity $\sigma$ is dimensionless, Eq.~\ref{eq:pressure} indeed gives the expected
dimensional scaling for $P(z_s)$. 
The electric field entering Eq.~\ref{eq:pressure} follows from Eq.~\ref{eq:e_spatial}. Taking the derivative on either side of the interface at $z=z_s$ gives
\begin{equation}
e'(z_s^\pm)
=j_0\int_{-\infty}^{\infty}dz_0
\left[\partial_z\tilde G(z,z_0)\right]_{z\to z_s^\pm}.
\end{equation}
Since the source permeates space, the integral over $z_0$ receives contributions from the three spatial regions $z_0<0,0<z_0<d$, and $z_0>d$, which must be treated separately.

Combining the corresponding Green's-function contributions, one may derive a closed analytical expression for the pressure at the interface $z=0$. 
Restricting to the normal-incidence configuration $k_\parallel=0$, it is convenient to introduce the phase factor $x \equiv e^{i\omega d}$
which describes propagation between the two interfaces separated by a distance $d$. 
In this limit, the combinations of terms arising from the Green's-function solution simplify considerably. In particular, several auxiliary quantities whose general expressions are given in Appendix~\ref{app:auxilary_Green_functions} reduce to compact analytical forms
\begin{equation} \label{eq:85}
\mathcal{A}=\omega(x-1)[2+\sigma(1-x)],
\end{equation}

\begin{equation}
\mathcal{I}_{+}
=-\frac{1}{\mathcal{D}}
\left[
(\sigma+2)+\sigma x^{2}
+\frac{\sigma+1}{\omega}\mathcal{A}
-2(\sigma+1)x
\right],
\end{equation}

\begin{equation}
\mathcal{I}_{-}
=\frac{1}{2\omega^{2}}
-\frac{1}{\mathcal{D}}
\left[
\frac{\sigma}{2}\big(-\sigma x^{2}+\sigma+2x^{2}+2\big)
+\frac{1}{\omega}\mathcal{A}
-2x
\right].
\end{equation}

\noindent The pressure at the interface $z=0$ is given by 
\begin{equation}
\begin{aligned}
P_z(z=0)
&=
-\frac{1}{2}\,
\Re\!\left[
\frac{\sigma(\omega)}{i\omega}\,
e_s(0)\,
\big(\bar{e}'(0)\big)^*
\right]
\\
&=
\frac{|j_0|^2}{4}\,
\Re \!\left[
\frac{\sigma(\omega) k_z^*}{\omega}\,
\mathcal{S}\,
\big(\mathcal{I}_+ + \mathcal{I}_-\big)^*
\right],
\end{aligned}
\end{equation}
where the surface-field coefficient $\mathcal S$, defined by
$e_s(0)=j_0\mathcal S$, is given by
\begin{equation}
\mathcal{S}
=\frac{2}{\mathcal{D}}\big[\sigma(x-1)-2\big],
\end{equation}
and the conductivity $\sigma(\omega)$ generally depends on the temperature $T$, chemical potential $\mu$, and damping parameter $\Gamma$. It is convenient to combine the coefficients $\mathcal{I}_\pm$ by defining
\begin{equation}\label{eq:90} 
\mathcal{T}
=
\frac{(\sigma+2)^2}{2}
+\left(2\sigma-\frac{\sigma^2}{2}\right)x^2
-2(\sigma+2)x
+\frac{\sigma+2}{\omega}\,\mathcal{A},
\end{equation}
so that
\begin{equation}
\mathcal{I}_+ + \mathcal{I}_- =
\frac{1}{2\omega^2} - \frac{\mathcal{T}}{\mathcal{D}}.
\end{equation}

\noindent Using this, together with the expression for $\mathcal{S}$, the pressure can be written in the compact form
\begin{equation}
P_z = \frac{|j_0|^2}{4}\,
\Re\!\Bigg[
\frac{\sigma(\sigma x-\sigma-2)}{\omega^2}
\Bigg(\frac{1}{\mathcal{D}} 
- \frac{2\omega^2\,\mathcal{T}^*}{|\mathcal{D}|^2}
\Bigg)
\Bigg],~ \omega = m_a,
\end{equation}
which naturally separates into two contributions: a single-pole term $\propto \Re(1/\mathcal{D})$ and a double-pole term $\propto 1/|\mathcal{D}|^{2}$.

\begin{figure}[h]
    \centering
    \includegraphics[width=1\linewidth]{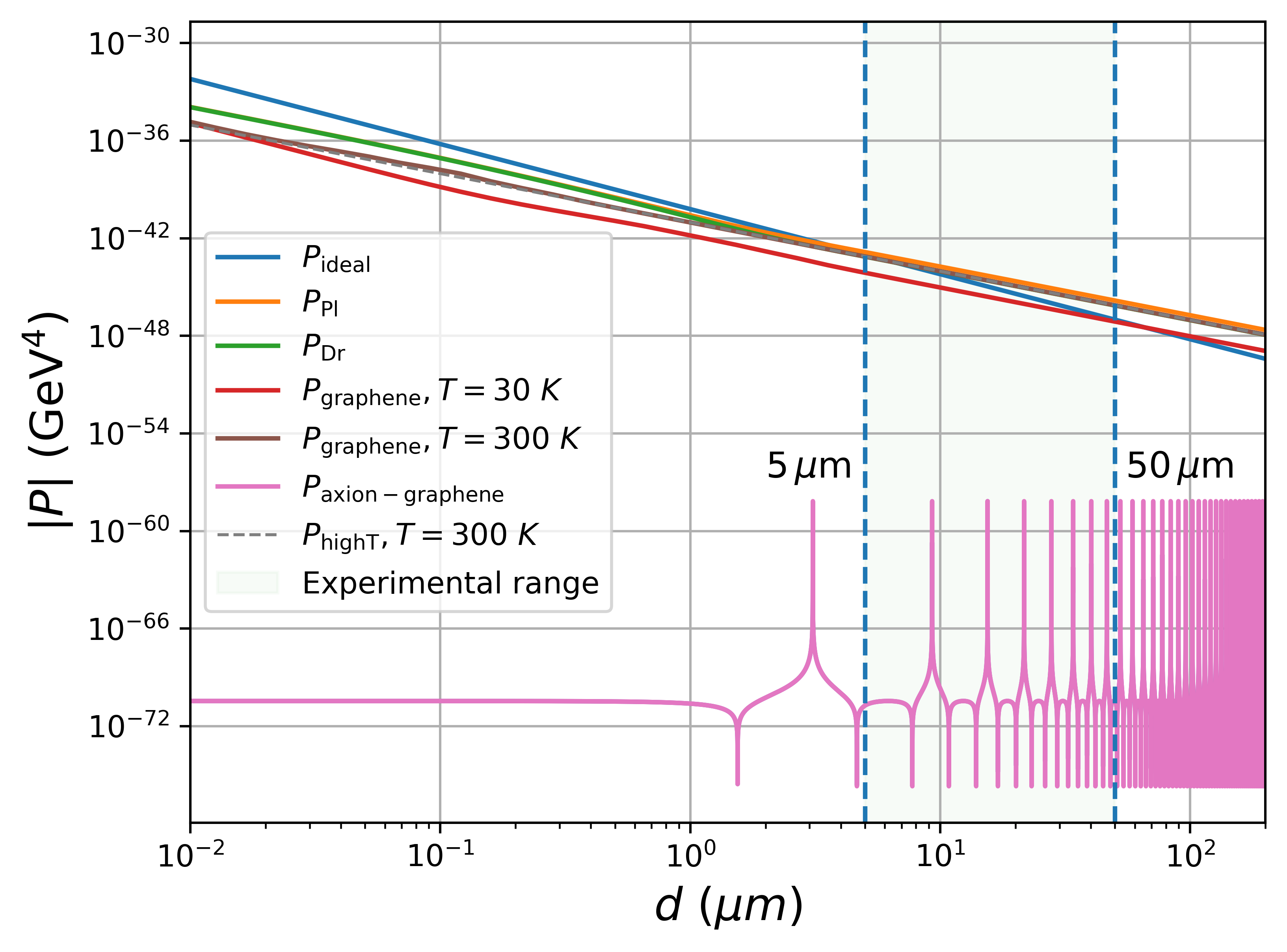}
    \caption{Magnitude of the pressure as a function of the separation $d$ for graphene plates in an external magnetic field. The axion-induced pressure is shown together with several Casimir backgrounds corresponding to ideal, plasma (Pl) and Drude (Dr) models, as well as the graphene Casimir pressure at $T=30\,\mathrm{K}$ and $T=300\,\mathrm{K}$. The shaded region indicates the experimentally accessible separation range. }    
    \label{fig:main_plot}
\end{figure}

The resulting pressure profile is shown in Fig.~\ref{fig:main_plot} as a function of the plate separation for an axion mass
$m_a=0.2\,\mathrm{eV}$, an external magnetic field $B_0=50\,\mathrm{T}$, comparable to the highest steady magnetic fields currently achievable in laboratory conditions \cite{miller2003nhmfl},
a chemical potential $\mu=10^4\,\mathrm{eV}$ and a dissipation rate $\Gamma = 10^{-5}\,\mathrm{eV}$. For comparison, the Casimir background pressure is shown for ideal metallic plates, and for the plasma and Drude models using parameters appropriate for copper,
$\omega_{\mathrm{Pl}}=2\,\mathrm{eV}$ and $\gamma_{\mathrm{Dr}}=0.066\,\mathrm{eV}$.
The graphene Casimir pressure is displayed for temperatures
$T=30\,\mathrm{K}$ and $T=300\,\mathrm{K}$ together with their high-temperature asymptotic limits. The ideal-conductor pressure is given by
\begin{equation}
P_{\mathrm{ideal}}
=
-\frac{\pi^2}{240d^4},
\end{equation}
while the Drude permittivity is
\begin{equation}
\varepsilon_{\mathrm{Dr}}(\omega)
=
1-\frac{\omega_{\mathrm{Pl}}^2}
{\omega(\omega+i\gamma_{\mathrm{Dr}})},
\end{equation}
with the plasma model recovered in the limit $\gamma_{\mathrm{Dr}}\to0$. The axion-induced contribution exhibits a sequence of resonant peaks associated with cavity modes originating from the pole structure of the Green’s function discussed above. The resonance positions are determined by the axion mass through the cavity resonance condition, while the shaded region indicates the experimentally accessible separation range.

\section{Resonances and experimental forecast}
\label{sec:resonances_and_experimental_forecast}

\subsection{Resonances}
\begin{figure}[h] 
    \centering
    \includegraphics[width=1\linewidth]{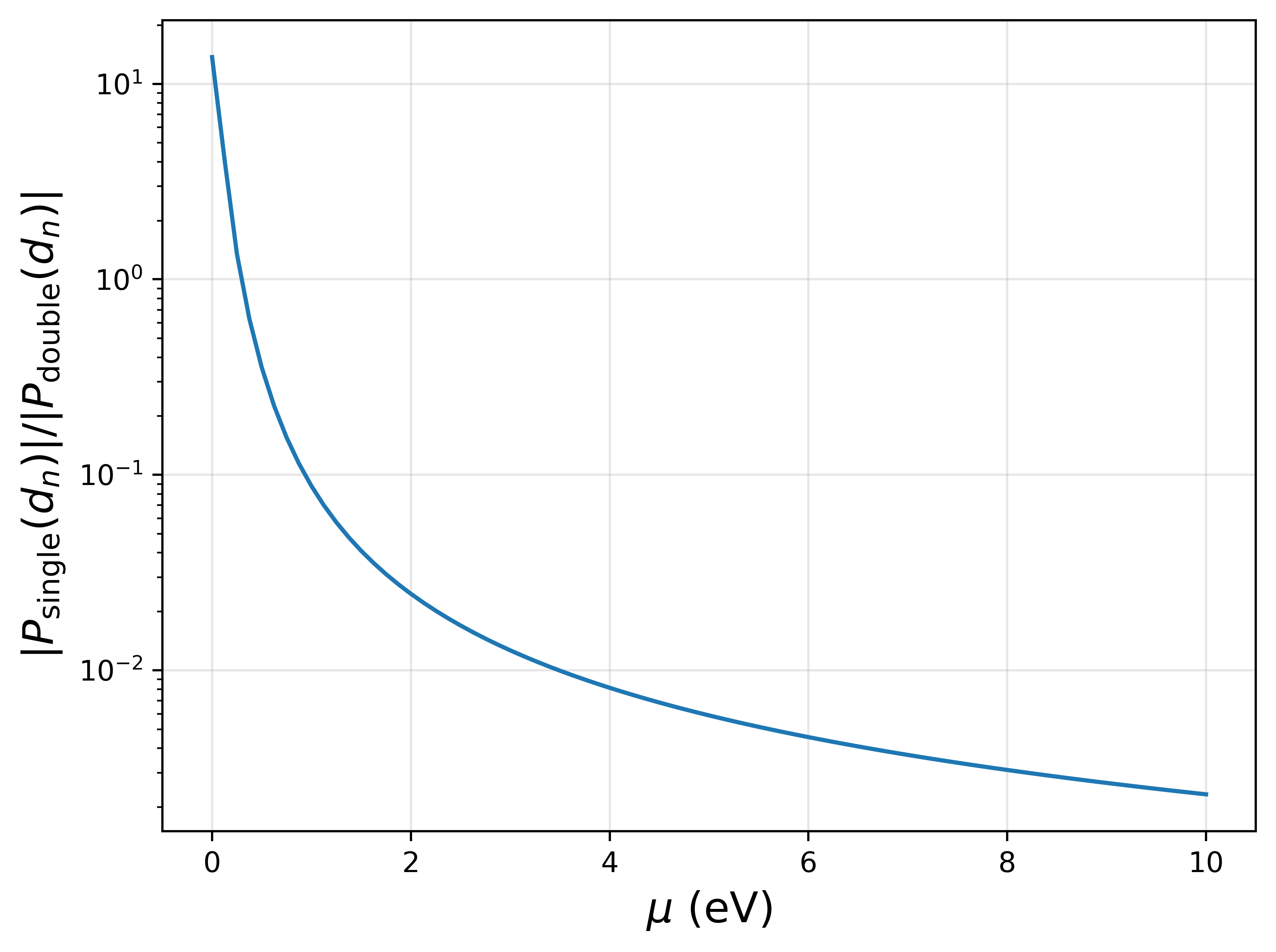}
    \caption{
Ratio of the single-pole to double-pole contributions to the resonantly enhanced axion-induced pressure as a function of the chemical potential $\mu$, for $m_a=0.01\,\mathrm{eV}$, $\Gamma=10^{-3}\,\mathrm{eV}$, and $T=300\,\mathrm{K}$.
}
   \label{Fig:ratio_vs_mu_original}
\end{figure}

For sufficiently large chemical potential, such that $\mu$ dominates over the relevant frequency scales set by the axion mass, the single-pole contribution becomes negligible, as illustrated in Fig.~\ref{Fig:ratio_vs_mu_original}, and the pressure near a resonance peak can be approximated locally by a Lorentzian profile in $d$. The resonance positions $d_n$ are given by the resonance condition in Eq.~\ref{eq:mass_resonance_condition} where the cavity is closest to a pole of the response function,
\begin{equation}
\Re[\mathcal{D}(d_n)] = 0,
\qquad
\Im[\mathcal{D}(d_n)] \neq 0,
\label{eq:res_condition_appendix}
\end{equation}
where the small imaginary part encodes dissipation and gives the resonance a finite width. Near $d=d_n$, the functions $\mathcal{F}(d)$ and $\mathcal{T}(d)$ vary slowly compared to the sharp variation of $\mathcal{D}(d)$, and may therefore be approximated by their values at resonance,
\[
\mathcal{F}(d)\approx \mathcal{F}_n \equiv \mathcal{F}(d_n),
\qquad
\mathcal{T}(d)\approx \mathcal{T}_n \equiv \mathcal{T}(d_n).
\]

\noindent Expanding $\mathcal{D}(d)$ to first order about the resonance position gives
\begin{equation}
\mathcal{D}(d)
\approx
i\mathcal{D}_I(d_n)
+
(d-d_n)\mathcal{D}'(d_n),
\end{equation}
where the resonance condition $\Re[\mathcal{D}(d_n)]=0$ has been used, such that $\mathcal{D}(d_n)=i\,\mathcal{D}_I(d_n)$.
Defining $
\alpha \equiv \mathcal{D}'(d_n)/m_a$, and
$\gamma_n \equiv \mathcal{D}_I(d_n)/\alpha$,
the denominator becomes
$
\mathcal{D}(d)
\approx
\alpha\left[m_a(d-d_n)+i\gamma_n\right].
$
Consequently,
\begin{equation}
|\mathcal{D}(d)|^2
\approx
|\alpha|^2
\left[
m_a^2(d-d_n)^2+\gamma_n^2
\right],
\end{equation}
which yields the characteristic Lorentzian structure of the resonance.
Substituting this into the pressure gives
\begin{equation} 
P_z(d)
\approx 
\frac{|j_0|^2}{4}
\frac{\mathcal{N}_n}{|\alpha|^2}
\frac{1}{m_a^2(d-d_n)^2+\gamma_n^2},
\label{eq:P_res high doping}
\end{equation}
where
\begin{equation}    
\alpha = 2im_a^2(2+\sigma)^2,
\end{equation}
and
\begin{equation}\label{eq:100}
\mathcal{N}_n
\equiv
-2\Re\!\left(\mathcal{F}_n\mathcal{T}_n^*\right)
=
-2\Re\!\left[
\sigma(\sigma x_n-\sigma-2)\mathcal{T}_n^*
\right],
\end{equation}
with $x_n=e^{i m_a d_n}$. The parameter $
\gamma_n \equiv -\ln|r|$
characterises the linewidth of the resonance, corresponding to a half-width at half-maximum $
\delta d_{\rm HWHM}= \gamma_n/m_a$.

The resonance width may equivalently be understood by allowing the cavity separation to become complex,
\begin{equation}
d_* = d_n - i\ell,
\end{equation}
while keeping the axion frequency fixed and real. Physically, the imaginary part encodes losses associated with the finite reflectivity of the cavity boundaries. Imposing the resonance condition
\begin{equation}
1-r^2 e^{2 i m_a d_*}=0,
\end{equation}
with $r=|r|e^{i\phi(r)}$, gives
\begin{equation}
d_*=
\frac{2\pi n-\phi(r)}{m_a}
-i\frac{-\ln|r|}{m_a}.
\end{equation}
The real part reproduces the resonance condition
\begin{equation}
m_a d_n=2\pi n-\phi(r),
\end{equation}
while the imaginary part determines the resonance width,
\begin{equation}
\gamma_n=-\ln|r|.
\end{equation}
Thus, highly reflective interfaces with $|r|\to1$ correspond to increasingly sharper resonances with $\gamma\to0$. Physically, the enhancement arises from the coherent excitation of a cavity mode by the axion-induced source. Near resonance, energy accumulates inside the cavity, while finite dissipation regulates the amplitude and gives the resonance a non-zero width.

\begin{figure}[h]
    \centering

    \begin{subfigure}{0.48\textwidth}
        \centering
        \includegraphics[width=\linewidth]{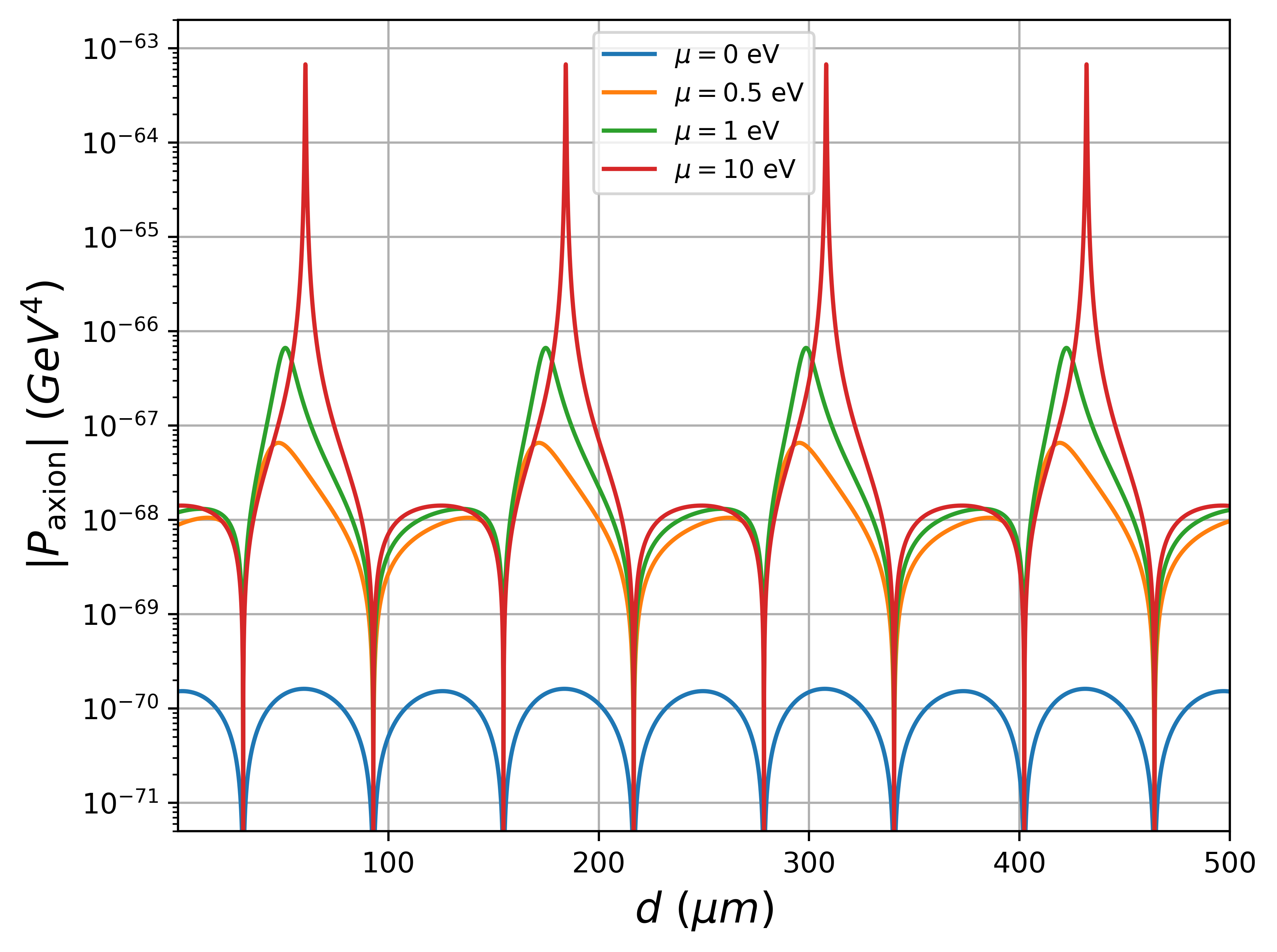}
        \label{fig:left}
    \end{subfigure}
    \hfill
    \begin{subfigure}{0.48\textwidth}
        \centering
        \includegraphics[width=\linewidth]{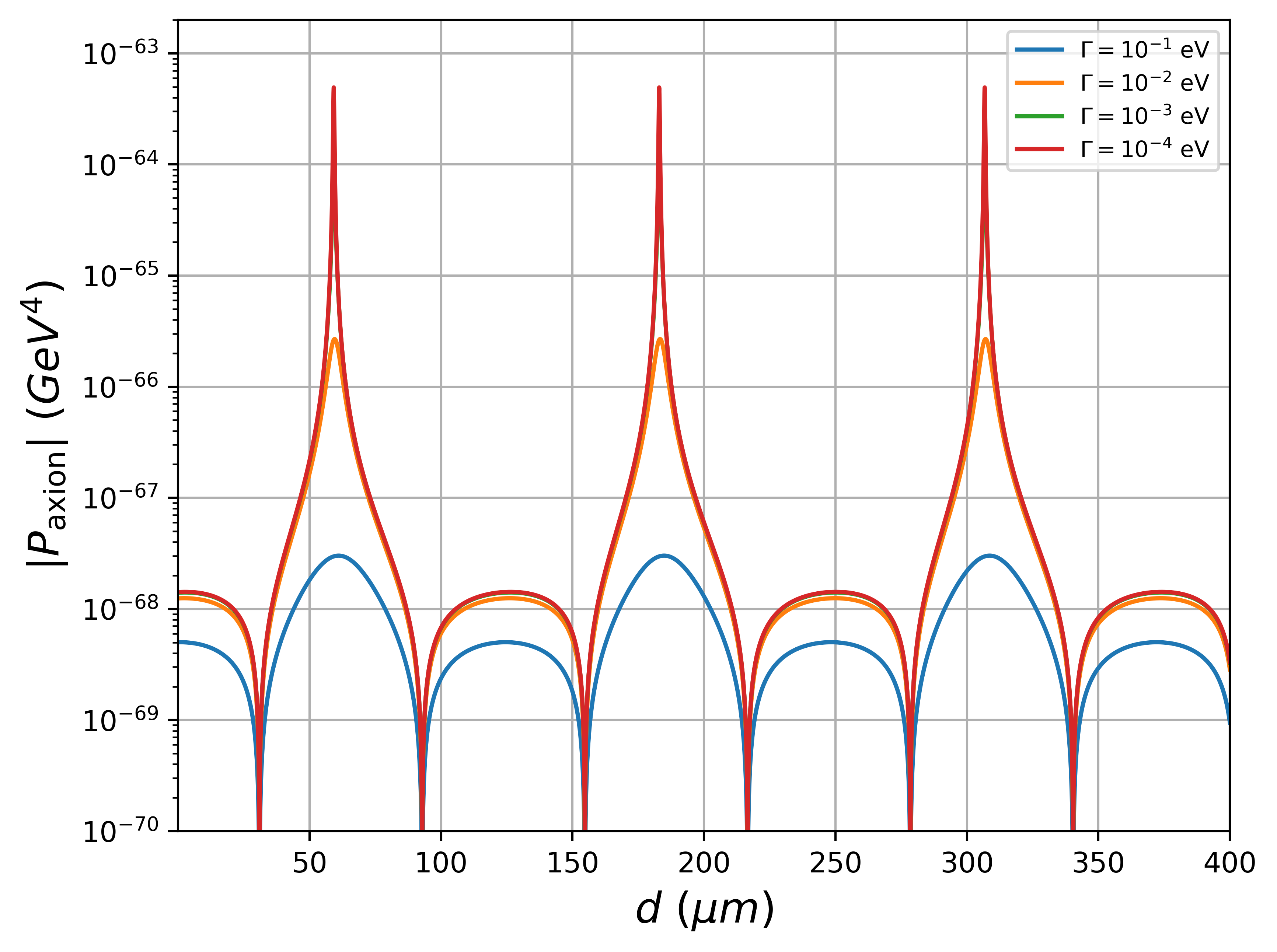}
        \label{fig:right}
    \end{subfigure}

    \caption{
    Axion-induced pressure as a function of plate separation for $m_a=0.01\,\mathrm{eV}$ and $T=300\,\mathrm{K}$. The upper panel shows the dependence of the resonant structure on the chemical potential $\mu$ for fixed $\Gamma=10^{-3}\,\mathrm{eV}$, while the lower panel shows the dependence on the dissipation parameter $\Gamma$ for fixed $\mu=5\,\mathrm{eV}$.
    }
    
    \label{fig:combined}
\end{figure}


\subsection{Doping and Dissipation}
The resonant enhancement of the axion-induced pressure is controlled by both the chemical potential $\mu$ and the dissipation rate $\Gamma$, which determine the electromagnetic response of graphene and therefore the quality of the cavity resonances. This is illustrated in Fig.~\ref{fig:combined}, where the pressure is shown as a function of plate separation for different values of $\mu$ (upper panel) and $\Gamma$ (lower panel). Increasing the chemical potential or decreasing the dissipation rate leads to sharper and more pronounced resonances, reflecting the enhanced reflectivity of the graphene sheets and the corresponding reduction in cavity losses. While the resonance positions exhibit a mild dependence on the graphene conductivity through the phase of the reflection coefficient, which is obtained from Eq.~\ref{eq:r_graphene}, their variation is relatively small compared with the dramatic changes in resonance height and width. The latter are governed by the magnitude of the reflection coefficient, or equivalently by the linewidth parameter $\gamma_n=-\ln|r|$, which decreases as the cavity becomes more reflective.

\begin{figure}[h] 
    \centering
    \includegraphics[width=1\linewidth]{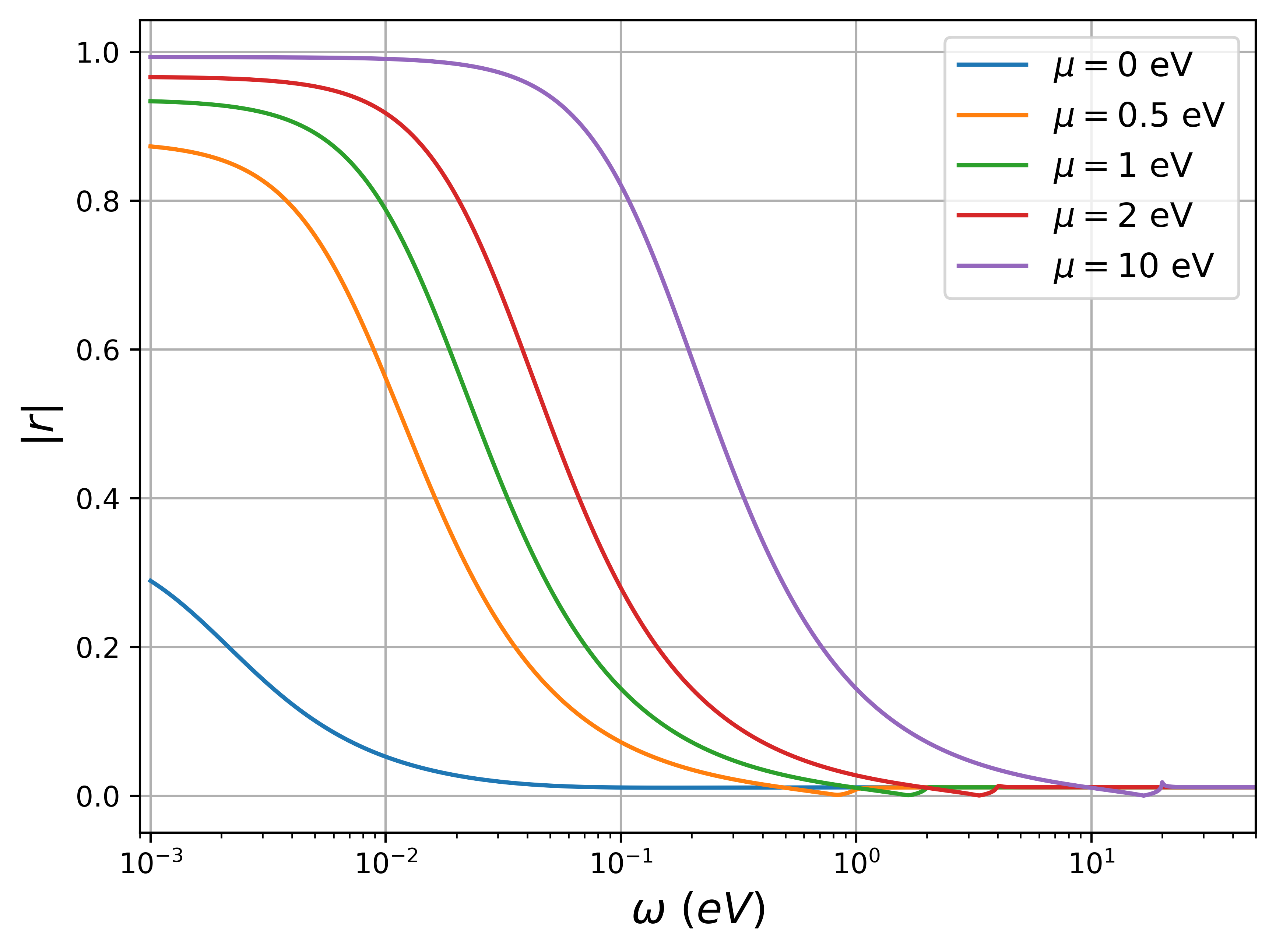}
    \caption{Magnitude of the reflection coefficient of graphene (conducting) sheet at normal incidence (Eq. \ref{eq:r_graphene}) as function of frequency, for a cavity at room temperature and damping parameter $\Gamma=10^{-3}$ eV in the Kubo conductivity model.}
   \label{Fig:mag_r_vs_omega_diff_mu}
\end{figure}


\begin{figure}[h] 
    \centering
    \includegraphics[width=1\linewidth]{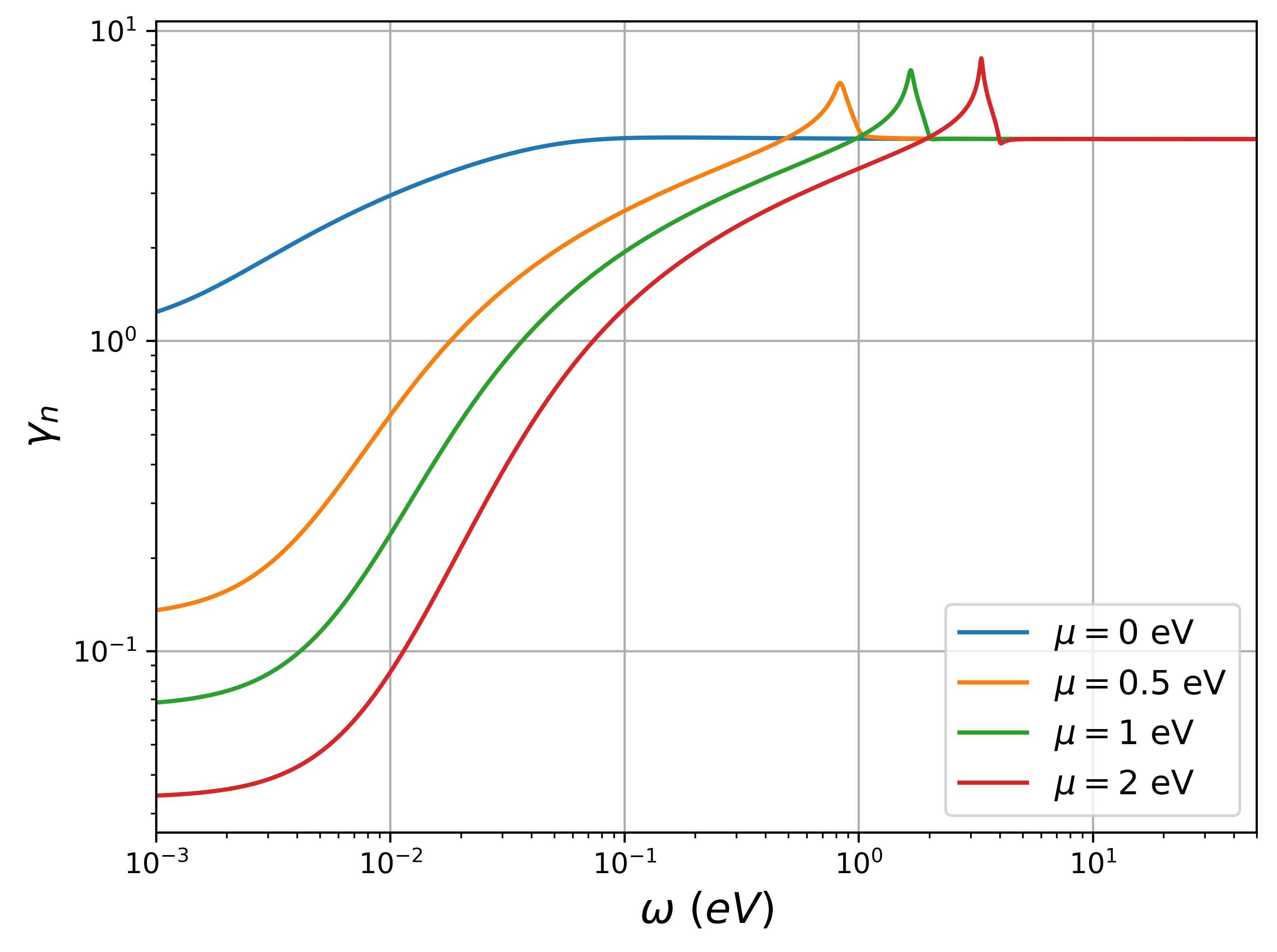}
    \caption{Resonance linewidth as a function of frequency, for graphene cavity at room temperature and damping parameter $\Gamma=10^{-3}$ eV in the Kubo conductivity model. }
   \label{Fig:gamma_vs_omega_diffmu}
\end{figure}

To understand this behaviour in more detail, Figure~\ref{Fig:mag_r_vs_omega_diff_mu} shows the magnitude of the reflection coefficient as a function of frequency for several values of $\mu$, while Fig.~\ref{Fig:gamma_vs_omega_diffmu} displays the corresponding linewidth parameter $\gamma_n$. At low frequencies, $\gamma_n$ depends strongly on $\mu$, as illustrated in Fig.~\ref{Fig:gamma_vs_omega_diffmu}. Increasing $\mu$ suppresses $\gamma_n$, leading to narrower and less dissipative resonances. As the frequency increases, all curves gradually approach a common asymptotic value. In this regime, the response becomes less sensitive to doping and is increasingly dominated by the universal high-frequency behaviour of graphene. The sharp peaks in $\gamma_n$ occur at frequencies where the graphene conductivity exhibits pronounced minima at the crossover between intraband-dominated response and interband/universal-conductivity behaviour, as also shown in Fig.~\ref{fig:modulus_sigma}. Since the cavity reflectivity is determined by the conductivity, a suppression of $|\sigma|$ reduces the reflectivity and increases the effective linewidth parameter $\gamma_n$, leading to broader and less efficient resonances. Smaller values of $\gamma_n$ therefore correspond to sharper, more coherent cavity resonances, while larger values correspond to broader and more weakly enhanced resonances. The plot illustrates how increasing $\mu$ suppresses the effect of dissipation, over a wider frequency range, allowing the cavity to maintain coherent resonant behaviour up to larger axion frequencies. It also shifts the resonance structure toward higher frequencies through the $\mu$-dependence of the graphene conductivity and cavity phase condition. 

\begin{figure}[h] 
    \centering
    \includegraphics[width=1\linewidth]{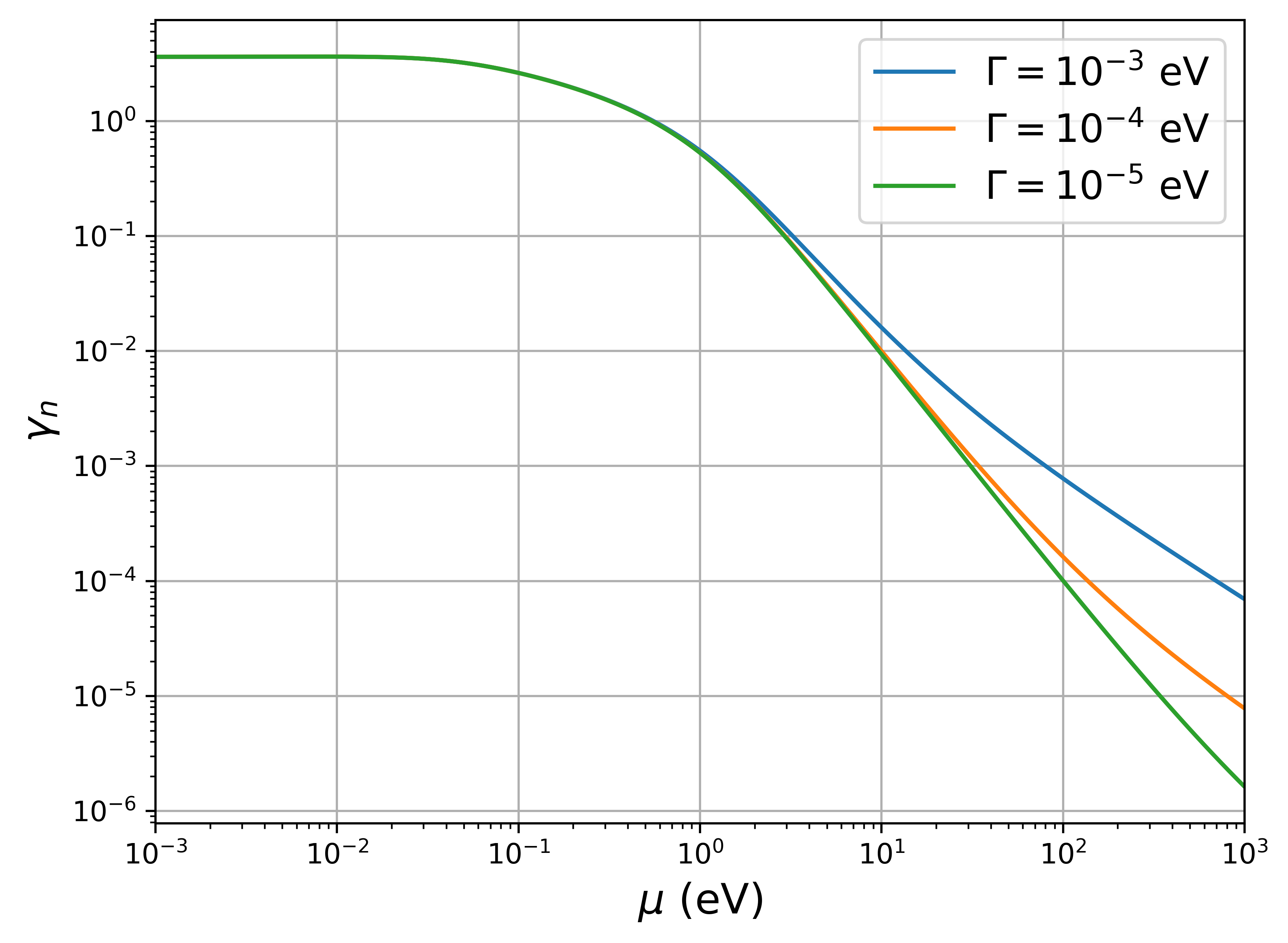}
    \caption{Resonance linewidth $\gamma_n$ as a function of the chemical potential $\mu$ for an axion mass $m_a=0.01\,\mathrm{eV}$ and several values of the dissipation rate $\Gamma$.}
   \label{fig:gamma_n_vs_mu}
\end{figure}

\begin{figure}[h]
    \centering
    \includegraphics[width=\linewidth]{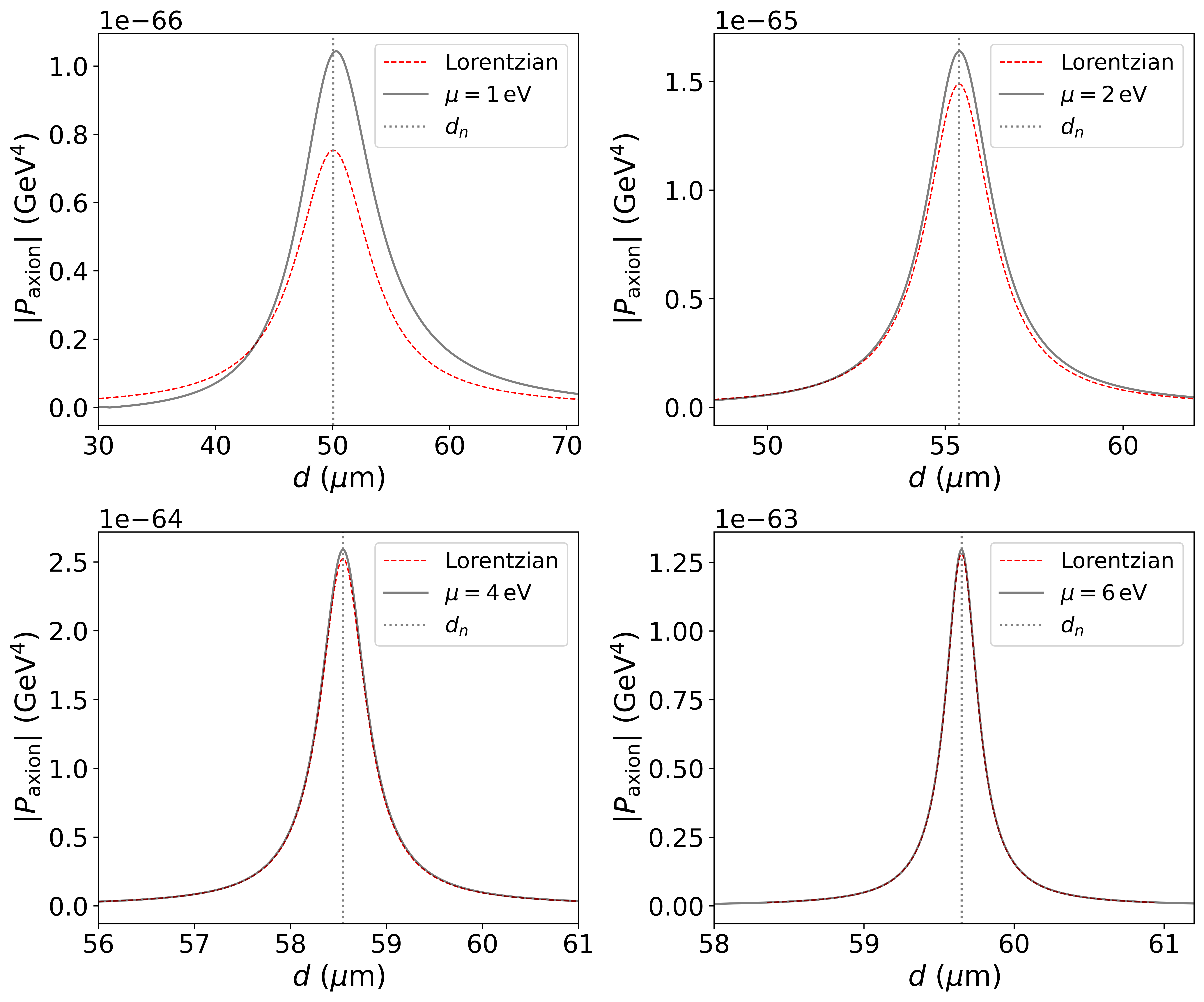}
    \caption{
   Resonant enhancement of the axion-induced pressure for different values of the graphene chemical potential. Increasing doping leads to sharper and more pronounced resonances, while also improving the agreement between the full numerical result and the local resonance approximation (the Lorentzian curve). The plots are shown for $m_a = 0.01~\mathrm{eV}$, $T = 300~\mathrm{K}$, and $\Gamma = 10^{-5}~\mathrm{eV}$.
    }
    \label{fig:fourplots}
\end{figure}

The dependence of the linewidth on the chemical potential is also illustrated in Fig.~\ref{fig:gamma_n_vs_mu} for a representative axion mass $m_a=0.01\,\mathrm{eV}$. For small chemical potentials, $\gamma_n$ is largely insensitive to $\mu$ and is controlled primarily by the dissipation rate $\Gamma$. As $\mu$ increases beyond the axion frequency scale, the linewidth decreases rapidly, reflecting the enhanced reflectivity of the graphene sheets and the corresponding reduction in cavity losses. This behaviour is more pronounced for smaller values of $\Gamma$, which allow significantly narrower resonances to be achieved at large chemical potential. Figure~\ref{fig:fourplots} illustrates the corresponding resonant pressure profiles for different values of the chemical potential. As $\mu$ increases, resonances become narrower and attain larger peak amplitudes, and the local Lorentzian approximation provides a progressively more accurate description of the full numerical result. This demonstrates that highly doped graphene cavities are particularly well suited to achieving strong resonant enhancement.

\begin{figure}[h]
    \centering
    \begin{subfigure}{0.48\textwidth} 
        \centering
        \includegraphics[width=\linewidth]{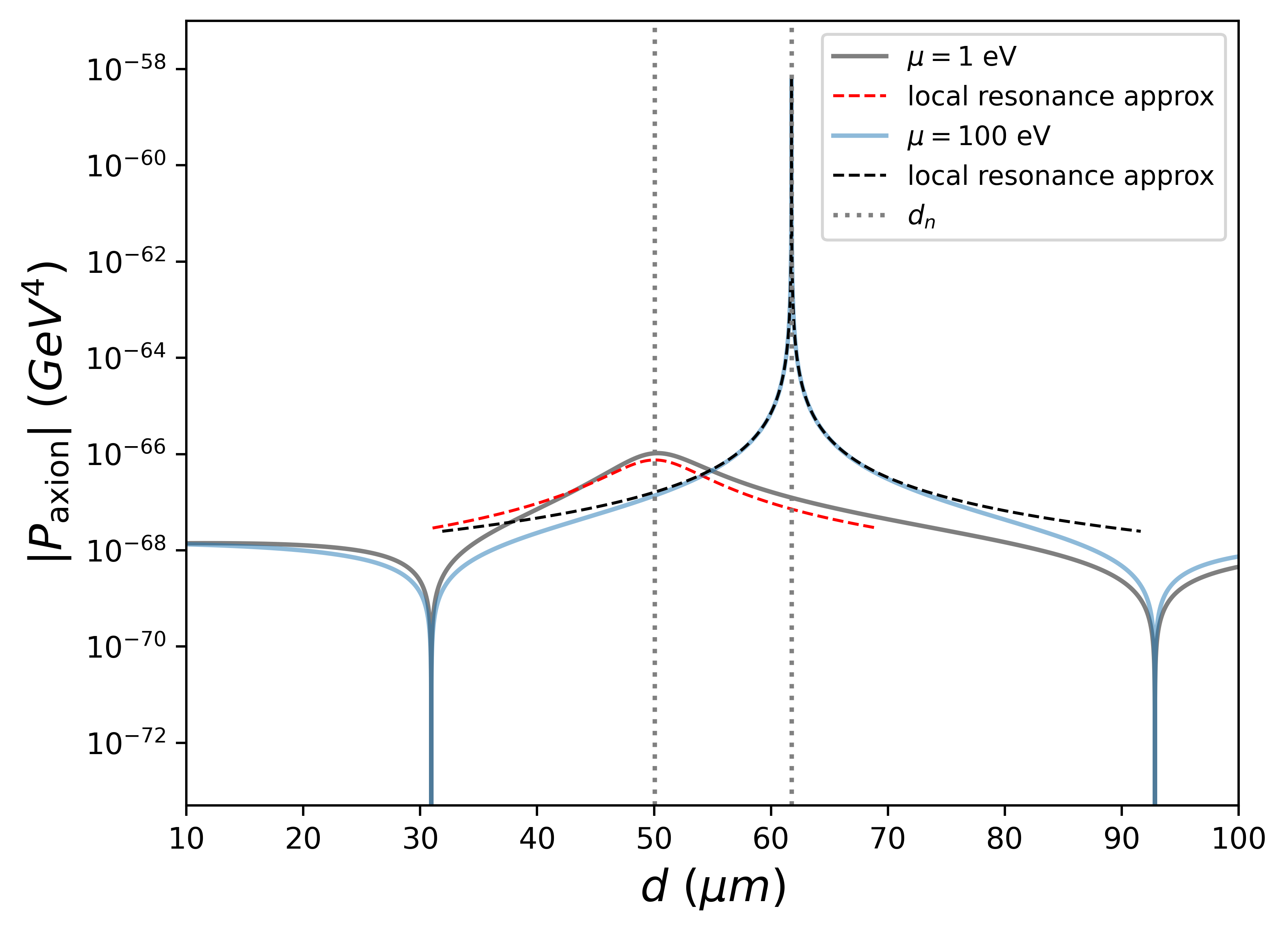}
    \end{subfigure}
    \hfill
    \begin{subfigure}{0.48\textwidth}
        \centering
        \includegraphics[width=\linewidth]{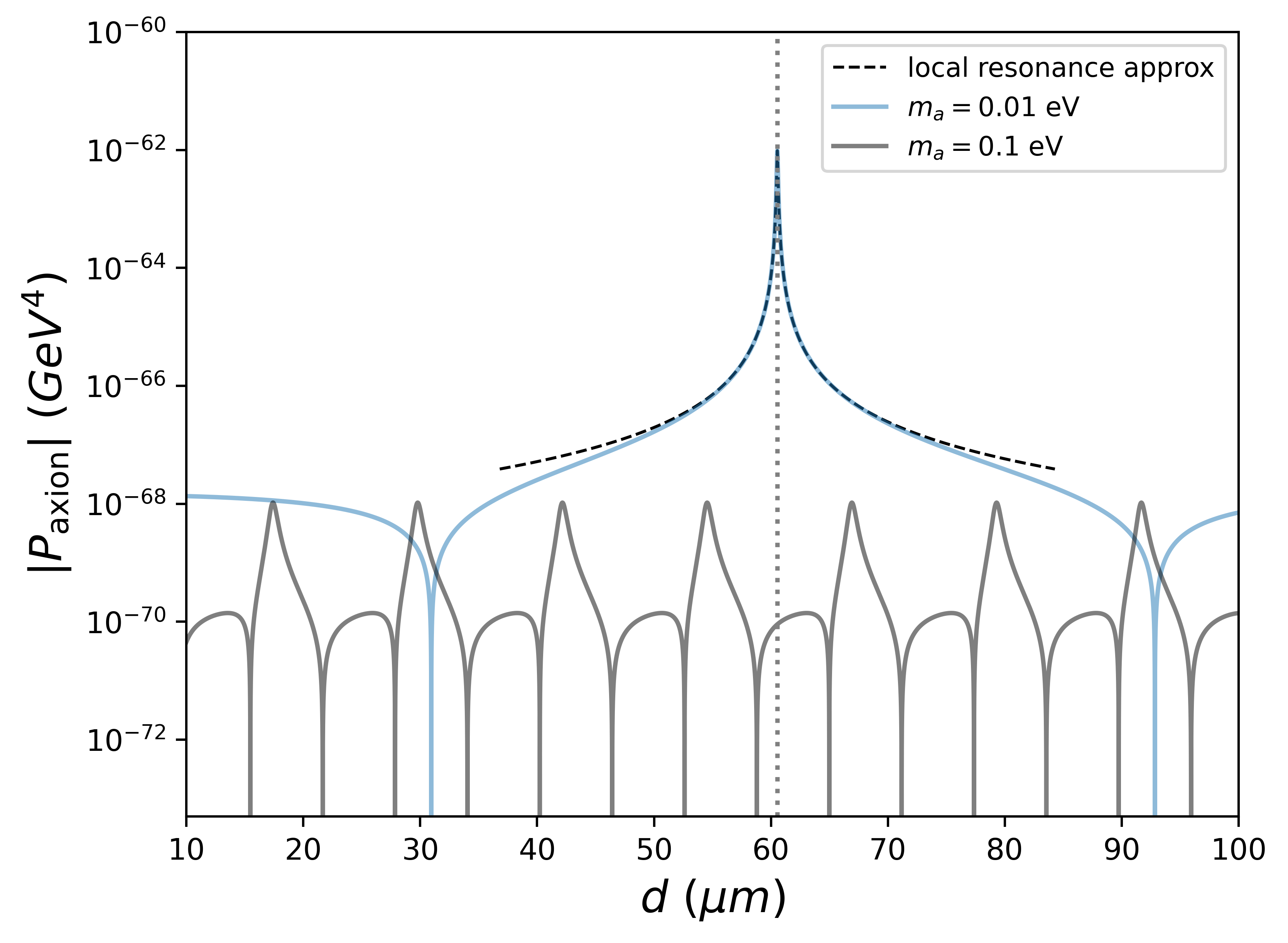}
    \end{subfigure}
    \caption{
Axion-induced pressure as a function of separation at $T=300$ K. The upper panel shows the effect of varying the graphene chemical potential for fixed $m_a=0.01~\mathrm{eV}$ and $\Gamma=10^{-5}~\mathrm{eV}$, while the lower panel shows the effect of varying the axion mass for fixed $\mu=10~\mathrm{eV}$ and $\Gamma=10^{-5}~\mathrm{eV}$. The dashed curves denote the local resonance approximation.
}
    \label{fig:changing_mass}  
\end{figure}

While the intrinsic properties of the cavity (finite conductivity) give rise to a finite resonance width 
\begin{equation}
\frac{\Delta \omega_{\mathrm{cav}}}{\omega} \sim \frac{\gamma_n}{m_a d},    
\end{equation}
the axion signal itself is also broadened by the finite velocity dispersion of dark matter in the Galactic halo. Since the axion energy is distributed over a range 
$\Delta E / E \sim v^2$, with $v \sim 10^{-3}$, the 
signal exhibits an irreducible fractional linewidth of order
\begin{equation}
\frac{\Delta\omega_{\rm DM}}{\omega}\sim v^2\sim10^{-6}.
\end{equation}
Consequently, once the intrinsic cavity linewidth becomes smaller than the astrophysical linewidth, 
\begin{equation}
\frac{\gamma_n}{m_a d} \lesssim v^2,
\end{equation}
further increases in the cavity quality factor no longer improve the resonance resolution, which is ultimately limited by the velocity dispersion of Galactic halo dark matter. For example, taking $m_a=0.01 \mathrm{eV}$ and $d=1 \mu \mathrm{m}$, this condition corresponds to $\gamma_n \lesssim 5 \times 10^{-8}$, or equivalently a reflectivity $|r| \gtrsim 1-5 \times 10^{-8}$, implying an almost perfectly reflecting cavity and losses several orders of magnitude smaller than those achievable in realistic graphene systems.



Figure~\ref{fig:changing_mass}  illustrates the dependence of the axion-induced pressure on both the graphene chemical potential and the axion mass. As the chemical potential is increased, the resonances become sharper and more pronounced, reflecting the enhanced conductivity of graphene and the corresponding increase in cavity reflectivity. The agreement between the full numerical result and the local resonance approximation also improves in this regime. In contrast, increasing the axion mass tends to reduce the magnitude of the pressure. Although the axion-induced current scales as $J_0\propto m_a$, larger masses probe higher frequencies where the graphene conductivity is reduced. The associated suppression of the cavity response outweighs the enhancement of the axion source over the parameter range considered in this work, indicating that the axion-induced pressure is primarily governed by the material response of the cavity rather than by the explicit mass dependence of the source current.

\subsection{Potential Sensitivity of Casimir Experiments }

To estimate the potential sensitivity of Casimir experiments in the presence of a magnetic field, we consider a setup capable of measuring the Casimir pressure at separations $d=5-50 ~ \mu m$ in the benchmark magnetic field $B_0 = 50$ T introduced above. While performing Casimir measurements under such extreme conditions is technically challenging, it remains within the realm of experimental feasibility. The chosen separation range reflects the fact that the conventional Casimir background increases rapidly at short distances (scaling as $d^{-4}$ in the ideal case), making small axion-induced deviations increasingly difficult to resolve. Moving to larger separations suppresses the background contribution and therefore provides a more favourable signal-to-background ratio for the resonantly enhanced axion-induced contribution.

We compare the axion-induced pressure, evaluated at resonances where it is maximally enhanced, to the conventional Casimir background, computed using the Lifshitz formalism. We assume a detection threshold of $\eta=1\%$, corresponding to a percent-level deviation from the Casimir pressure background $P_{\mathrm{bg}}$ being experimentally detectable. The sensitivity is then defined by the condition
\begin{equation}\label{eq:sensitivity}
P_{\mathrm{peak}}(g_{a\gamma\gamma},m_a) = \eta \, P_{\mathrm{bg}}(d_n),
\end{equation}
which we solve for the parameters $g_{a\gamma\gamma}$ and $m_a$. This benchmark is motivated by the precision achieved in modern Casimir-force measurements, where experimental uncertainties at the sub-percent to percent level have been reported over relevant separation ranges, depending on the experimental configuration. In particular, state-of-the-art micromechanical torsional oscillator experiments have achieved sub-percent precision in certain regimes, making the choice $\eta=1\%$ a conservative benchmark \cite{decca2007tests}. Precision Casimir measurements have been performed using torsion pendulums \cite{PhysRevLett.78.5}, atomic force microscopes \cite{mohideen1998precision,klimchitskaya1999complete}, and micromechanical torsional oscillators \cite{decca2007tests,decca2007novel,decca2005precise}, while additional experimental approaches have also been proposed \cite{grado1999possible,onofrio1995detecting}. 
Relevant sources of error include surface roughness, uncertainty in plate separation d, electrostatic patch potentials, finite conductivity corrections and finite-size effects (see Ref.~\cite{klimchitskaya2009casimir} for a comprehensive review).

To evaluate the peak pressure entering Eq.~\ref{eq:sensitivity}, we make use of the local resonance approximation. Within this approximation, Eq.~\ref{eq:P_res high doping}, the expression simplifies at the resonance peak $d = d_n$ to
\begin{equation}
P_{\mathrm{peak}}^{(n)}
=
\left|\frac{j_0^2}{4}\right|
\frac{\mathcal{N}_n}{|\alpha|^2 \gamma_n^2}.
\end{equation}

\noindent For axion dark matter, the effective current amplitude is
$
j_0 = g_{a\gamma\gamma}\, m_a^2 a_0 B_0,
$ where $a_0 = \sqrt{2\rho_{\mathrm{DM}}}/m_a$. Substituting this expression into the peak pressure yields
\begin{equation}
P_{\mathrm{peak}}^{(n)}
=
\frac{\rho_{\mathrm{DM}} B_0^2 m_a^2 g_{a\gamma\gamma}^2}{2}
\frac{\mathcal{N}_n}{|\alpha|^2 \gamma_n^2}.
\end{equation}

\noindent For temperatures $T\sim 30-300$ K and experimentally accessible range of separations, the system lies well within the highT regime, as shown in Fig \ref{fig:Casimirbkg_highT}, where the Casimir background can be described by Eq. \ref{eq:highTCasimir}, for a range of chemical potentials. In the large-separation (high-temperature) regime for graphene, the Casimir background is dominated by
\begin{equation}
|P_{\mathrm{bg}}(d)|
=
\frac{\zeta(3)\, T}{8\pi\, d^3}.
\end{equation}
Evaluating this at $d = d_n$ leads to 
\begin{equation}
\frac{\rho_{\mathrm{DM}} B_0^2 m_a^2 g_{a\gamma\gamma}^2}{2}
\frac{\mathcal{N}_n}{|\alpha|^2 \gamma_n^2}
=
\eta \,
\frac{\zeta(3)\, T}{8\pi\, d_n^3}.
\end{equation}
Solving for $g_{a\gamma\gamma}$ yields
\begin{equation}
g_{a\gamma\gamma,\mathrm{sens}}^{(n)}
=
\sqrt{
\frac{\eta\, \zeta(3)\, T}{4\pi\, \rho_{\mathrm{DM}} B_0^2\, m_a^2\, d_n^3}
\frac{|\alpha|^2 \gamma_n^2}{\mathcal{N}_n}
}.
\end{equation}

\begin{figure}
    \centering
    \includegraphics[width=1\linewidth]{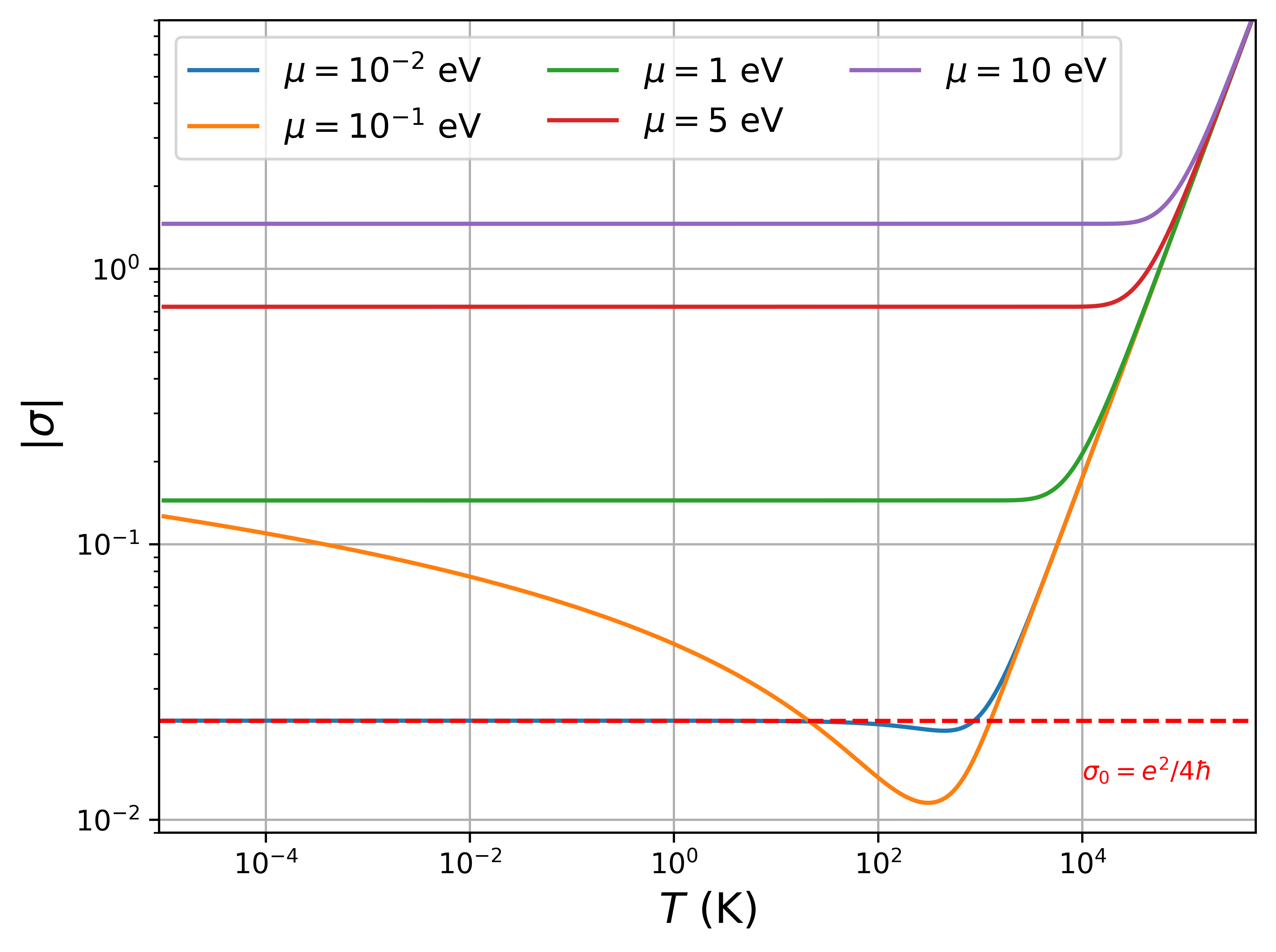}
    \caption{
Magnitude of the graphene conductivity $|\sigma|$ as a function of temperature for several values of the chemical potential $\mu$, evaluated at $\omega=m_a=0.2$ eV. The dashed line denotes the universal conductivity $\sigma_0=e^2/(4\hbar)$. Higher chemical potentials enhance the low-temperature response, while thermal excitations progressively wash out the dependence on $\mu$.
}
    \label{fig:mag_sigma_vs_T}
\end{figure}

While the above estimate assumes temperatures in the range $T \sim 30\text{--}300$ K, it is advantageous to consider lower temperatures. As discussed in Sec.~\ref{sec:casimir_effect} and illustrated in Fig.~\ref{fig:Casimirbkg_T4K}, in the regime $\mu \gg T$ the graphene response becomes effectively temperature-independent. This is clearly seen in Fig.~\ref{fig:mag_sigma_vs_T}, where the magnitude of the conductivity, computed within the Kubo formalism, remains approximately constant over a wide range of temperatures for sufficiently large chemical potential.  
This behaviour is consistent with the saturation of the static polarisation tensor entering the Lifshitz formalism, which in the same limit becomes only weakly dependent on temperature. 
In contrast, the Casimir background in the large-separation regime retains an explicit linear dependence on temperature, $P_{\mathrm{bg}} \sim T/d^3$. Consequently, lowering the temperature suppresses the background while leaving the axion-induced signal essentially unchanged. 
As a result, cryogenic operation provides an additional handle to improve the sensitivity, effectively enhancing the signal-to-background ratio in Eq.~\ref{eq:sensitivity}.

We fix the temperature at a cryogenic benchmark value of $T=4\,\mathrm{K}$, and compute the two pressures for a graphene sheet with chemical potential $\mu = 10\text{--}10^4\,\mathrm{eV}$. This parameter scan is used solely to characterise the dependence of the Casimir background on the graphene conductivity and to construct the fitting function below. As discussed below, the largest effective conductivities are subsequently reinterpreted within an effective-conductivity model for electronically decoupled multilayer graphene stacks. As shown in Fig.~\ref{fig:Casimirbkg_T4K}, increasing the chemical potential shifts the crossover separation between the $d^{-4}$ and $d^{-3}$ regimes to larger distances. To model the behaviour in the experimentally relevant range of separations $d = 5\text{--}500\,\mu\mathrm{m}$, we adopt the following fitting function:
\begin{equation}\label{eq:fit}
P(d,T=4~\text{K})=  \frac{d_0}{d^3} \left( 1+\frac{d_c}{d} \right),
\end{equation}
where $d_c$ characterises the crossover scale and $d_0$ sets the overall normalisation.

In contrast to the metallic case considered in Ref.~\cite{brax2024classical}, the graphene response is intrinsically frequency dependent through the Kubo conductivity $\sigma(\omega)$. Since the axion frequency is set by $\omega= m_a$, varying the axion mass modifies not only the resonance condition through $k_z=\omega$, but also the effective cavity reflectivity and phase shift. As a result, the resonance positions must be determined self-consistently as functions of $m_a$, rather than from a fixed relation $d_n \propto 1 / m_a$.

\begin{figure}[h]
    \centering

    \begin{subfigure}{\linewidth}
        \centering
        \includegraphics[width=\linewidth]{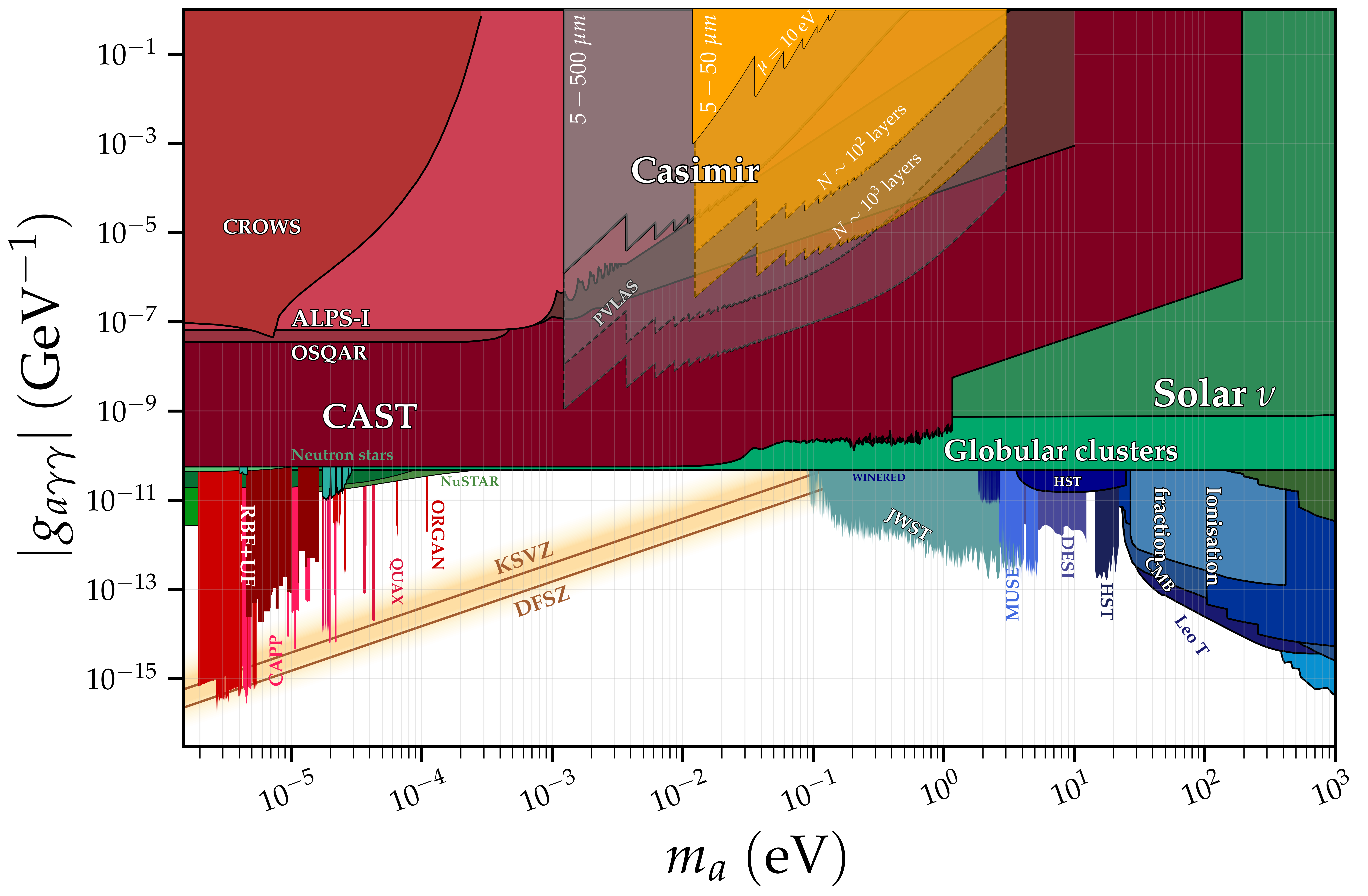}
    \end{subfigure}

    \vspace{0.3cm}

    \begin{subfigure}{\linewidth}
        \centering
        \includegraphics[width=\linewidth]{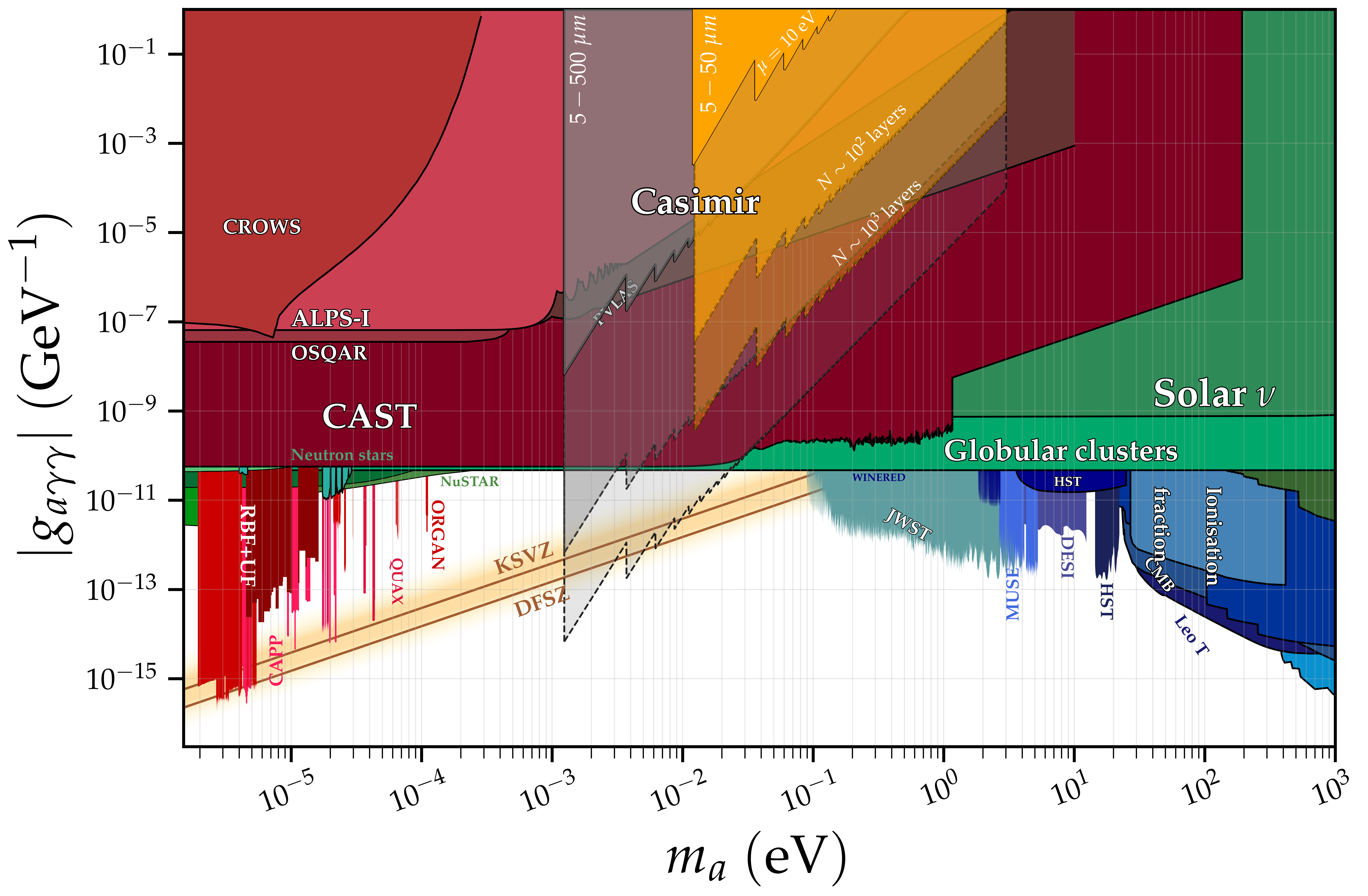}
    \end{subfigure}

    \caption{
    Sensitivity projections for the axion--photon coupling for the proposed graphene Casimir setup, for two different values of the graphene dissipation parameter, with the upper panel corresponding to $\Gamma = 10^{-3}\,\mathrm{eV}$ and the lower panel to $\Gamma = 10^{-5}\,\mathrm{eV}$. Solid curves correspond to the monolayer projection for $\mu=10\,\mathrm{eV}$, while the dashed $N \sim 10^2$ and $N \sim 10^3$ curves are  extrapolations obtained using the effective-conductivity model $\sigma\rightarrow N\sigma$.}
    \label{fig:sensitivity_gamma_comparison}
\end{figure}

The projected sensitivity of the present Casimir-based setup is shown in the two panels of Fig.~\ref{fig:sensitivity_gamma_comparison}, which display constraints on the axion--photon coupling $g_{a\gamma\gamma}$ as a function of the axion mass $m_a$, for a chemical potential $\mu =10\,\mathrm{eV}$. 
The strongest projected sensitivities are obtained by considering an effective stack of electronically decoupled graphene sheets, modelled by the replacement $\sigma \rightarrow N\sigma$, assuming that the layers are separated by distances much smaller than the relevant electromagnetic wavelength so that they experience essentially the same electric field, and, within linear response, their induced surface currents add linearly.
Since, in the low-temperature intraband regime, the graphene conductivity scales approximately linearly with the chemical potential, $
\sigma \propto \mu$, increasing the number of layers provides an effective means of reproducing the large conductivities associated with otherwise inaccessible chemical potentials. In this sense, stacks with $N\sim 10^2$ and $N \sim 10^3$ layers may be viewed as approximately mimicking the response of single graphene sheets with effective chemical potentials of order $\mu_{\rm eff}\sim10^3$ and $10^4\,\mathrm{eV}$, respectively, while remaining within a realistic single-layer doping regime. As discussed in Appendix~\ref{app:multilayer}, in the low-loss regime relevant here, the resonance linewidth decreases as $\gamma_n \propto N^{-2}$ and the peak resonant pressure scales as $P_z\propto N^4$, leading to a substantial enhancement of the signal. 
Accordingly, these benchmark projections may be regarded as extrapolations of the effective-conductivity model.
One possible experimental motivation for such enhanced effective conductivities may be provided by turbostratic multilayer graphene \cite{ uemura2018turbostratic}, in which adjacent graphene sheets are rotationally misaligned rather than Bernal (AB) stacked. The reduced interlayer overlap of the electronic orbitals largely preserves the linear Dirac-like band dispersion of individual graphene layers \cite{hass2008multilayer} while simultaneously increasing the overall conductivity through multiple parallel conduction channels.

Results are shown for $\Gamma=10^{-3}~\mathrm{eV}$ (upper panel) and $\Gamma=10^{-5}~\mathrm{eV}$ (lower panel). The two panels illustrate that the graphene dissipation rate $\Gamma$ is one of the primary parameters governing the projected sensitivity. Reducing $\Gamma$ sharpens the cavity resonances and enhances the resonant axion-induced pressure, leading to substantially stronger projected constraints over much of the parameter space.
Existing bounds from laboratory experiments (red), astrophysical observations (green), and cosmological probes (blue) are also shown for comparison \cite{AxionLimits}. Laboratory constraints include light-shining-through-a-wall (LSW) experiments, such as the CERN Resonant Weakly Interacting Sub-eV Particle (WISP) Search (CROWS) \cite{Betz_2013}, OSQAR at CERN, and the ALPS-I experiment \cite{Ehret_2010} at DESY, which probe axion-like particles through photon regeneration in strong magnetic fields. Astrophysical bounds arise from stellar cooling arguments, including globular clusters \cite{Ayala_2014} and solar neutrino observations \cite{Gondolo_2009}, while cosmological constraints are derived from large-scale structure, reionisation history, and cosmic microwave background (CMB) measurements, including data from HST \cite{Todarello_2026}, JWST \cite{pinetti2025first,roy2025sensitivity}, DESI, and dwarf galaxies such as Leo~T.

The orange region denotes the projected sensitivity for experimentally relevant separations $d = 5-50\,\mu\mathrm{m}$, while the grey region indicates the prospective sensitivity for an extended range $d = 5\text{--}500\,\mu\mathrm{m}$. The diagonal lines correspond to benchmark QCD axion models, namely the KSVZ \cite{kim1979weak,shifman1980can} and DFSZ \cite{zhitnitsky1980possible,dine1981simple} models. The maximum axion mass shown in the plots is determined by the range of validity of the Dirac description of graphene, which applies at characteristic energies below approximately 3 eV, where graphene can be considered a system of massless, or rather light, free electronic quasiparticles governed by the Dirac equation. This upper mass range also approximately coincides with the regime in which the resonant cavity enhancement begins to diminish: the sharp Lorentzian resonance structure gradually transitions into a pattern of interference fringes, such that the local Lorentzian approximation employed in Eq.~\ref{eq:P_res high doping} is no longer applicable. The marked improvement in the projected sensitivity from the upper to the lower panel demonstrates that the graphene dissipation rate is a key driver of the experimental reach. Reducing $\Gamma$ from $10^{-3}\,\mathrm{eV}$ to $10^{-5}\,\mathrm{eV}$ significantly enhances the resonant cavity response, strengthening the projected constraints by several orders of magnitude over much of the accessible axion mass range. In addition to improving the overall sensitivity, reducing $\Gamma$ modifies the mass dependence of the projected constraints. The exclusion contours for $\Gamma=10^{-3}\,\mathrm{eV}$ exhibit a continuously varying slope, whereas those for $\Gamma=10^{-5}\,\mathrm{eV}$ are both steeper and more nearly uniform across the accessible mass range. Consequently, the projected sensitivity for $\Gamma=10^{-5}\,\mathrm{eV}$ extends towards the KSVZ and DFSZ QCD axion models over part of the parameter space considered, highlighting the importance of achieving low graphene dissipation rates. More broadly, these results demonstrate that resonant enhancement can significantly extend the reach of Casimir-based searches for axion dark matter, motivating further investigation of experimentally realistic graphene platforms.

The detailed structure of the projected exclusion contours is determined by the resonant behaviour of the cavity, which gives rise to the characteristic sawtooth pattern.
This structure arises from the discrete set of cavity resonances. As the axion mass increases, a given resonance shifts to smaller separations, where the Casimir background is larger, leading to a gradual degradation in sensitivity. When a new resonance mode enters the experimentally accessible separation range, the corresponding larger separation reduces the background, resulting in a sudden improvement in sensitivity. This produces the characteristic sawtooth pattern. The overall behaviour of the projected sensitivity at large axion masses can be understood from the frequency dependence of the graphene response. Since the axion frequency is set by $\omega=m_a$, increasing the axion mass probes the graphene response at progressively higher frequencies.
In the high-frequency regime, the graphene conductivity typically decreases with increasing frequency, leading to a weaker induced current response and reduced reflectivity. This reduces the efficiency of resonant enhancement, causing the axion-induced pressure to decrease with increasing axion mass, as illustrated in the lower panel of Fig.~\ref{fig:changing_mass}. Simultaneously, increasing $\omega$ changes the phase accumulated by the cavity modes, modifying the interference condition and shifting the resonance positions.
Increasing $m_a$ modifies the resonance structure of the cavity itself. Since the resonance condition scales approximately as $d_n\sim 1/m_a$, the resonances shift to smaller separations and become more densely spaced within a fixed experimental separation range (since the spacing between neighbouring resonances also decreases as $\Delta d\sim 1/m_a$.). In addition, the reduced reflectivity modifies the resonance profiles and generally weakens the sharpness of the resonant peaks at large masses. These combined effects lead to the progressive degradation of the projected sensitivity at high axion masses visible in the foreseen exclusion contours. At sufficiently large masses, the resonant enhancement is substantially weakened and the sharp sawtooth structure gradually smooths into a more continuous curve as the individual resonance branches become less distinct.  

\begin{figure}[h]
    \centering
    \includegraphics[width=1\linewidth]{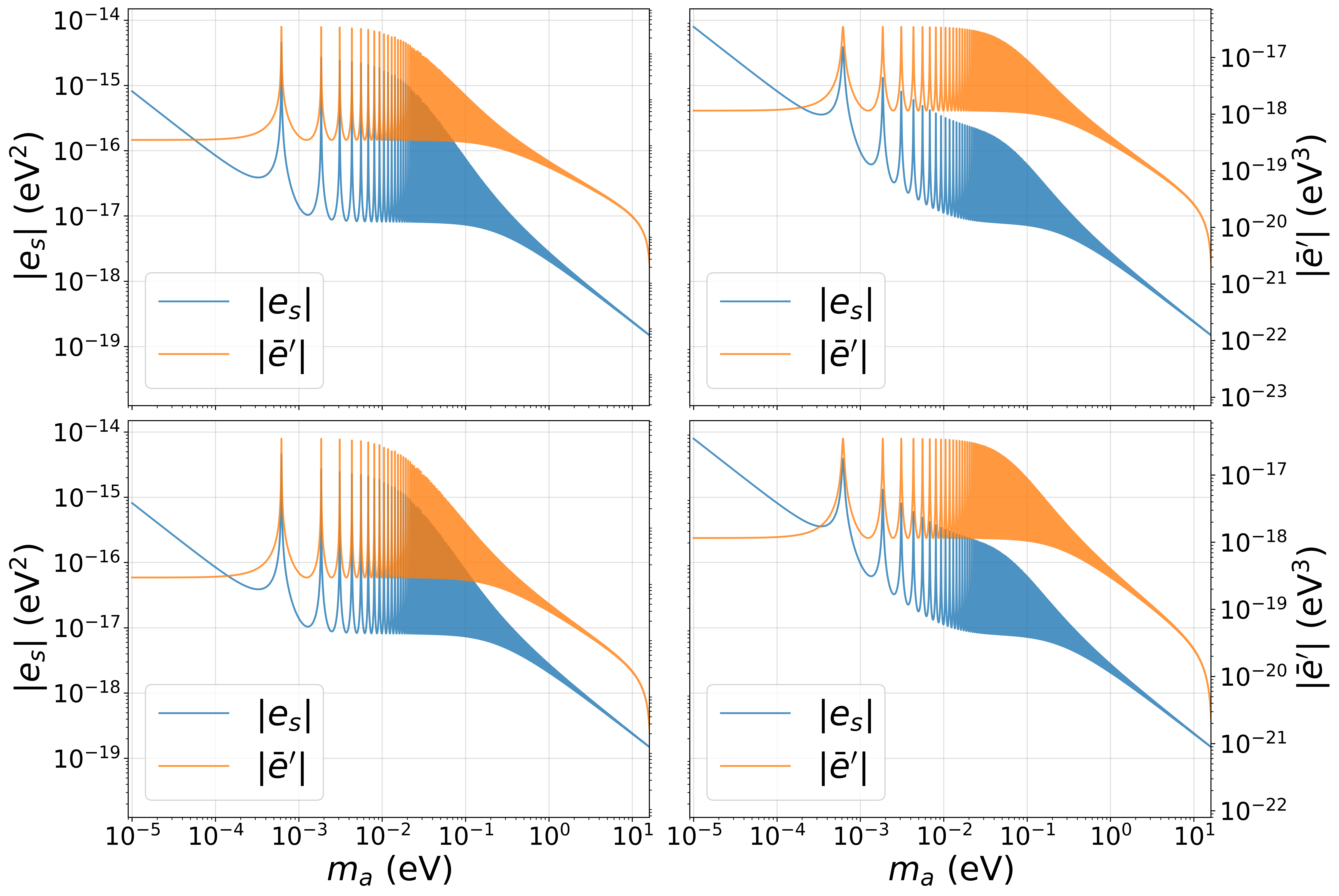}    
\caption{Magnitudes of the surface electric field $\left|e_s\right|$ and the averaged derivative field $\left|\bar{e}^{\prime}\right|$, as functions of the axion mass for $\mu=10\,\mathrm{eV}$. The panels, ordered from top left to bottom right, correspond to $(T,\Gamma)=\left(4\,\mathrm{K},10^{-3}\,\mathrm{eV}\right)$, $\left(4\,\mathrm{K},10^{-2}\,\mathrm{eV}\right)$,$\left(300\,\mathrm{K},10^{-3}\,\mathrm{eV}\right)$, and $\left(300\,\mathrm{K},10^{-2}\,\mathrm{eV}\right)$, respectively.}
    \label{fig:fields_vs_mass_four_panel}
\end{figure}

This behaviour is also illustrated in Fig. \ref{fig:fields_vs_mass_four_panel}, which shows the magnitudes of the surface electric field $|e_s|$ and the averaged derivative field $|\bar e'|$ as functions of the axion frequency $\omega=m_a$ for fixed cavity separation.  At low masses, the cavity exhibits a set of well-separated resonances corresponding to coherent excitation of cavity eigenmodes by the axion-induced source. As the axion mass increases, the resonances become progressively more densely spaced, consistent with the scaling in the resonance condition. The overall resonant enhancement decreases with increasing mass due to the weakening graphene response at high frequencies. Consequently, the efficiency of resonant energy build-up inside the cavity is suppressed, causing the resonance peaks to decrease in amplitude. At sufficiently large masses, the individual resonances begin to overlap and the sharp Lorentzian structure gradually transitions into a dense interference-fringe pattern. In this regime, the local resonance approximation is no longer valid and the response becomes increasingly non-resonant, approaching a smooth decaying envelope at high frequencies.

\section{Conclusions}
\label{sec:conclusions}

In the foregoing, we have investigated the axion-induced modification of the Casimir pressure in graphene cavities in the presence of an external magnetic field. The oscillating axion dark-matter background sources electromagnetic fields through the axion--photon coupling, generating induced surface currents on the graphene sheets and consequently a resonantly enhanced classical pressure. Using a Green's-function approach, together with the graphene conductivity derived within the Kubo formalism, we obtained analytic expressions for the induced electric field and the corresponding pressure, incorporating the effects of dissipation, chemical potential, and temperature. We showed that the system exhibits a tower of resonances determined by the cavity geometry and the axion Compton wavelength, with the resonant structure strongly controlled by the graphene conductivity and dissipation rate.

We further investigated the dependence of the resonant enhancement on the chemical potential and relaxation rate, demonstrating that graphene provides a tunable platform for optimising the axion-induced response. In particular, increasing the chemical potential enhances the resonant double-pole contribution, making it increasingly dominant relative to the single-pole term. We also estimated the potential sensitivity of Casimir experiments by comparing the resonantly enhanced axion-induced pressure to the conventional Casimir background, showing that detectable signals may arise within experimentally relevant parameter ranges.

While achieving the parameter regime with the strongest projected sensitivity remains experimentally challenging, particularly regarding precision measurements at large separations, and the realisation of sufficiently high effective conductivities, ongoing advances in graphene engineering and precision force measurements may help bring such configurations closer to experimental accessibility. In particular, effective stacks of electronically decoupled graphene sheets may provide a simple theoretical framework for exploring enhanced cavity responses. The present results therefore highlight the potential of tunable graphene cavities as a promising novel platform for probing axion dark matter through resonantly enhanced electromagnetic effects. Future work could include the study of finite mass-gap effects, explicit modelling of multilayer graphene configurations, spatially dispersive conductivity, and more realistic experimental geometries.



\acknowledgments 

This project has received funding from the French Alternative Energies and Atomic Energy Centre (CEA) under the AUDACE research programme.

\appendix

\section{Green's Function Solution}
\label{appendix:A}

In this appendix, we present the explicit Green's-function solution of the driven wave equation in the cavity geometry considered in the main text. 
The coefficients appearing in the piecewise solutions are obtained by imposing the electromagnetic boundary conditions at the interfaces. 
For completeness, we list below the expressions in the different spatial regions.

\subsection{Solution for $z_0 <0$: }
For a source located in the region $z_0<0$, the coefficients entering the Green's function solution are given by
\begin{equation}
A_0=0,
\end{equation}

\begin{equation}
\begin{aligned}
B_0=
\frac{-i e^{-ik_z z_0}}{2k_z\mathcal D}
\Big[
&k_z\sigma(k_z\sigma+2\omega)
+k_z\sigma(2\omega-k_z\sigma)e^{2ik_z d} \\
&\quad
+e^{2ik_z z_0}
\Big(
k_z^2\sigma^2 e^{2ik_z d}
-(k_z\sigma+2\omega)^2
\Big)
\Big].
\end{aligned}
\end{equation}

\begin{equation} \label{eq:C0}
C_0=-\frac{i}{2 k_z} e^{-i k_z z_0},
\end{equation}

\begin{equation}  \label{eq:D0}
D_0=
\frac{-i \sigma e^{-i k_z z_0}}{2 \mathcal{D}}
\left[
k_z\sigma+2\omega+(2\omega-k_z\sigma)e^{2ik_z d}
\right],
\end{equation}

\begin{equation}
E_0=\frac{i \omega(k_z \sigma+2 \omega)}{k_z \mathcal{D}} e^{-i k_z z_0},
\end{equation}

\begin{equation}
F_0=\frac{-i \omega \sigma}{ \mathcal{D}} e^{i k_z \left(2 d-z_0\right)},
\end{equation}

\begin{equation} 
\label{eq:G}
G_0=\frac{2 i \omega^2}{k_z \mathcal{D}} e^{-i k_z z_0},
\end{equation}

\begin{equation}
H_0=0.
\end{equation}

\subsection{Solution for $0<z_0<d$: }
For a source located inside the cavity region, $0<z_0<d$, the coefficients entering the Green's-function solution are given by

\begin{equation}
I_0=0,    
\end{equation}

\begin{equation}
J_0=\frac{i \omega e^{-i k_z z_0}}{ k_z \mathcal{D}} \left(-k_z \sigma e^{2 i k_z d}+ \left(k_z \sigma  +2 \omega \right) e^{2 i k_z z_0}\right)  ,
\end{equation}

\begin{equation}
K_0=\frac{-i \sigma e^{-i k_z z_0}}{2 \mathcal{D}}  
\left(-k_z \sigma e^{2 i k_z d}+ \left(k_z  \sigma +2 \omega \right) e^{2 i k_z z_0}\right) ,
\end{equation}

\begin{equation}
L_0=
\frac{i(k_z\sigma+2\omega)}
{2k_z\mathcal D}
\left[
(k_z\sigma+2\omega)e^{ik_z z_0}
-
k_z\sigma\,e^{ ik_z(2d-z_0)}
\right],
\end{equation}

\begin{equation}
M_0=
\frac{i(k_z\sigma+2\omega)}
{2k_z\mathcal D}
\left[
(k_z\sigma+2\omega)e^{-ik_z z_0}
-k_z\sigma e^{ik_z z_0}
\right],
\end{equation}

\begin{equation}
N_0=
\frac{-i\sigma}{2\mathcal D}
\left[
(k_z\sigma+2\omega)e^{ik_z(2d- z_0)}
-k_z\sigma e^{ik_z (2d+ z_0)}
\right],
\end{equation}

\begin{equation}
O_0=
\frac{i\omega}{k_z\mathcal D}
\left[
(k_z\sigma+2\omega)e^{-ik_z z_0}
-
k_z\sigma e^{ik_z z_0}
\right],
\end{equation}

\begin{equation}
P_0=0.
\end{equation}

\subsection{Solution for $z_0>d$: }

Finally, for a source located in the region $z_0>d$, the coefficients take the form

\begin{equation}
\begin{aligned}
Q_0=
\frac{-i e^{-ik_z(2d+z_0)}}{2k_z\mathcal D}
\Big[
&k_z^2\sigma^2 e^{4ik_z d}
-\left(k_z\sigma+2\omega\right)^2 e^{2ik_z d} \\
&+k_z\sigma\left(k_z\sigma+2\omega\right)e^{2ik_z z_0} \\
&-k_z\sigma\left(k_z\sigma-2\omega\right)e^{2ik_z(d+z_0)}
\Big],
\end{aligned}
\end{equation}

\begin{equation}
R_0=0,
\end{equation}

\begin{equation}
S_0=
\frac{-i\sigma e^{ik_z z_0}}{2\mathcal D}
\left[
2\omega-k_z\sigma
+(k_z\sigma+2\omega)e^{-2ik_z d}
\right],
\end{equation}

\begin{equation}
T_0=-\frac{i}{2 k_z} e^{i k_z z_0},
\end{equation}

\begin{equation}\label{eq:U0}
U_0=\frac{-i \omega \sigma}{\mathcal{D}} e^{i k_z z_0},
\end{equation}

\begin{equation}\label{eq:V0}
V_0=\frac{i \omega\left(k_z \sigma+2 \omega\right)}{k_z \mathcal{D}} e^{i k_z z_0},
\end{equation}

\begin{equation}
W_0=0,
\end{equation}

\begin{equation}\label{eq:X0}
X_0=\frac{2 i \omega^2}{k_z \mathcal{D}} e^{i k_z z_0}.
\end{equation}

\subsection{Sheet Derivatives $e'(0^\pm)$}
\label{app:leftsheet-sectorA}

\subsubsection{Contribution from $z_0<0$}
\noindent For sources located in the region $z_0<0$, the reduced Green's  function takes the form
\begin{equation}
\tilde G(z,z_0)=
\begin{cases}
C_0(z_0)e^{ik_zz}+D_0(z_0)e^{-ik_zz}, & z_0<z<0,\\[2pt]
E_0(z_0)e^{ik_zz}+F_0(z_0)e^{-ik_zz}, & 0<z<d.
\end{cases}
\end{equation}
The corresponding contributions to the field derivatives are therefore
\begin{equation}
\begin{aligned}
\left.e'(0^+)\right|_{z_0<0}
&= j_0 i k_z \int_{-\infty}^{0} dz_0 \left[E_0(z_0)-F_0(z_0)\right], \\
\left.e'(0^-)\right|_{z_0<0}
&= j_0 i k_z \int_{-\infty}^{0} dz_0 \left[C_0(z_0)-D_0(z_0)\right].
\end{aligned}
\end{equation}
Using Eq.~\ref{eq:C0} and \ref{eq:D0}, one finds
\begin{equation}
C_0(z_0)-D_0(z_0)=e^{-ik_zz_0}\,\mathcal C,
\end{equation}
where we have defined the $z_0-$independent factor

\begin{equation}
\mathcal C \equiv
-\frac{i}{2}
\left[
\frac{1}{k_z}
-
\frac{\sigma}{\mathcal D}
\left(
k_z\sigma(1-e^{2ik_z d})
+2\omega(1+e^{2ik_z d})
\right)
\right].
\end{equation}

\noindent With the retarded prescription $k_z\to k_z+i0^+$ (so that $\mathrm{Im}\,k_z>0$ and the integrand decays as $z_0\to-\infty$),
\begin{equation}\label{eq:neg_inf_zero_first}
\int_{-\infty}^{0}\!dz_0\,e^{-ik_zz_0}=\frac{i}{k_z},
\end{equation}
and hence
\begin{equation}
\int_{-\infty}^{0}\!dz_0\,\bigl[C_0(z_0)-D_0(z_0)\bigr]
=\frac{i}{k_z}\,\mathcal C. \\[2pt]
\label{eq:intCD-sectorA}
\end{equation}

\noindent Similarly,
\begin{equation}
E_0(z_0)-F_0(z_0)=e^{-ik_zz_0}\,\mathcal B,
\end{equation}
with

\begin{equation}
\mathcal B
=
\frac{i\omega}{\mathcal D}
\left[
\sigma\left(1+e^{2ik_z d}\right)
+
\frac{2\omega}{k_z}
\right].
\end{equation}

\noindent Using again Eq.~\ref{eq:neg_inf_zero_first}, one obtains

\begin{equation}
\int_{-\infty}^{0}\!dz_0\,\bigl[E_0(z_0)-F_0(z_0)\bigr]
=\frac{i}{k_z}\,\mathcal B.
\label{eq:intEF-sectorA}
\end{equation}

\subsubsection{Contribution from $0<z_0<d$}
\label{app:sectorB}

For sources located inside the cavity, $0<z_0<d$, the reduced Green's function takes the form
\begin{equation}
\tilde G(z,z_0)=
\begin{cases}
I_0 e^{ik_z z}+J_0(z_0)e^{-ik_z z}, & z<0,\\[2pt]
K_0(z_0)e^{ik_z z}+L_0(z_0)e^{-ik_z z}, & 0<z<z_0,
\end{cases}
\end{equation}
with $I_0=0$ from the radiation condition as $z\to -\infty$. Therefore,
\begin{equation}
\begin{aligned}
\partial_z \tilde G(0^-,z_0) &= - i k_z\, J_0(z_0), \\
\partial_z \tilde G(0^+,z_0) &= i k_z\!\left[K(z_0)-L(z_0)\right].
\end{aligned}
\end{equation}
The corresponding contributions to the field derivatives are
\begin{equation}
\begin{aligned}
\left.e'(0^-)\right|_{0<z_0<d}
&= -ik_z j_0\int_{0}^{d}\!dz_0\, \,J_0(z_0),\\
\left.e'(0^+)\right|_{0<z_0<d}
&= ik_z j_0\int_{0}^{d}\!dz_0\,\!\left[K_0(z_0)-L_0(z_0)\right].
\end{aligned}
\end{equation}

\noindent The coefficient $J_0(z_0)$ is given by,
\begin{equation}
J_0(z_0)=\frac{i\omega}{k_z\mathcal D}
\left[
\left(k_z\sigma+2\omega\right)e^{ik_z z_0}
-k_z\sigma\,e^{2ik_z d}\,e^{-ik_z z_0}
\right].
\end{equation}
Using
\begin{equation} \label{eq: 0d_integrals}
\begin{aligned}
\int_{0}^{d} dz_0\, e^{ik_z z_0} &= \frac{e^{ik_z d}-1}{ik_z}, \\
\int_{0}^{d} dz_0\, e^{-ik_z z_0} &= \frac{1-e^{-ik_z d}}{ik_z},
\end{aligned}
\end{equation}
one obtains
\begin{equation}
\begin{aligned}
\int_{0}^{d} dz_0\, J_0(z_0)
&=\frac{\omega}{k_z^{2}\mathcal D}
\Bigg[
\left(k_z\sigma+2\omega\right)\left(e^{ik_z d}-1\right) \\
&\qquad
-k_z\sigma e^{2ik_z d}\left(1-e^{-ik_z d}\right)
\Bigg].
\end{aligned}
\label{eq:intI-sectorB}
\end{equation}
Next, given 

\begin{equation}
\begin{aligned}
K_0(z_0)
&=\frac{-i\sigma}{2\mathcal D}
\Big( k_z\sigma \left(e^{2ik_z z_0}  -e^{2ik_z d}\right)  \\
&\qquad
+2\omega e^{2ik_z z_0}
\Big)e^{-ik_z z_0}, \\
L_0(z_0)
&=\frac{i\left(k_z\sigma+2\omega\right)}{2k_z\mathcal D}
\Big( k_z \sigma \left(e^{2ik_z z_0} -e^{2ik_z d}\right) \\
&\qquad +2\omega e^{2ik_z z_0}
\Big)e^{-ik_z z_0},
\end{aligned}
\end{equation}
we conveniently define
\begin{equation}
\mathcal F(z_0)\equiv
\left(
-k_z\sigma e^{2ik_z d}+ e^{2ik_z z_0}\left(  k_z\sigma +2\omega \right) \right)e^{-ik_z z_0}.
\end{equation}
Then

\begin{equation}
\begin{aligned}
K_0(z_0)-L_0(z_0)
&=\frac{-i}{\mathcal D}
\left(\sigma+\frac{\omega}{k_z}\right)\mathcal F(z_0).
\end{aligned}
\end{equation}

\noindent Moreover,
\begin{equation}
\begin{aligned}
\int_{0}^{d} \mathcal F(z_0)\,dz_0
&=
\left(k_z\sigma+2\omega\right)
\int_{0}^{d} dz_0\,e^{ik_z z_0}  \\
&\quad
-k_z\sigma e^{2ik_z d}
\int_{0}^{d} dz_0\,e^{-ik_z z_0},
\end{aligned}
\end{equation}
so that
\begin{equation}
\begin{aligned}
\int_{0}^{d} \mathcal F(z_0)\,dz_0
&=
\frac{1}{ik_z}
\Big[
\left(k_z\sigma+2\omega\right)\left(e^{ik_z d}-1\right) \\
&\qquad
-k_z\sigma e^{2ik_z d}\left(1-e^{-ik_z d}\right)
\Big].
\end{aligned}
\label{eq:intF-sectorB}
\end{equation}

\noindent Combining the above results yields

\begin{equation}
\begin{aligned}
\int_{0}^{d} dz_0\,\bigl[K_0(z_0)-L_0(z_0)\bigr]
&=
\frac{-i}{\mathcal D}
\left(\sigma+\frac{\omega}{k_z}\right)
\int_{0}^{d} dz_0\,\mathcal F(z_0) \\[4pt]
&=
\frac{-1}{k_z\mathcal D}
\left(\sigma+\frac{\omega}{k_z}\right) \\
&\quad\times
\Big[
\left(k_z\sigma+2\omega\right)\left(e^{ik_z d}-1\right) \\
&\qquad
-k_z\sigma e^{2ik_z d}\left(1-e^{-ik_z d}\right)
\Big].
\end{aligned}
\label{eq:intJK-sectorB}
\end{equation}

\subsubsection{Contribution from $z_0>d$}
\label{app:sectorC}

\noindent For sources located in the region $z_0>d$, the reduced Green's function takes the form
\begin{equation}
\tilde G(z,z_0)=
\begin{cases}
W_0 e^{ik_z z}+X_0(z_0)e^{-ik_z z}, & z<0,\\[2pt]
U_0(z_0)e^{ik_z z}+V_0(z_0)e^{-ik_z z}, & 0<z<d ,
\end{cases}
\end{equation}
with $W_0=0$ from the radiation condition as $z\to -\infty$. Hence
\begin{equation}
\begin{aligned}
\partial_z \tilde G(0^-,z_0) &= - i k_z\,X_0(z_0),\\
\partial_z \tilde G(0^+,z_0) &= i k_z\!\left[U_0(z_0)-V_0(z_0)\right].
\end{aligned}
\end{equation}
The corresponding contributions to the field derivatives are
\begin{equation}
\begin{aligned}
\left.e'(0^-)\right|_{z_0>d}
&= -ik_z j_0\!\int_{d}^{\infty}\!dz_0\, \,X_0(z_0),\\
\left.e'(0^+)\right|_{z_0>d}
&= i k_z j_0\!\int_{d}^{\infty}\!dz_0\,\!\left[U_0(z_0)-V_0(z_0)\right].
\end{aligned}
\end{equation}
In the coefficient $X_0(z_0)$, which is given in Eq.~\ref{eq:X0}, the prefactor is independent of $z_0$. Therefore
\begin{equation}
\int_{d}^{\infty} X_0(z_0)\,dz_0
=
\frac{2i\omega^2}{k_z\mathcal D}
\int_{d}^{\infty} dz_0\,e^{ik_z z_0}.
\end{equation}
For convergence we impose the retarded prescription
$k_z\to k_z+i0^+$ (so $\mathrm{Im}\,k_z>0$), yielding
\begin{equation}\label{eq: int_d_inf_e^ik_zz 0}
\int_{d}^{\infty} e^{ik_z z_0}\,dz_0
=
\left[\frac{e^{ik_z z_0}}{ik_z}\right]_{d}^{\infty}
=
\frac{i\,e^{ik_z d}}{k_z}.
\end{equation}
Hence
\begin{equation}
\int_{d}^{\infty} X_0(z_0)\,dz_0
=
-\frac{2\omega^2}{k_z^2\mathcal D}\,
e^{ik_z d}.
\label{eq:intW-sectorC}
\end{equation}
Next consider
\begin{equation}
\int_{d}^{\infty} dz_0\,
\bigl[U_0(z_0)-V_0(z_0)\bigr],
\end{equation}
where $U_0$ and $V_0$ are given by Eq.~\ref{eq:U0} and \ref{eq:V0}, respectively. We have
\begin{equation}
\begin{aligned}
\int_{d}^{\infty} 
\bigl[U_0(z_0)-V_0(z_0)\bigr] \,dz_0
&=
i\omega
\left[
\frac{-\sigma}{\mathcal D}
-
\frac{k_z\sigma+2\omega}{k_z\mathcal D}
\right]  \\
&\quad\times
\int_{d}^{\infty} e^{ik_z z_0}\,dz_0.
\end{aligned}
\end{equation}
Using again Eq.~\ref{eq: int_d_inf_e^ik_zz 0}, we obtain

\begin{equation}
\begin{aligned}
\int_{d}^{\infty} \bigl[U_0(z_0)-V_0(z_0)\bigr] \,dz_0
&=
\frac{2\omega}{\mathcal D}
\left(\sigma+\frac{\omega}{k_z}\right)
\frac{e^{ik_z d}}{k_z}.
\end{aligned}
\label{eq:intTU-sectorC}
\end{equation}

\subsection{Final Expressions for $e'(0^\pm)$}

\noindent Collecting the contributions from the three spatial regions,
the derivatives of the electric field at the left interface can be written as
\begin{equation}
\begin{aligned}
e'(0^-)
&= j_0\, i k_z
\Big[
\int_{-\infty}^{0} (C_0-D_0)\,dz_0 \\
&\quad
- \int_{0}^{d} J_0\,dz_0
- \int_{d}^{\infty} X_0\,dz_0
\Big]  \\
&\equiv j_0\, i k_z\,\mathcal I_- ,
\end{aligned}
\end{equation}

\begin{equation}
\begin{aligned}
e'(0^+)
&= j_0\, i k_z
\Big[
\int_{-\infty}^{0} (E_0-F_0)\,dz_0 \\
&\quad
+ \int_{0}^{d} (K_0-L_0)\,dz_0
+ \int_{d}^{\infty} (U_0-V_0)\,dz_0
\Big]  \\
&\equiv j_0\, i k_z\,\mathcal I_+ .
\end{aligned}
\end{equation}

\noindent The electric field at the interface follows from the Green's–function representation
\begin{equation}
\begin{aligned}
e(0)
&= j_0\int_{-\infty}^{\infty} dz_0\,\tilde G(0^+,z_0) \\
&= j_0
\Big[
\int_{-\infty}^{0} (E_0+F_0)\,dz_0  \\
&\quad
+ \int_{0}^{d} (K_0+L_0)\,dz_0
+ \int_{d}^{\infty} (U_0+V_0)\,dz_0
\Big],
\end{aligned}
\end{equation}
where the branch $0^+$ has been used; the same result follows from
the $0^-$ branch due to the continuity of $E_x$ across the interface.

\paragraph*{Region $z_0<0$:}

Combining the coeffecients $E_0$ and $F_0$ gives

\begin{equation}
\begin{aligned}
E_0+F_0
&= i\omega
\Big[
\frac{k_z\sigma+2\omega}{k_z\mathcal D} \\
&\quad
-\frac{\sigma e^{2ik_z d}}
{\mathcal D}
\Big] e^{-ik_z z_0}.
\end{aligned}
\end{equation}

\noindent With the retarded prescription in Eq.~\ref{eq:neg_inf_zero_first}, we obtain
\begin{equation}
\begin{aligned}
\int_{-\infty}^{0} (E_0+F_0)\,dz_0
&=
-\frac{\omega}{k_z\mathcal D}
\Bigg[
\frac{k_z\sigma+2\omega}{k_z}
-\sigma e^{2ik_z d}
\Bigg].
\end{aligned}
\end{equation}

\paragraph*{Region $0<z_0<d$:}

Combining the coefficients $K_0$ and $L_0$ gives

\begin{equation}
\begin{aligned}
K_0+L_0
&=
\frac{i\omega}{k_z\mathcal D}
\Big[
-k_z\sigma e^{2ik_z d}
\\
&\qquad
+
(k_z\sigma+2\omega)e^{2ik_z z_0}
\Big]
e^{-ik_z z_0}.
\end{aligned}
\end{equation}

\noindent Performing the integration yields

\begin{equation}
\begin{aligned}
\int_{0}^{d} (K_0+L_0)\,dz_0
&=
\frac{-\omega(e^{ik_z d}-1)}
{k_z^2\mathcal D}  \\
&\quad\times
\Big[
k_z\sigma(e^{ik_z d}-1)
-2\omega
\Big].
\end{aligned}
\end{equation}

\paragraph*{Region $z_0>d$:}

Combining $U_0$ and $V_0$,
one obtains

\begin{equation}
\begin{aligned}
\int_{d}^{\infty}(U_0+V_0)\,dz_0
&=
-\frac{2\omega^{2}}{k_z^{2}\mathcal D}
\,e^{ik_z d}. 
\end{aligned}
\end{equation}

\subsection{Right Sheet Derivatives $e'(d^\pm)$ }

\subsubsection{Contribution from $z_0<0$ }

For $0<z<d$ the Green's function takes the form
\begin{equation}
\tilde G = E_0 e^{ik_z z}+F_0 e^{-ik_z z},
\end{equation}
which gives
\begin{equation}
\partial_z \tilde G(d^-,z_0)
=
ik_z\!\left[E_0(z_0)e^{ik_z d}-F_0(z_0)e^{-ik_z d}\right].
\end{equation}

\noindent For $z>d$ one has
\begin{equation}
\tilde G = G_0 e^{ik_z z},
\end{equation}
leading to
\begin{equation}
\partial_z \tilde G(d^+,z_0)
=
ik_z\,G_0(z_0)e^{ik_z d}.
\end{equation}

\noindent The contribution of this sector to the derivatives of the electric field is therefore
\begin{equation}
\begin{aligned}
e'(d^-)|_{z_0<0}
&= j_0 \int_{-\infty}^{0} dz_0\,
ik_z\!\left[E_0(z_0)e^{ik_z d}-F_0(z_0)e^{-ik_z d}\right], \\
e'(d^+)|_{z_0<0}
&= j_0 \int_{-\infty}^{0} dz_0\,
ik_z\,G_0(z_0)e^{ik_z d}.
\end{aligned}
\end{equation}

\noindent The coefficients $E_0$ and $F_0$ can be written as
\begin{equation}
E_0(z_0)=C_E e^{-ik_z z_0},\qquad
F_0(z_0)=C_F e^{2ik_z d}e^{-ik_z z_0},
\end{equation}
where

\begin{equation}
C_E=\frac{i\omega(k_z\sigma+2\omega)}{k_z\mathcal D},
\qquad
C_F=\frac{-i\omega\sigma}{\mathcal D}.
\end{equation}

\noindent Using the retarded prescription in Eq. ~\ref{eq:neg_inf_zero_first},
one obtains
\begin{equation}
\begin{aligned}
e'(d^-)|_{z_0<0}
&=
j_0\left(C_F e^{2ik_z d}e^{-ik_z d}
-C_E e^{ik_z d}\right) \\
&=
-\frac{2i\omega j_0}{\mathcal D}
\left(\sigma+\frac{\omega}{k_z}\right)
e^{ik_z d}.
\end{aligned}
\end{equation}

\noindent For the derivative on the $d^+$ side we write
\begin{equation}
G_0(z_0)=\mathcal G_0 e^{-ik_z z_0},
\end{equation}
where
\begin{equation}
\mathcal G_0=\frac{2i\omega^2}{k_z\mathcal D}.
\end{equation}

Performing the integral gives
\begin{equation}
\begin{aligned}
e'(d^+)|_{z_0<0}
&=
j_0 i k_z e^{ik_z d}
\int_{-\infty}^{0} G(z_0)dz_0 \\
&=
-j_0 \mathcal G_0 e^{ik_z d}.
\end{aligned}
\end{equation}

\subsubsection{Contribution from $0<z_0<d$}

\noindent For $z_0<z<d$, the Green's function is
\begin{equation}
\tilde G = M_0 e^{ik_z z}+N_0 e^{-ik_z z},
\end{equation}
which gives
\begin{equation}
\partial_z \tilde G(d^-,z_0)
=
ik_z\!\left[M_0(z_0)e^{ik_z d}-N_0(z_0)e^{-ik_z d}\right].
\end{equation}
For $z>d$ one has
\begin{equation}
\tilde G(d^+,z_0)=O_0(z_0)e^{ik_z d},
\end{equation}
leading to
\begin{equation}
\partial_z \tilde G(d^+,z_0)
=
ik_z O_0(z_0)e^{ik_z d}.
\end{equation}

\noindent The contribution of this sector to the electric-field derivatives is therefore
\begin{equation}
\begin{aligned}
e'(d^-)\big|_{0<z_0<d}
&=
j_0 i k_z
\int_0^d
\Bigl[
M_0(z_0)e^{ik_z d}
\\
&\qquad\qquad
-
N_0(z_0)e^{-ik_z d}
\Bigr]
\,dz_0, \\
e'(d^+)\big|_{0<z_0<d}
&=
j_0 i k_z
\int_0^d
O_0(z_0)e^{ik_z d}
\,dz_0.
\end{aligned}
\end{equation}

\noindent The coefficients $M_0$ and $N_0$ can be written as
\begin{equation}
\begin{aligned}
M_0(z_0)
&=
\frac{i(k_z\sigma+2\omega)}
{2k_z\mathcal D}
\,\mathcal M(z_0)e^{-ik_z z_0}, \\
N_0(z_0)
&=
\frac{-i\sigma}{2\mathcal D}
\,\mathcal M(z_0)e^{ik_z(2 d- z_0)},
\end{aligned}
\end{equation}
where
\begin{equation}
\mathcal M(z_0)
= k_z\sigma\left(1-e^{2ik_z z_0}\right)
+2\omega .
\end{equation}

\noindent Combining the two contributions gives

\begin{equation}
M_0 e^{ik_z d}-N_0 e^{-ik_z d}=
\frac{i}{\mathcal D}
\left(
\sigma+\frac{\omega}{k_z}
\right)
\mathcal M(z_0)\,e^{ik_z(d-z_0)} .
\end{equation}

\noindent Substituting this into the expression for $e'(d^-)$ yields
\begin{equation}
\begin{aligned}
e'(d^-)|_{0<z_0<d}
&=-\frac{j_0(k_z\sigma+\omega)}
{\mathcal D}
\int_0^d dz_0\,
\mathcal M(z_0)
e^{ik_z(d-z_0)} .
\end{aligned}
\end{equation}
Using
\begin{equation}
\mathcal B(z_0)e^{ik_z(d-z_0)}
=
e^{ik_z d}
\left[
-k_z\sigma e^{ik_z z_0}
+(k_z\sigma+2\omega)e^{-ik_z z_0}
\right],
\end{equation}
the remaining integral becomes
\begin{equation}
\begin{aligned}
\int_0^d dz_0\,\mathcal M(z_0)e^{ik_z(d-z_0)}
&=
e^{ik_z d}
\Big[
-k_z\sigma\!\int_0^d e^{ik_z z_0}dz_0  \\
&\quad
+(k_z\sigma+2\omega)
\!\int_0^d e^{-ik_z z_0}dz_0
\Big].
\end{aligned}
\end{equation}

\noindent Using Eq.~\ref{eq: 0d_integrals}
one finally obtains
\begin{equation}
\begin{aligned}
e'(d^-)|_{0<z_0<d}
&=
i j_0
\frac{k_z\sigma+\omega}
{k_z\mathcal D}  \\
&\quad\times
(e^{ik_z d}-1)
\left[(k_z\sigma+2\omega)-k_z\sigma e^{ik_z d}\right].
\end{aligned}
\end{equation}

\noindent For the derivative on the $d^+$ side we write
\begin{equation}
O_0(z_0)
=
\frac{i\omega}{k_z\mathcal D}
\left[
-k_z\sigma e^{2ik_z z_0}
+(k_z\sigma+2\omega)
\right]e^{-ik_z z_0}.
\end{equation}
Expanding gives
\begin{equation}
O_0(z_0)
=
o_+ e^{ik_z z_0}+o_- e^{-ik_z z_0},
\end{equation}
where
\begin{equation}
o_+=-\frac{i\omega\sigma}{\mathcal D},
\qquad
o_-=\frac{i\omega(k_z\sigma+2\omega)}
{k_z\mathcal D}.
\end{equation}

\noindent Performing the integration yields
\begin{equation}
\begin{aligned}
e'(d^+)|_{0<z_0<d}
&=
j_0 e^{ik_z d}
\Big[
o_+(e^{ik_z d}-1)  \\
&\quad
+o_-(1-e^{-ik_z d})
\Big].
\end{aligned}
\end{equation}
Substituting the coefficients gives
\begin{equation}
\begin{aligned}
e'(d^+)|_{0<z_0<d}
&=
j_0 e^{ik_z d}
\Bigg[
-\frac{i\omega\sigma}{\mathcal D}
(e^{ik_z d}-1) \\
&\quad
+\frac{i\omega(k_z\sigma+2\omega)}
{k_z\mathcal D}
(1-e^{-ik_z d})
\Bigg].
\end{aligned}
\end{equation}

\subsubsection{Contribution from $z_0>d$}
For $0<z<d$ the Green's function takes the form
\begin{equation}
\tilde G = U_0 e^{ik_z z}+V_0 e^{-ik_z z},
\end{equation}
which implies
\begin{equation}
\partial_z \tilde G(d^-,z_0)
=
ik_z\!\left[
U_0(z_0)e^{ik_z d}-V_0(z_0)e^{-ik_z d}
\right].
\end{equation}
For $z>d$ one has
\begin{equation}
\tilde G = S_0 e^{ik_z z}+T_0 e^{-ik_z z},
\end{equation}
leading to
\begin{equation}
\partial_z \tilde G(d^+,z_0)
=
ik_z\!\left[
S_0(z_0)e^{ik_z d}-T_0(z_0)e^{-ik_z d}
\right].
\end{equation}
The contribution of this sector to the field derivatives is therefore
\begin{equation}
\begin{aligned}
e'(d^-)|_{z_0>d}
&=
j_0 i k_z
\int_d^{\infty} dz_0
\left[
U_0(z_0)e^{ik_z d}-V_0(z_0)e^{-ik_z d}
\right], \\
e'(d^+)|_{z_0>d}
&=
j_0 i k_z
\int_d^{\infty} dz_0
\left[
S_0(z_0)e^{ik_z d}-T_0(z_0)e^{-ik_z d}
\right].
\end{aligned}
\end{equation}
The coefficients share a common factor
$e^{ik_z z_0}$, so that
\begin{equation}
\begin{aligned}
U_0 &= u_0 e^{ik_z z_0}, \qquad
V_0 = v_0 e^{ik_z z_0}, \\
S_0 &= s_0 e^{ik_z z_0}, \qquad
T_0 = t_0 e^{ik_z z_0}.
\end{aligned}
\end{equation}
where the coefficients $u_0,v_0,s_0,t_0$ are independent of $z_0$.
The relevant elementary integral is
\begin{equation}\label{eq: integral_d_inf}
\int_d^{\infty} e^{ik_z z_0}dz_0
=
-\frac{e^{ik_z d}}{ik_z},
\end{equation}
obtained using the retarded prescription $\mathrm{Im}\,k_z>0$. 
Substituting the expressions for the coefficients and performing the integration yields
\begin{equation}
\begin{aligned}
e'(d^+)|_{z_0>d}
&=
i j_0
\Bigg[
\frac{\sigma}{2\mathcal D}
\Big(
k_z\sigma(1-e^{2ik_z d})  \\
&\qquad
+2\omega(e^{2ik_z d}+1)
\Big)
-\frac{1}{2k_z}
\Bigg].
\end{aligned}
\end{equation}

\noindent For the second derivative we obtain
\begin{equation}
e'(d^-)|_{z_0>d}
=
j_0 ik_z
\int_d^{\infty} 
\left( U_0 e^{ik_z d}
- V_0 e^{-ik_z d}
\right)\,dz_0.
\end{equation}

\noindent Using the explicit forms of $U_0$ and $V_0$ gives
\begin{equation}
\begin{aligned}
ik_z U_0 e^{ik_z d}
&=
\frac{\omega k_z \sigma}{\mathcal D}
e^{ik_z(z_0+d)},
\\
-ik_z V_0 e^{-ik_z d}
&=
\frac{\omega(k_z\sigma+2\omega)}{\mathcal D}
e^{ik_z(z_0-d)}.
\end{aligned}
\end{equation}

Factoring $e^{ik_z z_0}$ yields
\begin{equation}
\begin{aligned}
e'(d^-)|_{z_0>d}
&=
j_0
\int_d^{\infty} dz_0\, e^{ik_z z_0}
\Bigg[
\frac{k_z\omega\sigma}{\mathcal D}
e^{ik_z d}
\\
&\qquad
+\frac{\omega(k_z\sigma+2\omega)}{\mathcal D}
e^{-ik_z d}
\Bigg].
\end{aligned}
\end{equation}

\noindent Finally, performing the integration gives

\begin{equation}
\begin{aligned}
e'(d^-)|_{z_0>d}
&=
\frac{i\omega j_0}{\mathcal D}
\left[
\sigma\left(1+e^{2ik_z d}\right)
+\frac{2\omega}{k_z}
\right]. 
\end{aligned}
\end{equation}

\subsection{Final Expressions for $e'(d^\pm)$}

We now combine the contributions obtained in the previous
subsections. The derivative of the field at the $z=d$ interface
can be written as the sum of the three spatial sectors,
\begin{equation}
e'(d^-)
=
\left.e'(d^-)\right|_{z_0<0}
+
\left.e'(d^-)\right|_{0<z_0<d}
+
\left.e'(d^-)\right|_{z_0>d}.
\end{equation}

\noindent Collecting the corresponding expressions yields

\begin{equation}
\begin{aligned}
e'(d^-)
=\frac{i j_0}{k_z\mathcal D}
\Bigg[
&\omega k_z\sigma\,e^{2ik_z d}
+\omega(k_z\sigma+2\omega)
\\
&+(k_z\sigma+\omega)(e^{ik_z d}-1)
\\
&\times
\left(
k_z\sigma+2\omega
-k_z\sigma e^{ik_z d}
\right)
\\
&-\omega
\left(
2k_z\sigma+2\omega
\right)
e^{ik_z d}
\Bigg].
\end{aligned}
\end{equation}
\noindent Similarly,
\begin{equation}
e'(d^+)
=
\left.e'(d^+)\right|_{z_0<0}
+
\left.e'(d^+)\right|_{0<z_0<d}
+
\left.e'(d^+)\right|_{z_0>d},
\end{equation}
from which we obtain

\begin{equation}
\begin{aligned}
e'(d^+)
=
-i j_0
\left[
\frac{1}{2k_z}
+\frac{1}{\mathcal D}
\left(
\frac{2\omega^2}{k_z}
-2\omega\sigma e^{ik_z d}
\right.\right.
\\
\left.\left.
\qquad\qquad
+\frac{k_z\sigma^2}{2}
\left(e^{ik_z d}-1\right)
\left(e^{ik_z d}+1\right)
\right)
\right].
\end{aligned}
\end{equation}

\noindent The electric field evaluated at the surface can also be written
by collecting the Green's–function contributions from the three
regions,
\begin{equation}
\begin{aligned}
e(d)
&\equiv e_s(d)
\\
&=
j_0
\Bigg[
\int_{-\infty}^{0}
dz_0\,
G_0(z_0)\,e^{ik_z d}
\\
&\quad
+
\int_{0}^{d}
dz_0\,
\big(
M_0(z_0)e^{ik_z d}
+
N_0(z_0)e^{-ik_z d}
\big)
\\
&\quad
+
\int_{d}^{\infty}
dz_0\,
\big(
U_0(z_0)e^{ik_z d}
+
V_0(z_0)e^{-ik_z d}
\big)
\Bigg].
\end{aligned}
\end{equation}

\noindent Using the standard integrals obtained with the retarded
prescription ($\mathrm{Im}\,k_z>0$) in  Eq.~\ref{eq: 0d_integrals}, Eq.~\ref{eq:neg_inf_zero_first} and Eq.~\ref{eq: integral_d_inf},
the three contributions can be evaluated explicitly.

For $z_0<0$, using Eq.~\ref{eq:G} for 
$G_0(z_0)$,
we obtain
\begin{equation}
e_{<0}(d)
=
-\frac{2 j_0\omega^2}
{k_z^2\mathcal D}
\,e^{ik_z d}.
\end{equation}

\noindent For the region $0<z_0<d$ the finite-interval integral reduces to

\begin{equation}
e_{(0,d)}(d)
=
\frac{-j_0\omega}
{\mathcal D\,k_z^2}
(1-e^{ik_z d})
\left(
k_z\sigma \left( 1- e^{ik_z d} \right)+2\omega
\right).
\end{equation}

\noindent Finally, for $z_0>d$ we find

\begin{equation}
e_{>d}(d)
=
\frac{j_0\omega}{k_z^2\mathcal D}
\left[
k_z\sigma e^{2ik_z d}
-
\left(k_z\sigma+2\omega\right)
\right].
\end{equation}

\noindent Adding the three pieces and simplifying yields the final result

\begin{equation}
\begin{aligned}
e(d)
&=
\frac{2j_0\omega}{\mathcal D k_z^2}
\left[
k_z\sigma\left(e^{ik_z d}-1\right)
-2\omega
\right].
\end{aligned}
\end{equation}

\subsection{Auxiliary Functions}
\label{app:auxilary_Green_functions}
\begin{equation}
\begin{aligned}
\mathcal I_+
&= -\frac{1}{\mathcal D}
\Bigg[
\frac{\omega}{k_z}
\left(
\frac{k_z\sigma+2\omega}{k_z}
+\sigma e^{2ik_z d}
\right)
\\
&\quad
+\frac{\omega e^{ik_z d}}{k_z}
\left(
\sigma+\frac{k_z\sigma+2\omega}{k_z}
\right)
\\
&\quad
+\frac{\sigma+\omega/k_z}{k_z}
\,(e^{ik_z d}-1)
\big[k_z\sigma(1-e^{ik_z d})+2\omega\big]
\Bigg].
\end{aligned}
\end{equation}

\begin{equation}
\begin{aligned}
\mathcal I_-
&= \frac{1}{2k_z^2}
-\frac{1}{\mathcal D}
\Bigg[
\frac{\sigma}{2k_z}
\left(
k_z\sigma(1-e^{2ik_z d})
+2\omega(1+e^{2ik_z d})
\right)
\\
&\quad
+\frac{\omega}{k_z^2}
(e^{ik_z d}-1)
\big[k_z\sigma(1-e^{ik_z d})+2\omega\big]
\\
&\quad
-\frac{2\omega^2}{k_z^2}\,e^{ik_z d}
\Bigg].
\end{aligned}
\end{equation}

\noindent We introduce the quantity $\mathcal{S}$ that enters the surface-field factor
\begin{equation}
\begin{aligned}
\mathcal{S}
={}&
\int_{-\infty}^{0}(E_0+F_0)\,dz_0
+\int_{0}^{d}(K_0+L_0)\,dz_0 \\
&\quad
+\int_{d}^{\infty}(U_0+V_0)\,dz_0 .
\end{aligned}
\end{equation}
such that
\begin{equation}
e_s(0)=j_0\,\mathcal{S}.
\end{equation}

\noindent The three contributions evaluate to
\begin{equation}
\begin{aligned}
\mathcal{S}
&=
-\frac{1}{\mathcal D}
\Bigg[
\frac{\omega}{k_z}
\left(
\frac{k_z\sigma+2\omega}{k_z}
-\sigma e^{2ik_z d}
\right)
\\
&\quad
+\frac{\omega(e^{ik_z d}-1)}{k_z^2}
\left(
k_z\sigma(e^{ik_z d}-1)-2\omega
\right)
\\
&\quad
+\frac{2\omega^2}{k_z^2}e^{ik_z d}
\Bigg].
\end{aligned}
\end{equation}

\section{Effective Conductivity Description of Stacked Graphene}
\label{app:multilayer}

We model a stack of $N$ electronically decoupled graphene sheets by assuming that the layers are separated by distances much smaller than the relevant electromagnetic wavelength, so that each layer experiences essentially the same electric field. Neglecting interlayer tunnelling and hybridisation, linear response implies that the induced surface currents add linearly, yielding the effective sheet conductivity 
\begin{equation}
\sigma_{\mathrm{eff}}=N \sigma.
\end{equation}
This is analogous to $N$ identical resistors connected in parallel, for which the total conductance is the sum of the individual conductances. The reflection coefficient becomes
\begin{equation}
r_N = -\frac{N\sigma}{2+N\sigma}.
\end{equation}
\noindent For the parameter range relevant to the strongest projected sensitivities in Fig.~\ref{fig:sensitivity_gamma_comparison} (low-loss regime), the conductivity is predominantly reactive ($|\operatorname{Im} \sigma| \gg|\operatorname{Re} \sigma|$), justifying the approximation $\sigma \simeq i \sigma_I$. Therefore,
\begin{equation}
r_N \simeq -\frac{iN\sigma_I}{2+iN\sigma_I}.
\end{equation}
Taking the modulus gives
\begin{equation}
|r_N|
=
\frac{N|\sigma_I|}{\sqrt{4+N^2\sigma_I^2}},
\end{equation}
which may be equivalently written as 
\begin{equation}
|r_N|
=
\left(1+\frac{4}{N^2\sigma_I^2}\right)^{-1/2}.
\end{equation}
For \(N|\sigma_I|\gg 2\), we expand
\begin{equation}
|r_N|
\simeq
1-\frac{2}{N^2\sigma_I^2}.
\end{equation}
The linewidth parameter is
\begin{equation}
\gamma_N=-\ln |r_N|.
\end{equation}

\noindent Using \(-\ln(1-x)\simeq x\), we obtain
\begin{equation}
\gamma_N
\simeq
\frac{2}{N^2\sigma_I^2},
\end{equation}
and therefore,
\begin{equation}
\gamma_N \propto N^{-2}.
\end{equation}

\subsection{Scaling of the Resonant Pressure with the Number of Graphene Layers}
Recall the Lorentzian approximation at resonance, given in Eq.~\ref{eq:P_res high doping}. At large effective conductivity $\sigma_{\mathrm{eff}}$ corresponding to a stack of $N$ electronically decoupled graphene sheets,
the quantities entering the resonant pressure exhibit simple asymptotic scaling. From Eqs.~\ref{eq:85}, \ref{eq:90} and \ref{eq:100}, the auxiliary function satisfies
\begin{equation}
\mathcal{A}\propto N,
\end{equation}
while the dominant terms in \(\mathcal{T}\) are quadratic in the conductivity,
\begin{equation}
\mathcal{T}= \mathcal{O}(N^2).
\end{equation}
Consequently,
\begin{equation}
\mathcal{N}_n
=
-2\,\mathrm{Re}
\!\left[
\sigma(\sigma x_n-\sigma-2)\mathcal{T}_n^{*}
\right]
=\mathcal{O} (N^4).
\end{equation}

\noindent Similarly, we obtain, in the large-\(N\) limit,
\begin{equation}
\alpha = \mathcal{O} (N^2),
\end{equation}
and therefore
\begin{equation}
|\alpha|^2 =\mathcal{O}( N^4).
\end{equation}

\noindent Since the resonant pressure is proportional to
\begin{equation}
P_{\rm res}
\propto
\frac{\mathcal{N}_n}{|\alpha|^2\gamma_n^2},
\end{equation}
and the resonance linewidth scales as
\begin{equation}
\gamma_n\propto N^{-2},
\end{equation}
it follows that at resonance, the pressure increases quartically with the number of stacked graphene layers
\begin{equation}
P_{\rm res}
\propto
N^4.
\end{equation}

\noindent To illustrate this scaling numerically, Fig.~\ref{fig:combined} compares the resonant pressure for effective graphene stacks with $N=10^2$ and $N=10^3$. As expected from the analytical result above, increasing the number of layers by one order of magnitude enhances the peak pressure by four orders of magnitude.

\begin{figure}[h]
    \centering
    \includegraphics[width=1\linewidth]{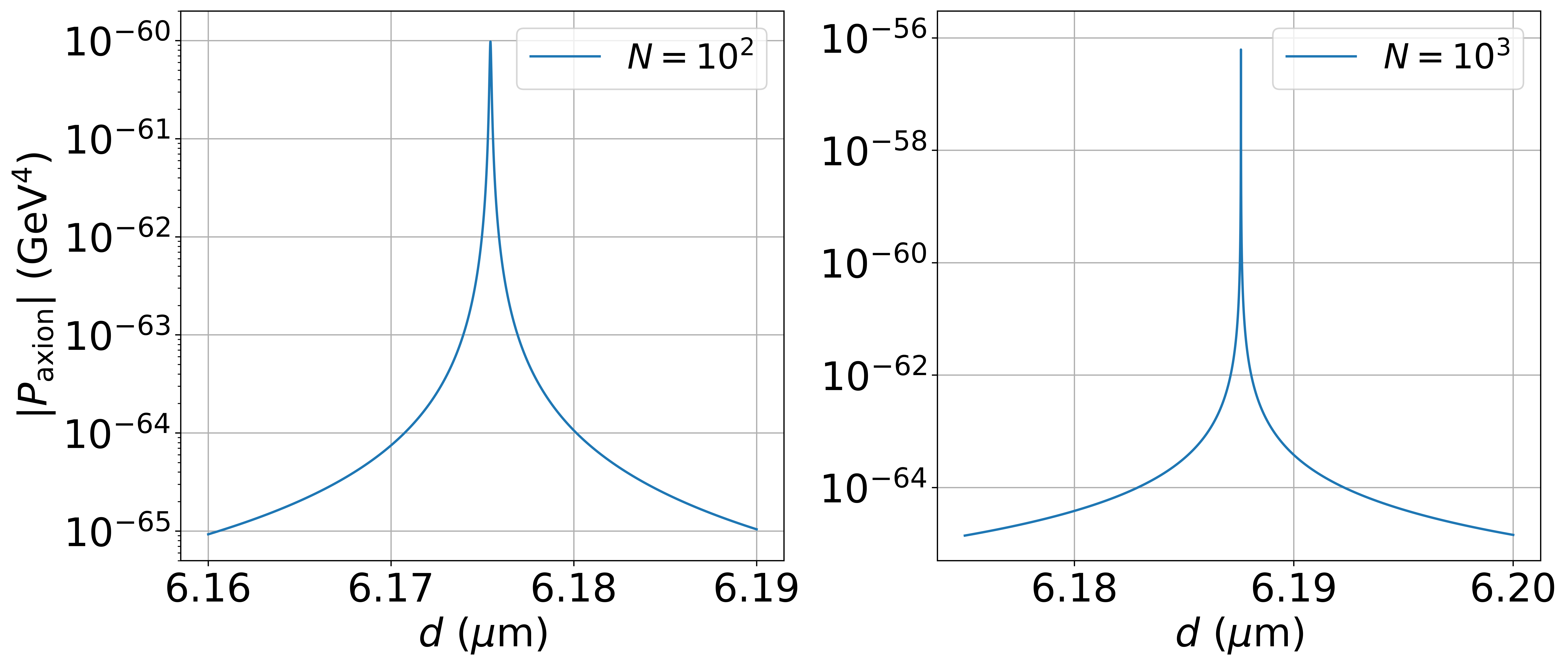}
    \caption{Comparison of the resonant pressure near a cavity resonance for effective graphene stacks with $N=10^2$ (left) and $N=10^3$ (right), with $m_a=0.1\,\mathrm{eV}$, $\Gamma=10^{-5}\,\mathrm{eV}$, $T=4\,\mathrm{K}$ and $\mu=10\,\mathrm{eV}$.}
    \label{fig:combined}
\end{figure}

\subsection{Relation to Large Chemical Potentials}
The low-temperature conductivity shown in Fig.~\ref{fig:mag_sigma_vs_T} motivates interpreting the large effective conductivities in terms of engineered multilayer graphene stacks. At $T=4\,\mathrm{K}$, the conductivity is approximately temperature independent and grows roughly linearly with the chemical potential,
\begin{equation}
|\sigma| \propto \mu .
\end{equation}

\noindent This behaviour is expected from the low-temperature intraband (Drude) contribution to the graphene conductivity, for which

\begin{equation}
\sigma_{\mathrm{intra}}
\sim
\frac{i\mu}{\omega+i\Gamma},
\end{equation}
\noindent so that, for fixed frequency and scattering rate, $|\sigma|\propto\mu$.

\noindent As discussed above, the effective conductivity also scales linearly with the number of layers,
\begin{equation}
\sigma_{\mathrm{eff}}\propto N.
\end{equation}

\noindent Combining the two scalings gives
\begin{equation}
\sigma_{\rm eff}\propto N\mu,
\end{equation}
suggesting that a stack of graphene sheets with chemical potential $\mu_0$ approximately reproduces the conductivity of a single graphene sheet with an effective chemical potential
\begin{equation}
\mu_{\rm eff}\simeq N\mu_0.
\end{equation}
For a single-layer doping of $\mu_0\simeq10\,\mathrm{eV}$, this corresponds to
\begin{equation}
\begin{aligned}
N &= 10^2 \qquad &\Rightarrow\qquad \mu_{\rm eff} &\sim 10^3\,\mathrm{eV},\\
N &= 10^3 \qquad &\Rightarrow\qquad \mu_{\rm eff} &\sim 10^4\,\mathrm{eV}.
\end{aligned}
\end{equation}

\noindent Consequently, the benchmark multilayer projections considered in this work may be interpreted in terms of enhanced effective conductivities arising from engineered stacks of electronically decoupled graphene sheets. This approach retains realistic single-layer doping while reproducing the large conductivities responsible for the enhanced resonant response.

\subsection{Validity and Mass Dependence}

The effective-conductivity model presented above provides only an approximate description of multilayer graphene. In particular, the graphene conductivity depends not only on the chemical potential, but also on the probing frequency, or equivalently the axion mass. In general, there is no universal correspondence between the number of graphene layers and an effective chemical potential. Instead, the equivalent number of layers required to reproduce the conductivity of a reference graphene sheet with chemical potential $\mu_0$ is, in principle, mass dependent,
\begin{equation}
N_{\rm eff}(m_a)\simeq
\frac{|\sigma(m_a,\mu_{\rm eff})|}
{|\sigma(m_a,\mu_0)|},
\end{equation}
where $\mu_0$ denotes the chemical potential of an individual graphene sheet. However, throughout the low-mass regime relevant to this work, approximately up to $m_a\sim\mathcal{O}(1\,\mathrm{eV})$, the conductivity is dominated by the intraband response and depends approximately linearly on the chemical potential, $|\sigma|\propto\mu$. 
Within this regime, the mass dependence largely cancels in the ratio above, yielding the simple approximation $
\mu_{\rm eff}\simeq N\mu_0.$
Thus, for the range of axion masses considered in our exclusion analysis, a stack of electronically decoupled graphene sheets provides a good approximation to the large effective conductivities obtained by extrapolating to very large chemical potentials.


\bibliographystyle{unsrt}
\bibliography{prl}


\end{document}